\documentclass[a4paper,fleqn,usenatbib]{mnras}
\usepackage{newtxtext,newtxmath}
\usepackage[T1]{fontenc}
\usepackage{ae,aecompl,bm,booktabs}
\usepackage{amsmath,color,xspace,multirow,longtable,graphicx,bigints,comment}
\usepackage{tabularx}

\newcommand\textlcsc[1]{\textsc{\MakeLowercase{#1}}}
\newcommand{\cii}{[C\,II]\xspace}
\newcommand{\BB}{$^{\rm 3D}$Barolo\xspace}

\title[Kinematic Diversity \& Rotation in SFGs at $z\sim4-6$]{The ALPINE-ALMA [CII] Survey: Kinematic Diversity \& Rotation in Massive Star Forming Galaxies at $\mathbf{z\sim4.4-5.9}$}

\author[G. C. Jones et al.]{
G. C. Jones$^{1,2}$\thanks{E-mail: gj283@cam.ac.uk},
D. Vergani$^{3}$,
M. Romano$^{4,5}$,
M. Ginolfi$^{6}$,
Y. Fudamoto$^{7,8}$,
M. B\'{e}thermin$^{9}$,\newauthor
S. Fujimoto$^{10,11}$,
B. C. Lemaux$^{12}$,
L. Morselli$^{4,5}$,
P. Capak$^{13}$,
P. Cassata$^{4}$,
A. Faisst$^{13}$,
O. Le F\`evre$^{9}$,\newauthor
D. Schaerer$^{14}$,
J. D. Silverman$^{15,16}$,
Lin Yan$^{17}$,
M. Boquien$^{18}$,
A. Cimatti$^{19,20}$,
M. Dessauges-Zavadsky$^{14}$,\newauthor
E. Ibar$^{21}$,
R. Maiolino$^{1,2}$,
F. Rizzo$^{10,11}$,
M. Talia$^{3,19}$,
G. Zamorani$^{3}$
\\
$^{1}$Cavendish Laboratory, University of Cambridge, 19 J. J. Thomson Ave., Cambridge CB3 0HE, UK\\
$^{2}$Kavli Institute for Cosmology, University of Cambridge, Madingley Road, Cambridge CB3 0HA, UK\\
$^{3}$INAF - Osservatorio di Astrofisica e Scienza dello Spazio, via Gobetti 93/3 - 40129, Bologna - Italy\\
$^{4}$Dipartimento di Fisica e Astronomia, Universit\'{a} di Padova, vicolo dell'Osservatorio 3, I-35122 Padova, Italy\\
$^{5}$INAF - Osservatorio Astrofisico di Padova, vicolo dell'Osservatorio 5, I-35122 Padova, Italy\\
$^{6}$European Southern Observatory, Karl-Schwarzschild-Str. 2, D-85748, Garching, Germany\\
$^{7}$Research Institute for Science and Engineering, Waseda University, 3-4-1 Okubo, Shinjuku, Tokyo 169-8555, Japan\\
$^{8}$National Astronomical Observatory of Japan, 2-21-1, Osawa, Mitaka, Tokyo, Japan\\
$^{9}$Aix-Marseille Universit\'{e}, CNRS, LAM, Laboratoire d'Astrophysique de Marseille, Marseille, France\\
$^{10}$Cosmic Dawn Center (DAWN), Jagtvej 128, DK2200, Copenhagen N, Denmark\\
$^{11}$Niels Bohr Institute, University of Copenhagen, Lyngbyvej 2, DK2100 Copenhagen \O, Denmark\\
$^{12}$Department of Physics \& Astronomy, University of California, Davis, One Shields Ave., Davis, CA 95616, USA\\
$^{13}$Infrared Processing and Analysis Center, California Institute of Technology, Pasadena, CA 91125, USA\\
$^{14}$Observatoire de Gen\`{e}ve, Universit\'{e} de Gen\`{e}ve, 51 Ch. des Maillettes, 1290 Versoix, Switzerland\\
$^{15}$Kavli Institute for the Physics and Mathematics of the Universe, The University of Tokyo  Kashiwa, Chiba 277-8583, Japan\\
$^{16}$Department of Astronomy, School of Science, The University of Tokyo, 7-3-1 Hongo, Bunkyo, Tokyo 113-0033, Japan\\
$^{17}$ The Caltech Optical Observatories, California Institute of Technology, Pasadena, CA 91125, USA\\
$^{18}$Centro de Astronom\'ia (CITEVA), Universidad de Antofagasta, Avenida Angamos 601, Antofagasta, Chile\\
$^{19}$University of Bologna, Department of Physics and Astronomy (DIFA), Via Gobetti 93/2, I-40129, Bologna, Italy\\
$^{20}$INAF - Osservatorio Astrofisico di Arcetri, Largo E. Fermi 5,I-50125, Firenze, Italy\\
$^{21}$Instituto de F\'{i}sica y Astronom\'{i}a, Universidad de Valpara\'{i}so, Avda. Gran Breta\~{m}a 1111, Valpara\'{i}so, Chile
}
\begin{document}
\label{firstpage}
\pagerange{\pageref{firstpage}--\pageref{lastpage}}
\maketitle

\begin{abstract}
While the kinematics of galaxies up to $z\sim3$ have been characterized in detail, only a handful of galaxies at high redshift ($z>4$) have been examined in such a way. 
The Atacama Large Millimeter/submillimeter Array (ALMA) Large Program to INvestigate [CII] at Early times (ALPINE) survey observed a statistically significant sample of 118 star-forming main sequence galaxies at $z=4.4-5.9$ in [CII]158$\mu$m emission, increasing the number of such observations by nearly 10x. 
A preliminary qualitative classification of these sources revealed a diversity of kinematic types (i.e., rotators, mergers, and dispersion-dominated systems).
In this work, we supplement the initial classification by applying quantitative analyses to the ALPINE data: a tilted ring model (TRM) fitting code (\BB), a morphological classification (Gini-M$_{\rm 20}$), and a set of disk identification criteria.
Of the 75 [CII]-detected ALPINE galaxies, 29 are detected at sufficient significance and spatial resolution to allow for TRM fitting and the derivation of morphological and kinematic parameters. 
These 29 sources constitute a high-mass subset of the ALPINE sample ($M_*>10^{9.5}\,M_{\odot}$).
We robustly classify 14 of these sources (six rotators, five mergers, and three dispersion-dominated systems); the remaining sources showing complex behaviour.  
By exploring the G-M$_{20}$ of $z>4$ rest-frame FIR and [CII] data for the first time, we find that our $1''\sim6$\,kpc resolution data alone are insufficient to separate galaxy types.
We compare the rotation curves and dynamical mass profiles of the six ALPINE rotators to the two previously detected $z\sim4-6$ unlensed main sequence rotators, finding high rotational velocities ($\sim50-250$\,km\,s$^{-1}$) and a diversity of rotation curve shapes.
\end{abstract}

\section{Introduction}\label{intro}

%Stuff's been done from z=0-3
The past century has seen a massive broadening of scope in the study of galaxy kinematics. Early works focused on spectral line emission from single nearby galaxies (e.g., \citealt{peas18,burb59}), while current studies are able to probe the kinematics of many local objects (e.g., 
\citealt{garr02,cons05,debl08,shap08,puec10,gome19,kors19,denb20}) or single objects in much greater detail (e.g., \citealt{cari06,chem09,cram19,nort19,brai20}). This detailed analysis has also been extended to galaxies at intermediate redshift (i.e., $z\sim1-3$; t$_{\rm H}\sim6-2$\,Gyr) with integral field spectroscopy, mostly in the rest-frame optical and near-infrared (e.g., \citealt{fors06,epin12,burk16,turn17,harr17,moli17,swin17,wisn19,loia19}). These studies have revealed galaxies with a variety of kinematic types (i.e., rotators, mergers, dispersion-dominated sources), as well as inflows and outflows. 
%Updated from fors06,epin12,wisn15,burk16,loia19

One major finding from studies up to $z\sim2$ is that most of these galaxies show evidence for significant dark matter halos. By combining KMOS and MUSE data for $\sim1500$ star-forming galaxies (SFGs) at $z\sim0.6-2.2$, \citet{tile19} find that galaxies in this epoch mainly feature flat or rising rotation curves at large radii. This is quite similar to the rotation curves of local galaxies, which show nearly constant or only slightly declining velocities at large radii (e.g., \citealt{debl08}), indicating the additional gravitational force from an underlying dark matter halo (e.g., \citealt{rubi70}).

While a multitude of galaxies at $z<3$ have been well characterized kinematically, it is only recently, with the advent of the Atacama Large Millimeter/submillimeter Array (ALMA) and the Jansky Very Large Array (JVLA), that we are able to characterize the kinematics of high redshift ($z>4$) galaxies, using rest-frame far-infrared (FIR) emission lines. This includes the clumpy nature of $z\sim5-7$ galaxies observed with ALMA in [CII] emission (e.g., \citealt{carn18}), rotational model fitting of $z\sim4-6$ galaxies (e.g., \citealt{debr14,shao17,jone17,rizz20,frat20}), and even evidence for ordered rotation of galaxies at $z>6$ (\citealt{smit18,bakx20}). However, the number of galaxies at $z>4$ that are kinematically characterizable (i.e., observed with great enough S/N and resolution) is still much less than at $z<4$. These high-redshift galaxies represent some of the first galaxies to form in the Universe and sit on the ramp-up of star formation rate density (e.g., \citealt{mada14}), so their morphological and kinematic (i.e., morpho-kinematic) state  is crucial for informing theories of early galaxy formation evolution, as well as constraining ongoing cosmological simulations (e.g., \citealt{pall17,koha19}).

Addressing this lack of kinematically characterized galaxies at $z>4$ was one of the main driving goals of the ALMA Large Program to INvestigate CII at Early Times (ALPINE; \citealt{lefe20,beth20,fais20}). This survey observed 118 galaxies at $z=4.4-5.9$ in [CII]\,158$\mu$m and the surrounding rest-frame FIR emission, following the success of the pilot program \citep{capa15} and increasing the number of such observations by an order of magnitude. Source selection was based on UV luminosity (L$_{\rm UV}>0.6$\,L*), pre-existing spectroscopic redshifts, and lack of type 1 AGN. ALPINE sources lie on the star forming main sequence (e.g., \citealt{noes07,fais20}) and are thus broadly representative of the underlying population of galaxies at these redshifts.

As the first large spectroscopic survey of $z>4$ star-forming galaxies, ALPINE probes a unique era of cosmic history, and provides the first measure of the morpho-kinematic diversity in the $z\sim4-6$ epoch. Using a custom line search algorithm, \citet{beth20} found that 75 of the 118 targets were detected in  [CII] emitting emission. The morpho-kinematic diversity of this sample was then assessed by a team within the ALPINE collaboration \citep{lefe20} who examined the [CII] channel maps, integrated intensity (moment 0) maps, velocity fields (moment 1), position-velocity diagrams (PVDs) along the major and minor axes, integrated spectra, and ancillary photometry \citep{fais20} . Based on this information, each member independently classified each galaxy as rotating (class 1), merging (class 2), extended dispersion-dominated (class 3), compact dispersion-dominated (class 4), or too weak to characterize (class 5), and the class for each galaxy was agreed upon.

This initial qualitative classification was an invaluable first step towards characterization of the sample, and we may add further measures to contextualize and interpret these classes. This addition is crucial, as the [CII] observations of this sample are relatively limited. That is, the spatial resolution of this dataset is low, with at best only $\sim2-3$ beams across each source. In addition, while these sources are ``normal'' for this redshift range (i.e., on the star-forming main sequence), the modest integration times (i.e., $<1$\,hour) result in limited sensitivity. These effects cause the true nature of the galaxy to be partially obfuscated by random noise peaks, and to be spatially smoothed.

However, by investigating the morpho-kinematics of each source using different methods, it is possible to refine their classification, adding evidence for their intrinsic behaviour. To this end, we conduct a quantitative analysis of the morpho-kinematics of the sample by fitting each galaxy with a tilted ring fitting code and applying well-tested local morpho-kinematic classification criteria.

While tilted ring models (TRM) fitting codes have mainly been implemented to derive the physical parameters of rotating disk galaxies at low- and high-redshift (e.g., \citealt{fan19,shar21}), they may also be used to classify the morpho-kinematic class of galaxies. That is, if the galaxy is truly a rotating disk, then the models will be well fit and yield rotation curves, morphological parameters, and kinematic parameters. On the other hand, early stage mergers will be poorly fit (but see discussions of \citealt{simo19}). Dispersion-dominated galaxies will be well fit with a TRM, but since their intrinsic morphologies are not necessarily flat disks, the best-fit physical parameters may not be physical. So by fitting a TRM to each of the [CII] data cubes in ALPINE, we may add additional evidence to each kinematic classification.

In this work, we examine the 75 [CII]-detected galaxies in the ALPINE survey using tilted ring models and morpho-kinematic classification criteria, with the goals of studying the kinematic diversity of this unique sample, testing the applicability of low-redshift classification criteria to high-redshift observations, and characterizing the properties (e.g., rotation curves, morphological parameters, velocity dispersion) of  rotation-dominated ALPINE galaxies. We begin by describing the ALPINE survey and the creation of the [CII] data cubes in Section \ref{mommaker}. The \BB modelling procedure and results are then presented in Section \ref{methsec}. We test the applicability of two quantitative morpho-kinematic classifiers from low-redshift studies in Section \ref{gm20w15}. Following a discussion of these results in Section \ref{DISCU}, we conclude in Section \ref{CONC}. We assume a standard concordance cosmology ($\Omega_{\Lambda}$,$\Omega_m$,h)=(0.7,0.3,0.7) throughout.

\section{Observations \& Data Cube Creation}\label{mommaker}
The overview of the ALPINE survey is presented in \citet{lefe20}, while the observations, data reduction and the creation of the public catalog\footnote{Catalog and data products are available at  \url{https://cesam.lam.fr/a2c2s/data_release.php}} are detailed in \citet{beth20}, and the multiwavelength photometry analysis of the ALPINE sources is described in \citet{fais20}. The [CII]$158$\,$\mu$m and the surrounding rest-frame FIR continuum emission of each source was observed with ALMA in cycles 5 and 6. Each observation was calibrated using the standard heuristic-based CASA (Common Astronomy Software Applications; \citealt{mcmu07}) pipeline, and the results were inspected for quality. After some minor additional flagging, these data were used to create preliminary data cubes, which are now publicly available. 

Each cube was continuum-subtracted in the \textit{uv}-plane using the CASA task \textlcsc{uvcontsub}, resulting in a line-only, continuum-free data cube. Since the restoring beam of each observation was comparable (average of $1.13''\times0.85''$, \citealt{beth20}), each image was created using a uniform cell size of $0.15''$. The line cubes were constructed with channels of $25\,$km\,s$^{-1}$, or $\sim30$\,MHz. All cubes were cleaned down to $3\sigma$ (CASA \textlcsc{tclean}) and were created using natural weighting.

\section{Tilted Ring Model Fitting}\label{methsec}
While there is a multitude of codes that extract the physical parameters and kinematic details of a galaxy from observational data, they may be divided into two broad classes: codes that fit models to the velocity field (e.g.; GIPSY \textlcsc{rotcur}, \citealt{vand92}; NEMO \textlcsc{rotcurshape}, \citealt{teub95}; \textlcsc{\textlcsc{reswri}}, \citealt{scho99}; \textlcsc{ringfit}, \citealt{simo03}; \textlcsc{kinemetry}, \citealt{kraj06}; \textlcsc{diskfit}, \citealt{sell15}) and codes that use the entire data cube (e.g.; \textlcsc{tirific}, \citealt{jozs07}; \textlcsc{dysmal}, \citealt{cres09}; \textlcsc{kinms}, \citealt{davi13}; \BB, \citealt{dite15}; \textlcsc{galpak 3d}, \citealt{bouc15}; \textlcsc{Blobby3D}, \citealt{vari19}). 

We choose to apply the 3-D tilted ring model fitting code \BB to the [CII] emission of the ALPINE sample. This  code has already been well-tested on data cubes with relatively low resolution and sensitivity (\citealt{dite15}) and has been successfully applied to high-redshift ($z>2$) observations (e.g., \citealt{tali18,fan19,neel20,frat20}). 

\subsection{Signal Isolation}\label{SI}

To fit our data with a tilted ring model, we must first mask the [CII] emission and exclude sources that are detected at too low resolution or sensitivity for kinematic modelling. First, the data cubes for each [CII]-detected ALPINE source were trimmed to only include the sideband (i.e., two spectral windows; SPWs) containing [CII] emission, and the rest frequency of each was set to that of [CII] at the redshift derived by \citet{beth20}. In some cases, the spectral setup created a gap between the two SPWs, which we masked.

Next, the line emission in each continuum-subtracted cube was identified using the SEARCH algorithm in \BB, which is based on the code DUCHAMP \citep{whit12}. After automatically determining the median noise level of each channel, this algorithm searches for pixels with intensities above a user-provided SNR (SNR$_{\rm upper}=3.0$). Here, SNR is defined as the value of a given pixel divided by the noise level of its channel. A three-dimensional (i.e., RA, DEC, velocity) search is then conducted around these peaks for emission above a second user-provided SNR (SNR$_{\rm lower}=2.5$), creating a three-dimensional mask of the signal in the cube. Decreasing SNR$_{\rm lower}$ (e.g., to 2.0) results in the inclusion of more pixels for every source, but in many cases also encloses noise peaks that complicate kinematic characterization. On the other hand, SNR$_{\rm upper}$ is simply used to find the bright pixels at the centre of each mask, so slightly varying this value (e.g., from $2.5-3.0$) has little effect for most our [CII]-detected sources ($>3.5\sigma$; \citealt{beth20}).

To avoid a contribution from non-target sources (i.e., noise peaks and serendipitous sources distant from the phase centre; see \citealt{loia20} for analysis of this latter type of objects), we remove all sources that are $\ge5$\,beams from the centre from our signal mask. In addition, we exclude all unresolved sources by removing sources that feature a masked area (i.e., $2.5\sigma$ contour) smaller than the half-power area of the synthesized beam of the input [CII] data cube.

\subsubsection{SEARCH Results}\label{searchstuff}
Of the 118 galaxies observed in ALPINE, \citet{beth20} found [CII] emission ($>3.5\sigma$) in 75 at the expected frequency from the UV spectroscopy results \citep{fais20}. By applying \BB SEARCH to the continuum subtracted cubes of these 75 [CII]-detected ALPINE galaxies, we identify emission in 40 galaxies, given the constraints outlined above. 

The disparity in the number of detected galaxies is mainly due to the fact that the line search algorithm used by \citet{beth20} is more sensitive to low-level, broad emission than SEARCH. That is, the prior algorithm is able to search the cube using kernels of different velocity widths, while SEARCH searches for pixels above a single value and then merges them. Since our morpho-kinematic analysis will examine each spectral channel of the data cube (see Section \ref{3dfd}), we require significant (i.e., $>2\sigma$) emission in all spectral channels and thus must exclude broad, low-level emission.

An additional important difference is our criterion that the identified signal must have a $2.5\sigma$ contour that is larger than the half power contour of the restoring beam. This limit on compactness, which is not present in the \citet{beth20} search, is necessary to avoid fitting marginally- or un-resolved galaxies. It excludes all compact dispersion dominated sources (morpho-kinematic class 4 in \citealt{lefe20}).

The five morpho-kinematic classes of \citet{lefe20} contain 9, 31, 15, 8, and 12 galaxies (75 total) for classes 1 through 5, respectively. Of these sources, we recover 6, 21, 13, 0, and 0 galaxies, respectively (40 total). The total lack of compact or weak (class 4 or 5, respectively) galaxies is explained by the algorithm criteria noted above. However, we do recover $\sim67-87\%$ of the sources originally classified into classes 1-3.

\subsection{3DFIT Details}\label{3dfd}

Once the [CII] signal from resolved sources has been isolated using \BB SEARCH, the main function of \BB (i.e., 3DFIT) is then used to fit a tilted ring model to the line emission. In general, tilted ring models assume that a galaxy may be approximated as a set of concentric rings, each with its own geometric and kinematic  parameters (e.g., \citealt{rogs74}). Three-dimensional model fitting considers some variables that are included in two-dimensional fits: inclination (\textit{i}), position angle (\textit{PA}), rotational velocity ($\rm v_{rot}$), systemic redshift ($z$ or v$_{\rm sys}$), and the spatial centroid (x$_0$, y$_0$). By expanding to the spectral dimension, we may also consider velocity dispersion ($\sigma_v$), radial brightness profile ($I(r)$), and scale height ($Z_{\circ}$). While this task is able to automatically predict many parameters and is well-tested for low-SNR and low-resolution observations, its performance is dependent on the initial parameter estimates and the overall model geometry, including the radius of the model and the width of each ring. Before we detail the fitting process, we will discuss the methods used to derive physically motivated estimates for each parameter.

\subsubsection{Parameter Estimation}\label{PE}

To begin, each cube is collapsed over all channels containing line emission (as identified by \BB SEARCH) using the CASA toolkit task \textlcsc{im.moments}. The resulting integrated intensity (i.e., moment 0) map is fit with a two-dimensional Gaussian using the CASA toolkit task \textlcsc{im.fitcomponents}, yielding a best-fit peak intensity, integrated flux density, beam-deconvolved FWHM of the major and minor axes, position angle, and central position. In addition, a velocity field (moment 1) and velocity dispersion (moment 2) map are created using the same task, including only pixels identified through the SEARCH algorithm.

To avoid under- or over-sampling the data, the ring geometry is determined in an automatic fashion. If the emission in the moment zero map is sufficiently resolved to allow CASA to fit a 2-D Gaussian and deconvolve the intrinsic emission from the synthesized beam, then we set the maximum model radius to $0.8\times$(the deconvolved FWHM of the major axis of the 2-D Gaussian), based on the high-resolution kinematic analysis of \citet{neel20}. If the source is not sufficiently resolved, we exclude it from further analysis. 

The minimum radius of each ring is fixed to be the FWHM of the minor axis of the restoring beam divided by 2.5. This factor of 2.5 is comparable to other beam/ring width ratios (e.g.; 2.2, \citealt{shao17}; 2.5, \citealt{tali18,fan19}) and is chosen as a middle-ground between the number of pixels per beam FWHM ($\sim5$) and the lower factors adopted by other studies of galaxy morpho-kinematics (e.g.; $\lesssim1$, \citealt{shel20,pina20,sala20}; $\sim2$, \citealt{debl08}). Note that while this high ratio may oversample the data in the case of 2-D (i.e., velocity field) modelling, our use of 3-D modelling reduces the effects of beam smearing, allowing us to sample spatial scales slightly smaller than the synthesized beam (e.g., \citealt{dite15}).

The number of rings is then determined by dividing the maximum model radius by the minimum ring radius and rounding down to the closest integer. If this number is $\ge2$, then the width of each ring is set to the maximum model radius divided by the number of rings. We do not consider models containing only one ring.

While \BB allows the user to fit the central position of each ring at the same time as the inclination and position angle, 
we fix the central position of all model rings to the morphological central position, as found through a 2-D Gaussian fit to the moment zero map, in order to reduce the free parameters of the fit. Since we wish to test how well the data are fit by a rotating disk model, this is tantamount to assuming that the morphological and kinematic centres of the galaxy are coincident. 

The initial guess for the velocity dispersion is based on the moment two map, and a thin disk is assumed ($Z_{\circ}=0.01''\sim60$\,pc at $z\sim5$). Each galaxy is given an initial systemic velocity estimate of $\rm 0\,km\,s^{-1}$ with respect to the assumed redshift. The inclination is allowed to vary between $10-80^{\circ}$, with an initial guess of $45^{\circ}$. \BB provides an initial guess for the rotational velocity based on the data cube, while we estimate the position angle by examining the high-velocity channel maps.

\subsubsection{Fitting Procedure}

Using these initial morphological and kinematic estimates and a list of ring radii and widths, \BB first creates a model of the innermost ring by populating a physical volume with discrete clouds, such that all estimated parameters are recreated. This model is then converted into an observational cube (i.e., with axes of RA, Dec, velocity), convolved with the synthesized beam of the input data cube, normalized by setting the integrated flux per spaxel in the model and input data cubes equal, and compared to the input cube. Each parameter (i.e., rotational velocity, velocity dispersion, systemic velocity, inclination, and position angle) is varied, until the absolute residual (i.e., $\rm |model-observation|$ over all pixels) is minimized. When the residual is minimized, the fitting stops, and the next ring is analysed. After all rings have been fit, the best-fit spatial parameters (i, PA) and v$_{\rm sys}$ across all rings are averaged and fixed. The fitting is then repeated for the remaining parameters (i.e., rotational velocity and velocity dispersion).

The final outputs of this process are morphological parameters from a 2-D Gaussian fit to the moment zero map (i.e., inclination, position angle, central position, radius), morpho-kinematic parameters from the \BB fit (i.e.,  inclination, position angle, velocity dispersion profile, rotation curve, systemic velocity/redshift), and a best-fit model data cube that may be directly compared to the data. For further discussion of this procedure, see Appendix \ref{morebb}.

\subsubsection{Tilted Ring Model Results}
\begin{figure*}
\centering
\includegraphics[width=0.49\textwidth]{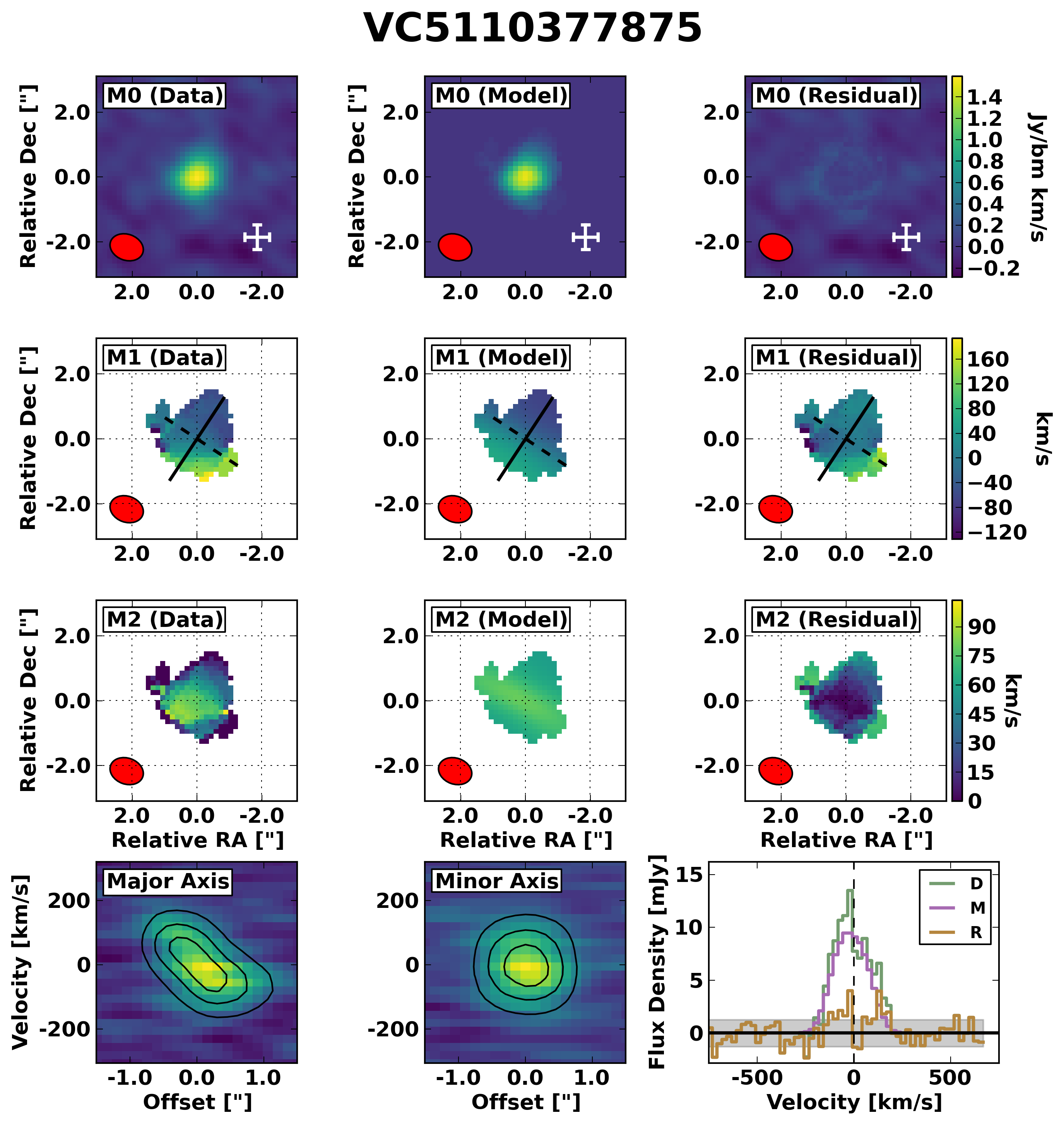}
\includegraphics[width=0.49\textwidth]{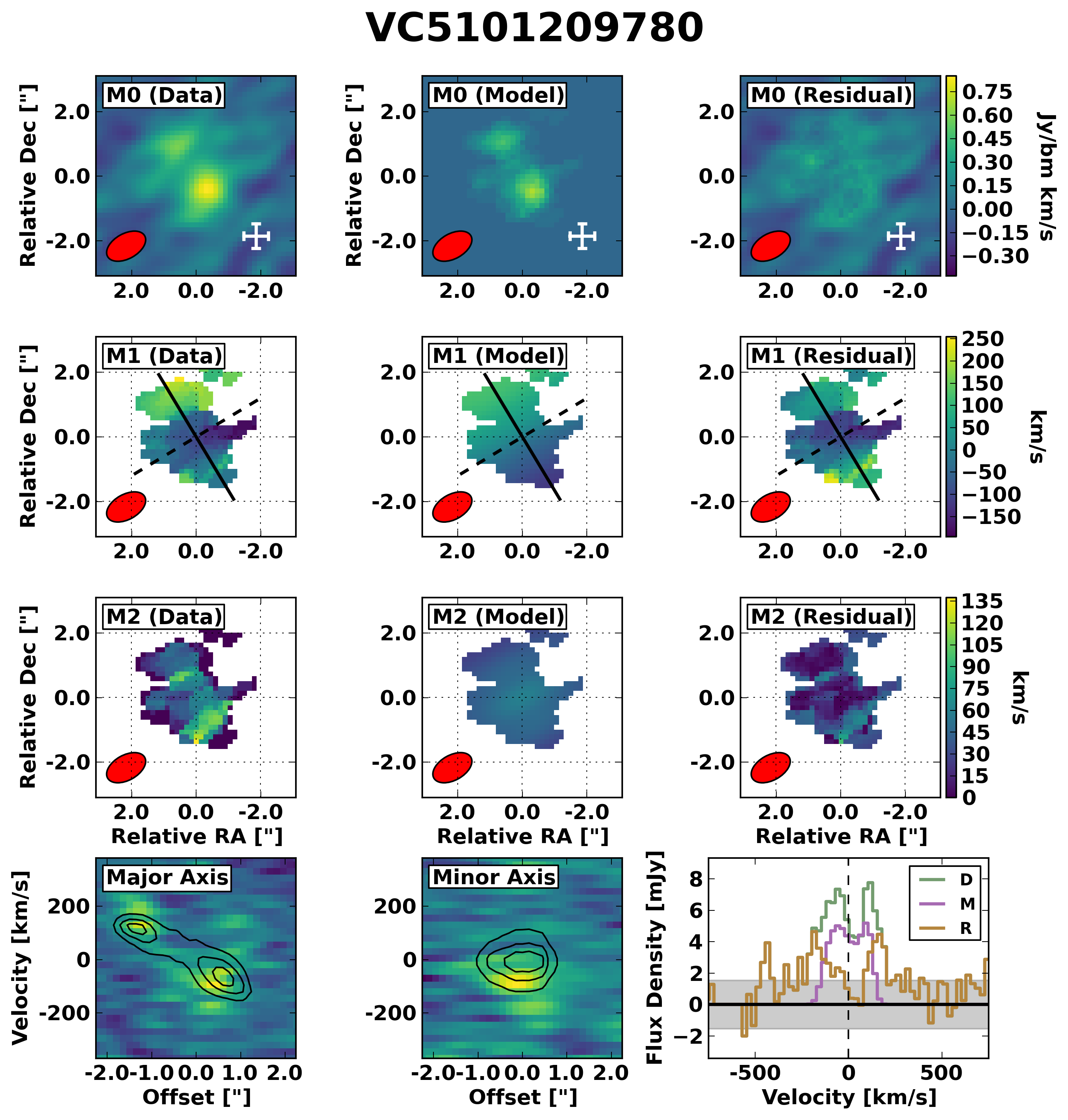}
\includegraphics[width=0.49\textwidth]{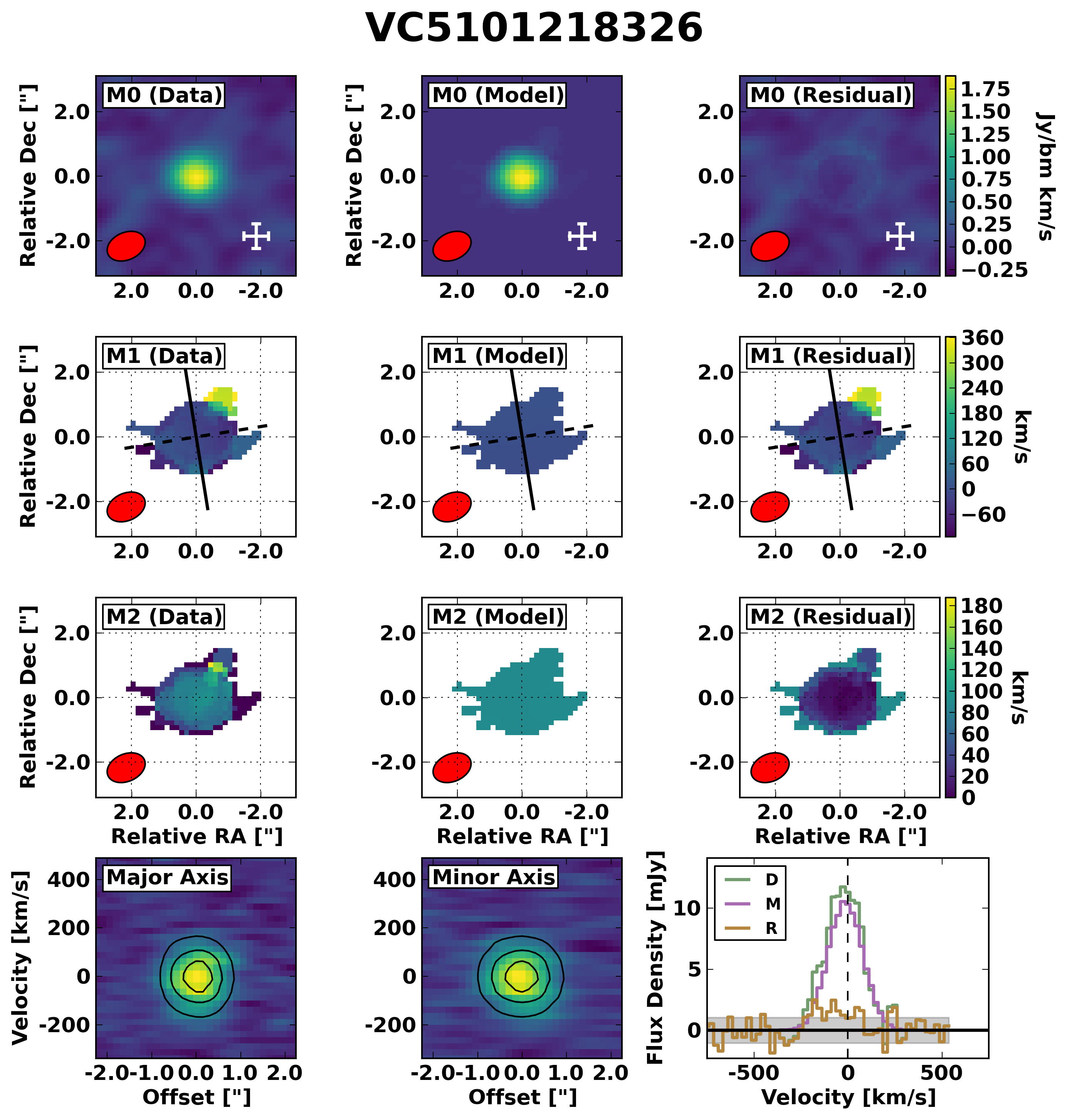}
\includegraphics[width=0.49\textwidth]{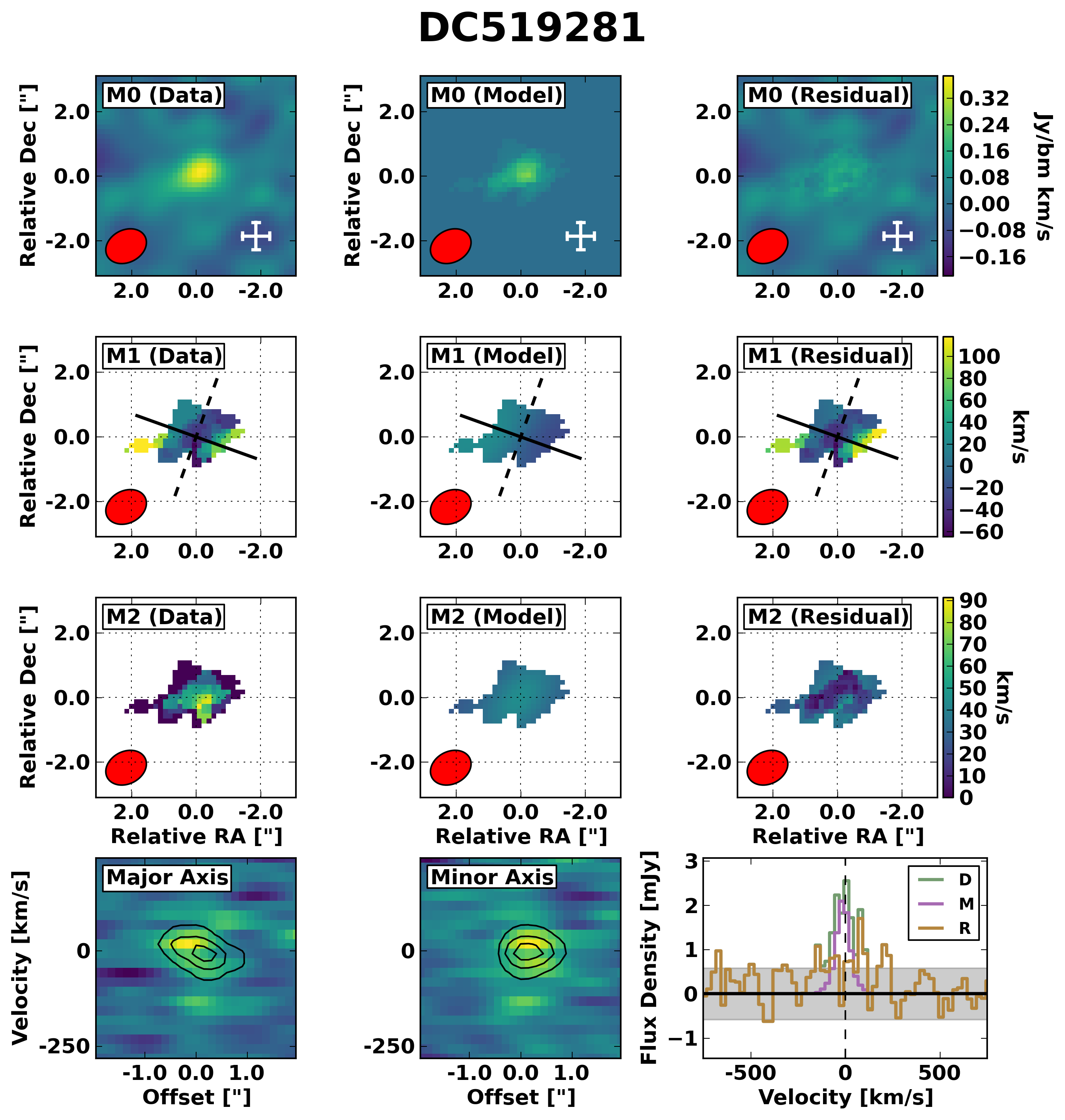}
\caption{Moment maps, PVDs, and spectra for observed data, model, and residual for an example rotator (upper left), merger (upper right), dispersion dominated (lower left), and uncertain class (lower right) galaxy (see Section \ref{synth} for details of each kinematic class). In each of these figures, the first three rows show (from top to bottom) the moment 0 (integrated intensity), moment 1 (velocity field), and moment 2 (velocity dispersion field; see \ref{PE} for details of moment map creation). For these rows, the three columns denote (from left to right) the observed data cubes, model cubes, and the corresponding residuals. The white crosses in the lower right corner of each panel in the first row show a 5\,kpc$\times$5\,kpc physical scale. Solid lines in the second row represent the kinematic major axis, while the dashed lines represent the minor axis. The restoring beam is shown as a red ellipse. The bottom row shows (from left to right) the major axis PVD, minor axis PVD, and integrated spectra. For each PVD, the observed data are shown by the background colour, while the contours represent the model at $20\%$, $50\%$, and $80\%$ of its maximum value. The data (D), model (M), and residual (R) spectra are depicted by the green, purple, and orange lines, respectively. The data spectrum is extracted from the continuum-subtracted [CII] cube over all spaxels where line emission was identified by \BB SEARCH. The $1\sigma$ uncertainty, calculated as (average RMS noise level per channel)$\rm \times\sqrt{number\,\,of\,\,beams\,\,in\,\,source}$, is shown by the shaded grey area. See Appendix \ref{appdesc} for plots of other sources.}
\label{BBresults1A}
\end{figure*}

When \BB 3DFIT is run on the 40 recovered sources (see Section \ref{searchstuff} for details of signal isolation), $15\%$ (six sources) are not successfully modelled due to being only marginally resolved (i.e., it was not possible to deconvolve the moment zero map from the synthesized beam), while another $\sim13\%$ (five sources) are resolved but are excluded due to only featuring 1\,ring. For the 29 galaxies that are successfully fit, we present comparisons of moment maps, PVDs, and spectra for the data and models in Figures \ref{BBresults1A} and \ref{BBresults1} through \ref{BBresults6}, list the best-fit parameters in Table \ref{bbtab}, and show the best-fit rotation curves and velocity dispersion profiles in Figure \ref{rcs}. Each source is also discussed in Appendix \ref{appdesc}.

\begin{table*} 
\centering 
\caption{Morphological and kinematic parameter values for ALPINE galaxies successfully fit by \BB, as well as the two previously detected $4<z<6$ rotating unlensed SFGs (see Section \ref{MSR}). Reported uncertainties on redshift are the greater of the fit uncertainty and the width of a velocity channel. The morphological position angles and inclinations are taken from two-dimensional Gaussian fits to moment zero maps created from SEARCH-identified [CII] emission, while the kinematic values are the best-fit values from \BB. The W15 criteria correspond to the five tests of disk-like behaviour suggested by \citet{wisn15}, as explored in Section \ref{w15sec}. We list both the morpho-kinematic class derived by \citet{lefe20} (L20) and the class derived by combining the results of our analyses (J21, see Section \ref{synth}). }
\label{bbtab} 
\begin{tabular}{c||ccccc|ccccc|cc} 
\toprule 
Name & $z$ & $\rm PA_M$ & $\rm PA_K$ & $\rm i_M$ & $\rm i_K$ & & & W15 & & & KC1 & KC2 \\
 &  & $\rm [^{\circ}]$ & $\rm [^{\circ}]$ & $\rm [^{\circ}]$ & $\rm [^{\circ}]$ & 1 & 2 & 3 & 4 & 5 & L20 & J21  \\ \hline 
CG32 & $4.4105\pm0.0005$ & $90\pm60$ & $238\pm14$ & $50\pm20$ & $55\pm19$ & $\checkmark$ & $\checkmark$ & $\checkmark$ & $\times$ & $\checkmark$ & 2 & ROT\\
DC396844 & $4.5424\pm0.0005$ & $161\pm28$ & $197\pm14$ & $60\pm19$ & $57\pm31$ & $\checkmark$ & $\checkmark$ & $\times$ & $\times$ & $\times$ & 1 & ROT\\
DC417567 & $5.6700\pm0.0006$ & $160\pm80$ & $49\pm4$ & $40\pm50$ & $68\pm13$ & $\checkmark$ & $\checkmark$ & $\checkmark$ & $\times$ & $\checkmark$ & 2 & UNC\\
DC432340 & $4.4045\pm0.0005$ & $11\pm18$ & $187\pm8$ & $59\pm19$ & $68\pm12$ & $\checkmark$ & $\checkmark$ & $\checkmark$ & $\checkmark$ & $\checkmark$ & 2 & UNC\\
DC434239 & $4.4883\pm0.0005$ & $28\pm8$ & $217\pm3$ & $63\pm8$ & $73\pm13$ & $\checkmark$ & $\checkmark$ & $\times$ & $\checkmark$ & $\times$ & 2 & MER\\
DC454608 & $4.5835\pm0.0005$ & $140\pm40$ & $193\pm9$ & $67\pm25$ & $57\pm8$ & $\times$ & $\checkmark$ & $\times$ & $\times$ & $\times$ & 2 & UNC\\
DC494057 & $5.5446\pm0.0005$ & $60\pm90$ & $185\pm13$ & $31\pm25$ & $46\pm4$ & $\checkmark$ & $\checkmark$ & $\times$ & $\times$ & $\times$ & 1 & ROT\\
DC519281 & $5.58\pm0.02$ & $100\pm80$ & $70\pm10$ & $60\pm30$ & $53\pm9$ & $\times$ & $\checkmark$ & $\times$ & $\times$ & $\times$ & 2 & UNC\\
DC552206 & $5.5016\pm0.0005$ & $138\pm23$ & $322\pm7$ & $52\pm13$ & $61\pm17$ & $\checkmark$ & $\checkmark$ & $\times$ & $\checkmark$ & $\times$ & 2 & ROT\\
DC627939 & $4.5333\pm0.0006$ & $170\pm90$ & $278\pm4$ & $50\pm30$ & $66\pm8$ & $\checkmark$ & $\checkmark$ & $\times$ & $\times$ & $\times$ & 2 & UNC\\
DC683613 & $5.5421\pm0.0005$ & $170\pm40$ & $353\pm2$ & $58\pm30$ & $49\pm19$ & $\checkmark$ & $\times$ & $\times$ & $\checkmark$ & $\times$ & 3 & UNC\\
DC733857 & $4.5445\pm0.0005$ & $87\pm27$ & $84\pm16$ & $84\pm19$ & $74\pm1$ & $\times$ & $\checkmark$ & $\times$ & $\checkmark$ & $\times$ & 3 & UNC\\
DC773957 & $5.6773\pm0.0023$ & $101\pm11$ & $35\pm5$ & $76\pm11$ & $66\pm19$ & $\checkmark$ & $\checkmark$ & $\times$ & $\times$ & $\times$ & 2 & UNC\\
DC818760 & $4.5611\pm0.0005$ & $94\pm2$ & $90\pm5$ & $74\pm2$ & $62\pm11$ & $\times$ & $\checkmark$ & $\checkmark$ & $\checkmark$ & $\checkmark$ & 2 & MER\\
DC848185 & $5.2930\pm0.0005$ & $149\pm11$ & $320\pm9$ & $53\pm9$ & $52\pm0$ & $\checkmark$ & $\times$ & $\times$ & $\checkmark$ & $\times$ & 3 & DIS\\
DC873321 & $5.1544\pm0.0005$ & $118\pm5$ & $281\pm10$ & $73\pm6$ & $65\pm11$ & $\checkmark$ & $\times$ & $\times$ & $\checkmark$ & $\times$ & 2 & MER\\
DC873756 & $4.5457\pm0.0005$ & $121\pm19$ & $298\pm0$ & $42\pm10$ & $39\pm28$ & $\checkmark$ & $\times$ & $\times$ & $\checkmark$ & $\times$ & 2 & DIS\\
DC881725 & $4.5778\pm0.0005$ & $140\pm40$ & $323\pm9$ & $76\pm33$ & $48\pm11$ & $\checkmark$ & $\checkmark$ & $\times$ & $\checkmark$ & $\times$ & 1 & ROT\\
VC5100537582 & $4.5502\pm0.0005$ & $60\pm90$ & $78\pm6$ & $40\pm50$ & $49\pm3$ & $\checkmark$ & $\checkmark$ & $\times$ & $\checkmark$ & $\times$ & 3 & UNC\\
VC5100541407 & $4.5632\pm0.0005$ & $69\pm15$ & $69\pm7$ & $59\pm11$ & $66\pm9$ & $\times$ & $\times$ & $\times$ & $\checkmark$ & $\times$ & 2 & UNC\\
VC5100559223 & $4.5626\pm0.0005$ & $167\pm28$ & $12\pm6$ & $75\pm26$ & $49\pm4$ & $\checkmark$ & $\times$ & $\times$ & $\times$ & $\times$ & 3 & UNC\\
VC5100822662 & $4.5205\pm0.0005$ & $0\pm80$ & $11\pm15$ & $40\pm50$ & $53\pm7$ & $\times$ & $\checkmark$ & $\times$ & $\checkmark$ & $\times$ & 2 & MER\\
VC5100994794 & $4.5800\pm0.0005$ & $55\pm27$ & $243\pm9$ & $69\pm16$ & $49\pm17$ & $\checkmark$ & $\checkmark$ & $\times$ & $\checkmark$ & $\times$ & 3 & UNC\\
VC5101209780 & $4.5701\pm0.0016$ & $23\pm16$ & $31\pm13$ & $57\pm12$ & $72\pm8$ & $\checkmark$ & $\checkmark$ & $\times$ & $\checkmark$ & $\times$ & 2 & MER\\
VC5101218326 & $4.5741\pm0.0006$ & $51\pm25$ & $9\pm2$ & $54\pm27$ & $44\pm4$ & $\times$ & $\times$ & $\times$ & $\times$ & $\times$ & 3 & DIS\\
VC510786441 & $4.463\pm0.004$ & $4\pm6$ & $1\pm2$ & $65\pm7$ & $66\pm10$ & $\times$ & $\checkmark$ & $\times$ & $\checkmark$ & $\times$ & 2 & UNC\\
VC5110377875 & $4.5506\pm0.0005$ & $146\pm18$ & $147\pm7$ & $56\pm15$ & $50\pm6$ & $\checkmark$ & $\checkmark$ & $\times$ & $\checkmark$ & $\times$ & 1 & ROT\\
VC5180966608 & $4.5296\pm0.0005$ & $125\pm34$ & $311\pm9$ & $45\pm22$ & $52\pm5$ & $\checkmark$ & $\times$ & $\times$ & $\checkmark$ & $\times$ & 2 & UNC\\
VE530029038 & $4.4297\pm0.0005$ & $130\pm60$ & $306\pm7$ & $50\pm50$ & $47\pm15$ & $\checkmark$ & $\checkmark$ & $\times$ & $\checkmark$ & $\times$ & 1 & UNC\\ \hline
J0817, Low-Res & $4.2605\pm0.0009$ & $65\pm24$ & $93\pm24$ & $58\pm23$ & $48\pm5$ & $\checkmark$ & $\checkmark$ & $\times$ & $\checkmark$ & $\times$ & --- & ROT\\
J0817, High-Res & $4.2600\pm0.0005$ & $109\pm9$ & $108\pm7$ & $46\pm6$ & $64\pm7$ & $\checkmark$ & $\checkmark$ & $\times$ & $\checkmark$ & $\times$ & --- & ROT\\
HZ9 & $5.5413\pm0.0004$ & $83\pm22$ & $17\pm10$ & $57\pm14$ & $58\pm7$ & $\checkmark$ & $\checkmark$ & $\times$ & $\times$ & $\times$ & --- & ROT\\
\bottomrule 
\end{tabular} 
\end{table*} 

These figures show that the morpho-kinematic diversity noted by \citet{lefe20} is indeed present. Some galaxies depict perfect velocity gradients and are well fitted with tilted ring models, while others show significant residuals, and others seem dispersion-dominated. 

We apply \BB to each source with the intent of examining the residuals in each moment map, PVD, and the integrated spectrum, in order to refine the morpho-kinematic classification for the 40 ALPINE galaxies whose emission was identified though \BB SEARCH (see Section \ref{SI}). That is, we do not pre-select rotators from the ALPINE sample before fitting the data with tilted ring models. Due to its optimization for low-SNR and low-resolution observations and large number of free parameters, it easily fits mergers and dispersion-dominated sources with rotating disk models. 
%Indeed, the only sources fit with $>2$\,rings are those with peculiar morphologies (e.g., DC432340) or mergers (e.g., DC818760). 
Because of this, the kinematic values presented in Table \ref{bbtab} and curves in Figure \ref{rcs} should be interpreted with care. Only the results for galaxies that are robustly classified as rotators (see Section \ref{MSR}) are fully reliable.

\begin{figure*}
\centering
\includegraphics[width=\textwidth]{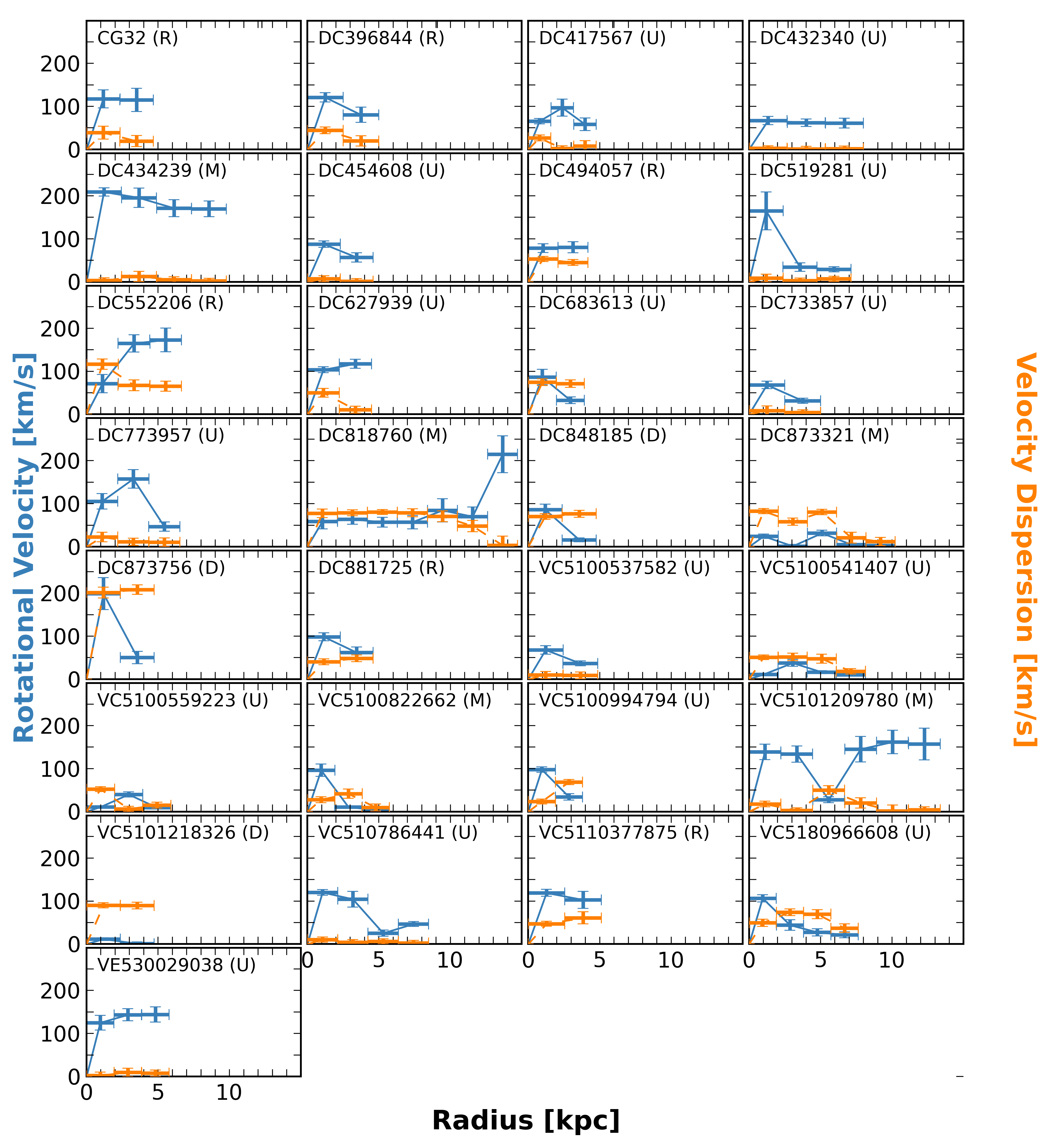}
\caption{Best-fit rotation velocity (blue) and velocity dispersion (orange) curves for each galaxy. Vertical extent of each point shows the fit uncertainty output by \BB (i.e., not including systematic uncertainty on the inclination), while the horizontal extent shows the width of each ring. The letters after each galaxy name denote the morpho-kinematic classification we assign in Section \ref{synth}: ROT (R), MER (M), DIS (D), or UNC (U).}
\label{rcs}
\end{figure*}

\subsection{Subsample Properties}\label{subpro}
The previous subsections have detailed the methods used to isolate [CII] line emission in the ALPINE data cubes and fit this signal with the tilted ring model fitting code \BB. Before proceeding, it is worth briefly discussing the features of our 29 \BB -fit galaxies.

Our signal isolation and model fitting routine has successfully fit the majority of the bright [CII]-detected ALPINE sources. Of the 21 sources detected at SNR$_{\rm [CII]}\ge10$ by \citet{beth20}, we only exclude four sources due to weak  (i.e., spectrally broad) and/or marginally resolved [CII] emission. While this does result in the exclusion of some strong but compact [CII] emitters, such as DC488399 (SNR$_{\rm [CII]}=26.2$) and DC630954 (SNR$_{\rm [CII]}=11.2$, \citealt{beth20}), we are exploring the three-dimensional (i.e., RA, Dec, velocity) distribution of [CII] emission in these galaxies, and so resolved data are required. 
%The kinematics of these sources may be explored by finding the [CII] linewidth (e.g., \citealt{dess20}), and future, high-resolution observations will enable their morpho-kinematic analysis.

\begin{figure}
\centering
\includegraphics[width=\columnwidth]{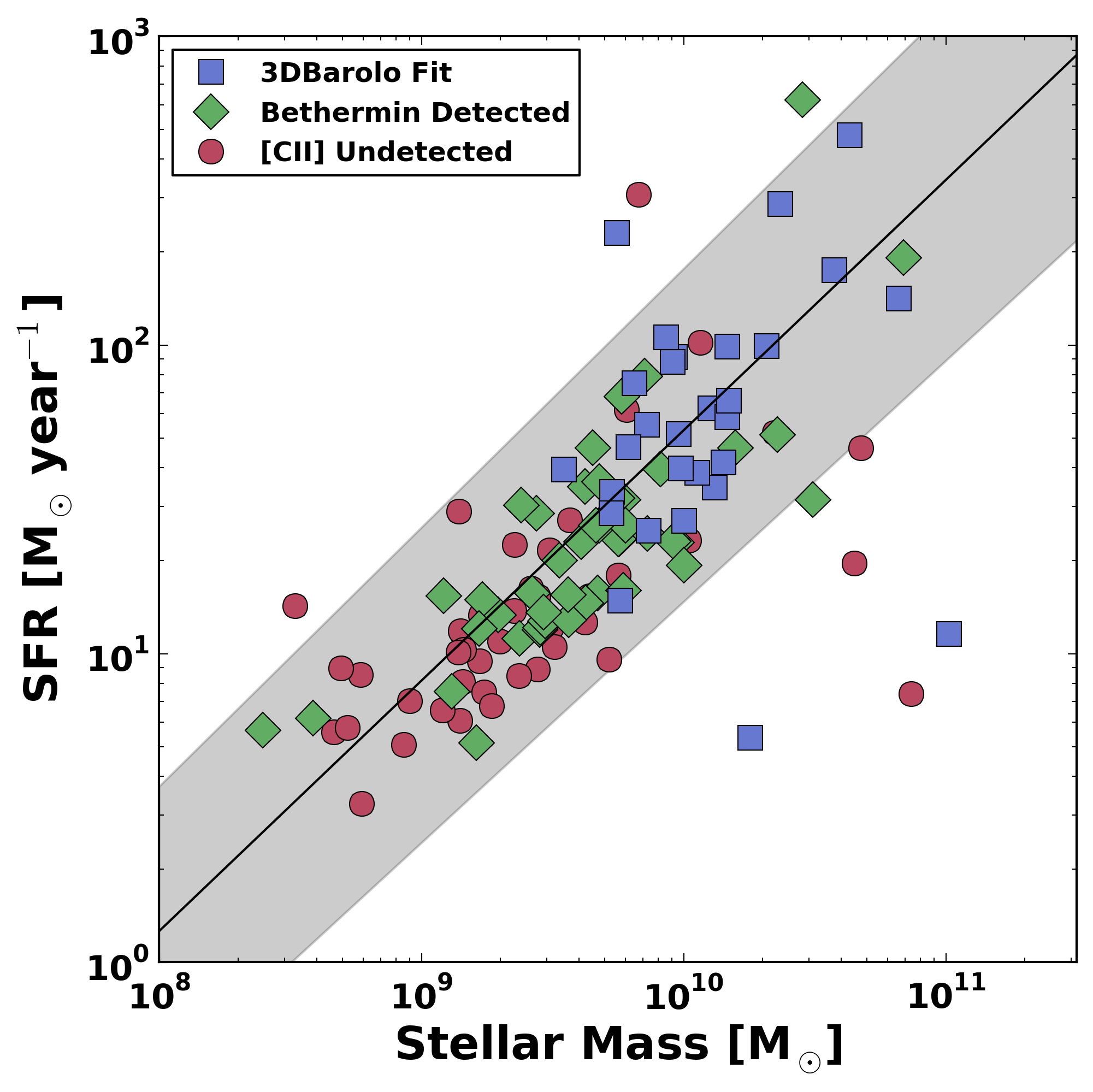}
\includegraphics[width=\columnwidth]{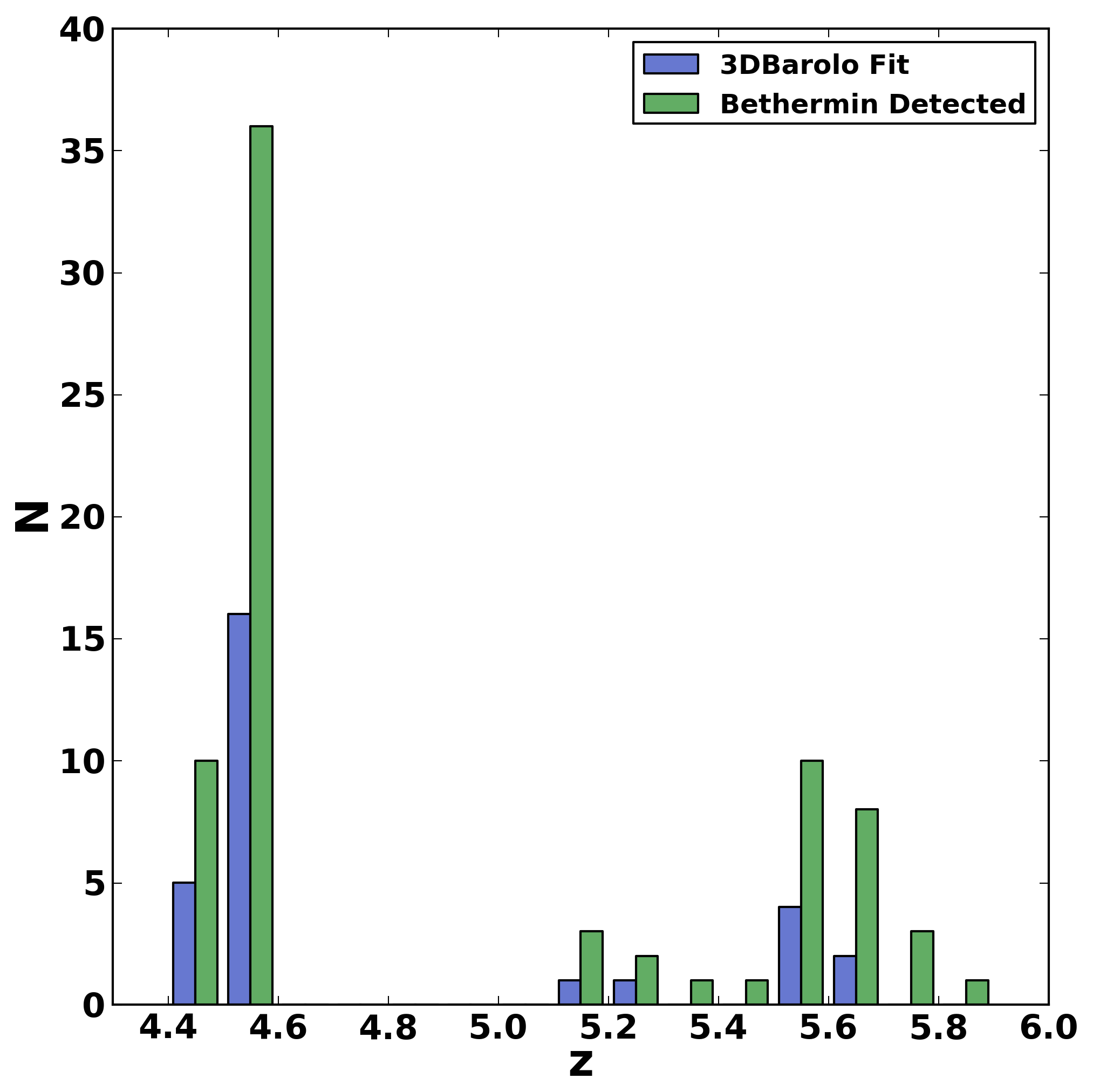}
\caption{Properties of the 29 \BB -fit galaxies, compared to the 75 [CII]-detected galaxies and full sample. 
Top: SFR and M$_*$ for all 118 ALPINE galaxies, based on modelling of ancillary photometry with LePhare \citep{fais20}. Sources undetected (i.e., $<3.5\sigma$) by the line search algorithm of \citet{beth20} are shown as red circles. The sources identified by \BB SEARCH are shown by blue squares, while the \citet{beth20} detected sources that are not identified by SEARCH are shown by green diamonds. The star-forming main sequence relation of \citet{spea14} for a representative ALPINE redshift ($z=5.15$) is shown by the solid black line, while the grey shaded region shows the scatter induced by varying each variable in the Speagle relation by $1\sigma$, as well as varying $z$ over the redshift range of the ALPINE sample ($z=4.4-5.9$).
Bottom: Redshift distribution for the 29 \BB -fit sources (blue) compared to that of the parent sample of 75 sources detected in [CII] emission by \citet{beth20} (green).
}
\label{fig:MS}
\end{figure}

To further examine the subset of ALPINE sources studied in this work, we compare their location on the SFR-M$_*$ plane (see top panel of Figure \ref{fig:MS}, blue squares) to the ALPINE sources excluded by this analysis but detected by the line search algorithm of \citet{beth20} (green diamonds), and the [CII]-undetected sources (red circles). Note that the ALPINE sample was chosen to contain SFGs, so the vast majority of these sources are within the $1\sigma$ scatter of the main sequence relation. This plot shows that the subset fit here contains massive systems (i.e., M$_*\gtrsim10^{9.5}$\,M$_{\odot}$) that exhibit a wide range of SFR, including a few sources above and below the star-forming main sequence. 

We may also examine the redshift distribution of the \BB -fit galaxies compared to the 75 [CII]-detected ALPINE sources \citep{beth20} (see bottom panel of Figure \ref{fig:MS}). While each redshift bin contains $\gtrsim2\times$ as many [CII]-detected sources as \BB -fit sources, we can see that the \BB -fit galaxies cover nearly the same redshift range as the full sample. Thus, our sample of 29 \BB- fit ALPINE galaxies contains massive SFGs at a range of SFR/M$_*$ and redshifts.

\section{Additional Classification Methods}\label{gm20w15}

Thus far, we have presented the best-fit tilted ring models for a subset of the ALPINE sample. To further examine the morpho-kinematics of these sources, we apply two objective, quantitative classification methods tested at $z<4$ to the sample: the Gini-M$_{\rm 20}$ method of \citet{lotz04} and the five disk-like criteria of \citet{wisn15}. Each of these methods has not been applied to rest-frame FIR and [CII]158\,$\mu$m observations of $z>4$ SFGs, so this Section represents an investigation of their applicability to our data.

\subsection{Gini - M$_{20}$ Analysis}\label{gm20}

First, we test whether the ALPINE sample may be classified based solely on its observed morphology by examining the distribution of our sources in the ``\textit{G}-M$_{\rm 20}$'' plot (e.g., \citealt{lotz04}). This approach has been used to distinguish different morphological types in the local universe.

The Gini coefficient (or \textit{G}) is a measure of how uniformly the brightness is distributed. This value approaches 0 when all pixels have the same value (e.g., diffuse emission), and approaches 1 when all brightness is focused in a few pixels (e.g., central nuclei). On the other hand, the moment of light (M$_{20}$) is the normalized second order moment of the pixels that make up the brightest 20\% of the galaxy, and is high when non-axisymmetric features (e.g., bars, clumps, spiral arms) are present. If both of these measures are calculated for each galaxy, then it is possible that mergers and single sources will occupy separate parameter spaces \citep{lotz08}.

\subsubsection{Implementation}

This analysis is applicable to two-dimensional images (i.e., not data cubes), so we apply it to both rest-frame FIR continuum (i.e., $\sim160\,\mu$m) maps and [CII] moment zero maps from the ALPINE observations. Both sets of maps are taken from the ALPINE Data Release 1 (DR1), and the details of their creation are available in \citet{beth20}. This set of maps contains 75 [CII] moment zero maps for the [CII]-detected set of sources and 23 continuum images for the FIR-detected set. Since our goal is to compare the \textit{G} and M$_{\rm 20}$ distribution of these maps to their [CII] morpho-kinematic diversity, we exclude two sources that are detected in FIR emission but not [CII] (CANDELS$\_$GOODSS$\_$19 and DEIMOS$\_$COSMOS$\_$460378), bringing the number of analysed continuum maps to 21.

To isolate the signal in each map, we follow a procedure similar to that of \BB SEARCH. The noise level of each map is determined by sigma clipping above and below $3\sigma$ and taking the standard deviation of the resulting pixels (see astropy task \textit{sigma$\_$clipped$\_$stats}). All pixels above $3\times$ this standard deviation are identified. All such peaks outside of 20\,pixels (i.e., $\sim4$\,beams) from the centre are rejected as noise or serendipitous sources. A mask is then created by extrapolating out from the central peaks to their surrounding $1\sigma$ contours\footnote{A higher threshold was tested ($2\sigma$) which shifted each source to a higher $G$ value, but was not used because it excluded low-level emission.}. By applying this mask to the original map, we are able to isolate the pixels that are associated with the target source. We choose to mask the data in this way, rather than analysing the full image or using an aperture, to ensure that each masked map contains only signal.

To calculate \textit{G}, it is first necessary to sort the value of each pixel in the map ($f_i$) into increasing order (i.e., $f_1$ for the smallest value, $f_n$ for the largest). \textit{G} may then be calculated as:
\begin{equation}
\rm G=\frac{\sum_{i=1}^{n}(2i-n-1)f_i}{\bar{f}n(n-1)}
\end{equation}
where n is the number of pixels and $\bar{f}$ is the mean pixel value.

M$_{\rm 20}$ requires the values of all pixels sorted in decreasing order (i.e., $f_1$ for the largest value, $f_n$ for the smallest). We then calculate the second spatial moment of each pixel ($M_i$) and their sum ($M_{tot}$):
\begin{equation}
M_i=f_i\left[ (x_i-x_c)^2+(y_i-y_c)^2 \right]
\end{equation}
\begin{equation}
M_{tot}=\sum_{i=1}^nM_i
\end{equation}
where ($x_c,y_c$) is the source centre computed by minimizing $M_{tot}$. Next, we find the index $i_{20}$ such that $\sum_{i=1}^{i_{20}}f_i=0.2\sum_{i=1}^nf_i=0.2f_{tot}$, and find $M_{20}$ using:
\begin{equation}
\rm M_{20}=log_{10}\left(\frac{\sum_{i=1}^{i_{20}}M_i}{M_{tot}}\right)
\end{equation}

Our approach does not allow for the calculation of uncertainties on $G$ or $M_{20}$, but we do note that these values are affected by imaging parameters (e.g., image weighting, cell size, cleaning threshold), as well as the choice of 3-D mask. 

\subsubsection{Results}

%We present the \textit{G} and M$_{\rm 20}$ distributions for the ALPINE FIR continuum maps and [CII] moment zero maps in Figure \ref{fig:gm20}. Each point is coloured by the kinematic classification from \citet{lefe20}. For the continuum maps of the 21 sources detected in [CII] and FIR emission (top panel), the majority of the galaxies lie at low $G$ (i.e., $\sim0.2-0.3$), and there is no obvious separation between the kinematic classes. There are two galaxies at high-$G$, indicating bright nuclei. Note that no `compact dispersion-dominated' or `weak' objects (i.e., class 4 or 5, respectively) were detected in continuum emission.

\begin{figure}
\centering
\includegraphics[width=\columnwidth]{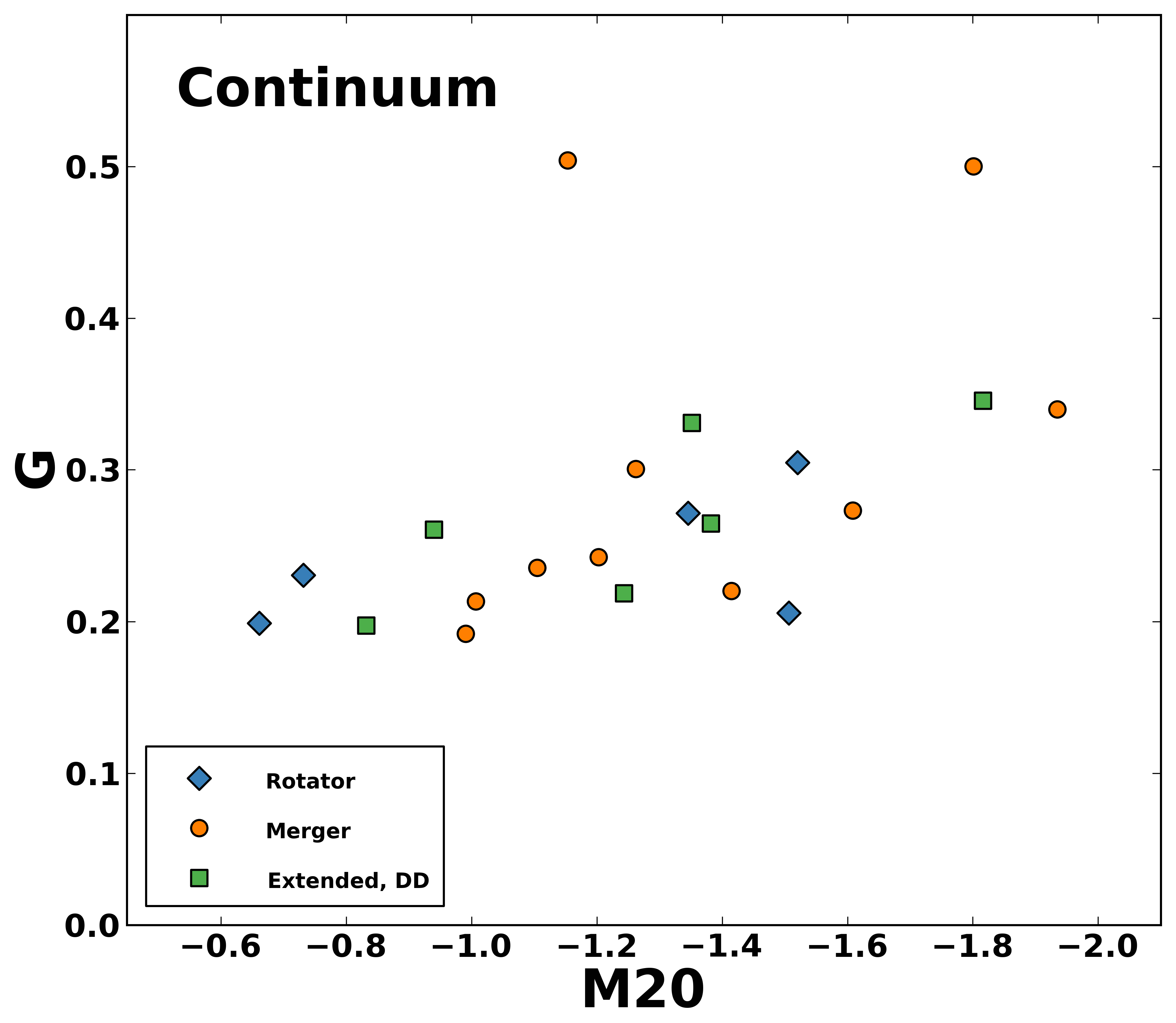}
\includegraphics[width=\columnwidth]{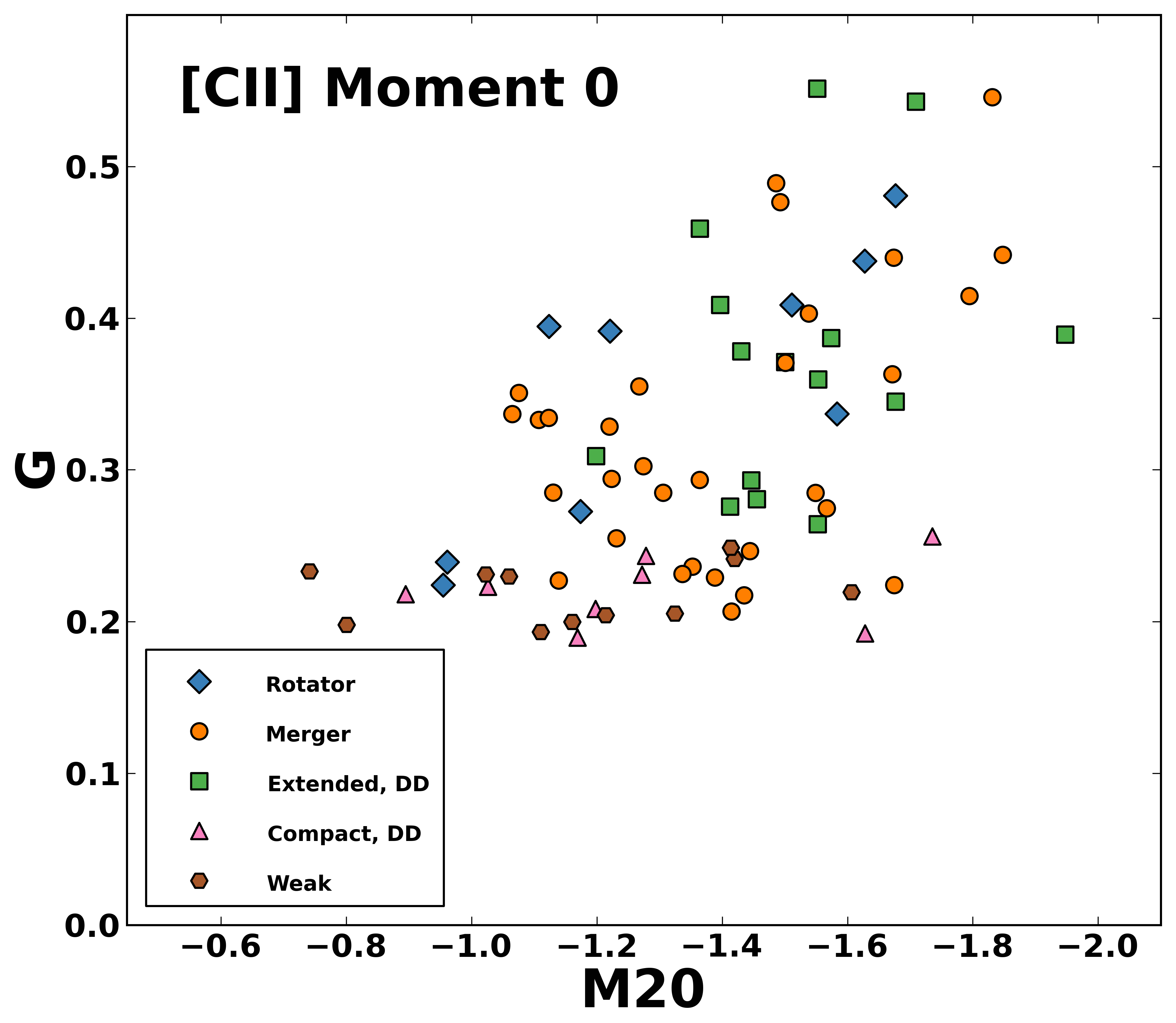}
\caption{Gini-M$_{\rm 20}$ for the continuum (top) and [CII] moment zero (bottom) maps. Points are coloured by morpho-kinematic classification of \citet{lefe20}.}
\label{fig:gm20}
\end{figure}

The bottom panel of Figure \ref{fig:gm20} shows that the [CII] moment zero maps feature a distinct distribution compared to the continuum maps. All of the `compact dispersion-dominated' and `weak' objects lie at low $G$ (i.e., $\sim0.2-0.3$). While the class 1-3 objects spread to higher $G$ (i.e., $\sim0.2-0.55$), there again is no separation between the kinematic classes.

Since this is the first application of this technique to ALMA FIR continuum and [CII] emission of $z>4$ SFGs, there is no comparison sample to which we may compare. One successful application of this procedure was that of \citet{lotz08}, which analysed the stellar distribution of lower redshift ($0.2<z<1.2$) galaxies as observed with HST and found that mergers featured high $G$ values (i.e., clumpy emission), E/S0/Sa had low M$_{\rm 20}$ values (i.e., symmetric morphology), and Sb-Irr galaxies featured low $G$ values (i.e., significant diffuse emission). This analysis of stellar morphology is not directly applicable to FIR and [CII] emission at $4\lesssim z\lesssim6$, since the latter trace warm dust, star formation regions, and molecular gas. 

This lack of separation between morphological types of galaxies was also seen in a sample of 494 SFGs at $2.5\lesssim z\lesssim3.0$, as observed with HST H160 \citep{tali14}. Indeed, this HST study found that ellipticals, compact, disk-like, and irregular galaxies were focused between $-2< \rm{M_{20}}<-1$, with significant overlap between ellipticals, compact galaxies, disk-like galaxies, and irregulars.

Since the galaxies in the ALPINE sample are marginally resolved (i.e., $<5$\,beams across the source) and feature a variety of SNR, it is conceivable that the observed \textit{G}-M$_{20}$ scatter could be affected by observational properties. To determine the effects of the degree of resolution on the recovered \textit{G} and M$_{\rm 20}$, we assume  the extreme case of  each source being completely unresolved, and convolve a point source with the set of ALPINE synthesized beams \citep{beth20}. Using these PSFs, we explore the effects of a range of SNR values on the recovered quantities. In practice, we take each of the 118 restoring beam axis ratios and explore five upper SNR values (3, 5, 10, 30, 50), assuming that the lower SNR threshold is 1. We construct a spatial model grid identical to that of the data (i.e., $0.15''$ cells) and calculate \textit{G} and M$_{\rm 20}$ for each SNR$_{\rm upper}$.

\begin{figure}
\centering
\includegraphics[width=\columnwidth]{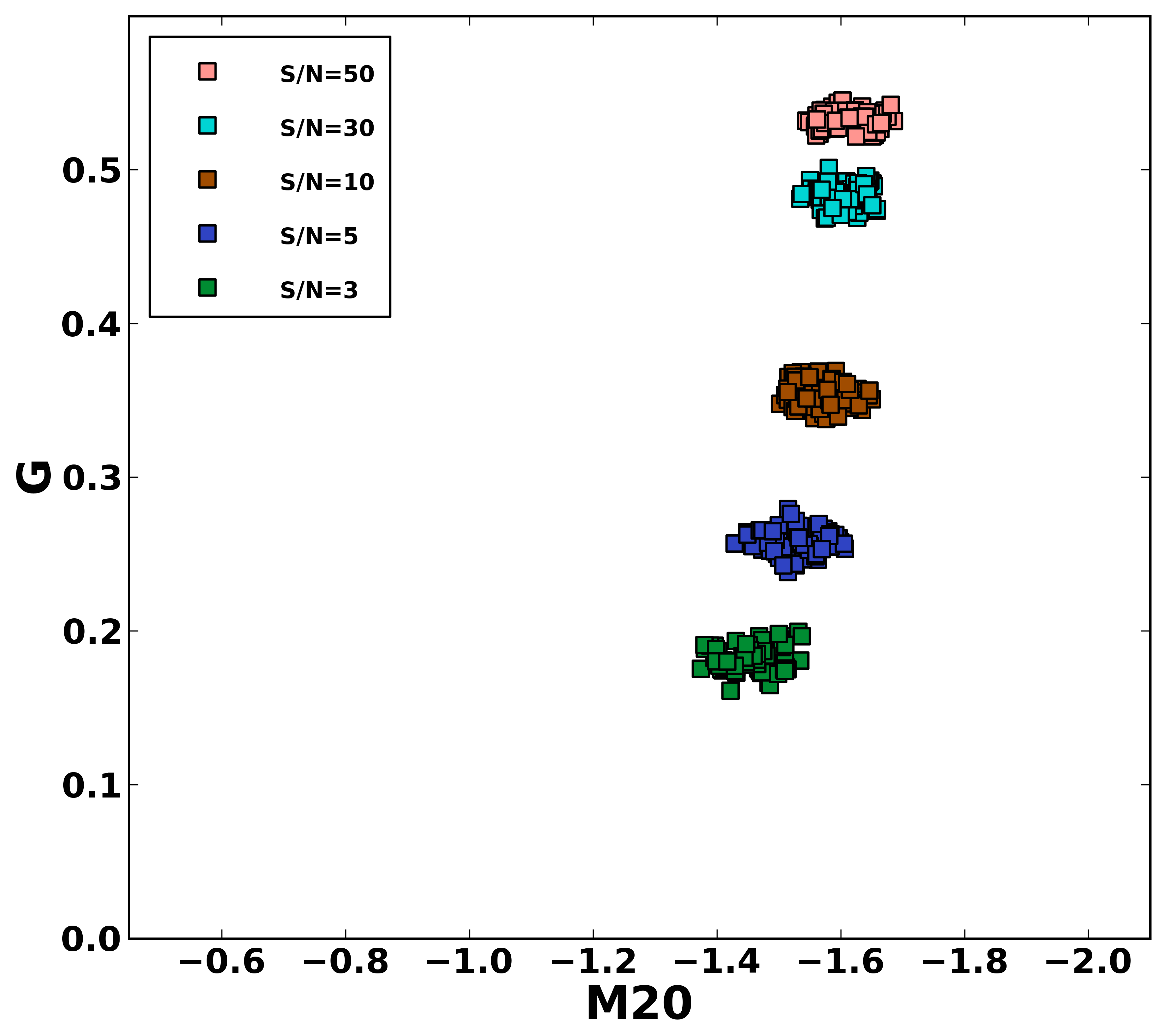}
\caption{G-M$_{\rm 20}$ for a set of model point sources convolved with a 2-D Gaussian beam. For each of the 118 ALPINE synthesized beams, we explore five upper SNR thresholds (3, 5, 10, 30, 50) for SNR$_{\rm lower}=1.0$.}
\label{fig:gm20U}
\end{figure}

As seen in Figure \ref{fig:gm20U}, the 118 beams of ALPINE are similar, and thus generate only a small scatter in the plane for a given SNR$_{\rm upper}$. The SNR threshold generates a significant effect, as stronger point sources result in higher \textit{G} values. The ALPINE continuum and [CII] moment zero maps feature peak SNRs of 3.6-22.3 and 3.5-32.6, respectively. This suggests that the scatter in \textit{G} values of the ALPINE sample may be explained by a point source convolved with a Gaussian. On the other hand, the point source models in Figure \ref{fig:gm20U} occupy a narrow range of M$_{\rm 20}$, so the observed range of ALPINE M$_{\rm 20}$ sources is likely dominated by intrinsic source morphology. This simple test demonstrates that the observed data are not sufficient for classification on morphology alone. However, this is among the first applications of the Gini-M$_{20}$ morphological classification technique to a statistical sample of FIR observations of high-redshift ($z>4$) SFGs and will be useful for future investigations of galaxy morphology in the early Universe.

%The effects of low resolution and sensitivity on the recovered G and M$_{\rm 20}$ values were explored more fully in \citet{lotz04}. Both values are well recovered when the average SNR is greater than 2, while large deviations are apparent in both when the spatial resolution is $\ge1$\,kpc/px. Our SNR detection threshold is higher ($3\sigma$), but our approximate spatial resolution is comparable to this value (1\,kpc/px$\sim0.15''$/px at $z\sim5$). From these criteria, and the investigation shown in Figure \ref{fig:gm20U}, our data are not sufficient for morphological classification using this method due to a lack of spatial resolution. 

Future high-resolution spectral and continuum observations will likely enable this classification to be performed unambiguously. These include lines observable with ALMA (e.g., [CII]\,$158\mu$m, [OIII]\,$88\mu$m) or future observatories such as JWST (e.g., H$\alpha$, H$\beta$, [OIII]\,$5007$\,\AA, [OII]\,$3727$\,\AA), as well as the underlying rest-frame FIR and IR continuum emission.

\subsection{Disk Identification}\label{w15sec}

One of the strengths of this survey is the three-dimensional nature of our data (i.e., RA, Dec, velocity). This allows us to move beyond the two-dimensional morphological classification described above to examine the morpho-kinematics of each [CII] data cube. This was done in Section \ref{methsec} using \BB, and here we add modified versions of the five disk identification criteria of \citet{wisn15}: 
\begin{enumerate}
\item If a slice of the velocity field is taken along the kinematic major axis, does it show a slope that is significant (i.e., $>3\sigma$)?
\item Is the average rotational velocity greater than the average velocity dispersion, across all rings?
\item Is the average location of the two extreme values of the velocity field (moment 1 map) within one beam width of the peak of velocity dispersion (moment 2)?
\item Is the difference between the morphological and kinematic position angles less than $30^{\circ}$?
\item Is the average location of the two extreme values of the velocity field (moment 1 map) within one beam width of the peak of intensity (moment 0)?
\end{enumerate}
If some or all of these criteria are met for a given source, then it is likely a rotating disk galaxy. Conversely, if none are met, the source is likely disturbed or a merger. Each criterion may be assessed using the three moment maps (i.e., integrated intensity, line of sight velocity, velocity dispersion) and the results of a \BB tilted ring model fit. The results for each galaxy are presented in Table \ref{bbtab}.

The first criterion, which originally was the presence of a continuous velocity gradient, was made quantitative by requiring a significant slope in position-velocity space along the kinematic major axis. This replicates the standard ``rotating disk'' criterion used by many high-redshift investigations (e.g., \citealt{smit18,bakx20}), and agrees with a by-eye inspection that some sources show obvious gradients (e.g., VE.9038) and others do not (e.g., DC519281).

The second criterion requires the best-fit model to be strongly rotation-dominated. Perhaps surprisingly, this is met by almost all sources.

Both the third and fifth criteria relate to how well ordered the galaxy is. For an ideal disk, the extreme values of velocity would be at the extreme edges of the major axis, while both the integrated intensity and velocity dispersion would peak at the galaxy centre. On the other hand, the intensity peak in a merging system may be located in one of the component galaxies, resulting in a failure of criterion three. Similarly, the velocity dispersion map may be more complex, due to tidal features and shocks introduced by ongoing merging activity, resulting in a failure of criterion five.

The fourth criterion is a measure of how well the morphology and rotation pattern agree. Because the morphological position angle ($\rm PA_M$) is derived by fitting a 2D Gaussian to the [CII] moment zero map, a marginally resolved source will return a $\pm PA_M$ that is strongly influenced by the synthesized beam. The most illustrative case is DC396844, where the restoring beam is perpendicular to the kinematic axis, resulting in a failure of this criterion. On the other hand, CG32 is barely resolved (as seen by the best-fit $\rm PA_M=90\pm60^{\circ}$), but the kinematic angle happens to agree ($\rm PA_K=238\pm14^{\circ}$), so this criterion returns a positive value.

Only two sources in our subset meet all five criteria: DC432340 (classified as `uncertain') and DC818760 (classified as `merger', see Section \ref{synth}). Since neither of these are robustly determined to be rotating disk galaxies, we find that the resolution and sensitivity of our data are insufficient for this specific analysis alone.

While the W15 criteria only highlight one measure of the data each (e.g., agreement between moment maps, the presence of a velocity gradient), they are of use when combined with the detailed best-fit source characteristics from \BB (Figure \ref{BBresults1A} and \ref{BBresults1} through \ref{BBresults6}), as discussed in the next Section.

\section{Discussion}\label{DISCU}

\subsection{Synthesis of ALPINE Morpho-kinematic Classification}\label{synth}

By combining our new \BB fits and \citet{wisn15} disk criteria, we are now able to describe the morphology and kinematics of a massive subset of the galaxies in the ALPINE sample in unprecedented detail. Individual descriptions of each galaxy fitted with \BB are included in Appendix \ref{appdesc}.

We classify each source into one of three robust classes or a fourth `uncertain' classification:
\begin{itemize}
\item ROT: Rotators
\item MER: Mergers
\item DIS: Dispersion-dominated
\item UNC: Uncertain
\end{itemize}
The ROT and MER classes are identical to classes 1 and 2 of \citet{lefe20} respectively, while DIS is essentially a combination of their classes 3 and 4 (i.e., `extended dispersion-dominated' and `compact dispersion-dominated', respectively)\footnote{Since our signal isolation procedure (see Section \ref{SI}) has excluded all compact sources, no class 4 objects are included in our analysis.}. The `UNC' class contains sources that we are unable to classify due to low SNR, low spectral resolution, and/or conflicting evidence from our combination of analyses. Note that the UNC class is quite different from class 5 (i.e., `weak') of \citet{lefe20}, as it also contains strongly detected lines that we are unable to classify into ROT, MER, or DIS due to complex morpho-kinematics.

When classifying galaxies, we choose to be conservative. That is, a galaxy must clearly show multiple spatial and/or spectral components to be a merger, while it must exhibit a well-ordered velocity gradient and a single resolved source to be a rotator. Similarly, dispersion-dominated sources must show identical PVDs along orthogonal slices and be well-fit by a dispersion-dominated \BB model. If a source does not fit into these classes for any reason, it is classified as `UNC', regardless of its strength. 

Of course, this approach is not perfect, as it is still possible that close-separation mergers may be misidentified as rotators (\citealt{simo19}), and face-on rotators may be interpreted as dispersion-dominated galaxies (e.g., \citealt{koha19}). More importantly, these classes are a vast oversimplification of the true morpho-kinematics of each source, as each galaxy may contain ordered rotation, minor and major mergers, star-formation and AGN-driven outflows, filamentary accretion, and a number of other evolutionary processes that affect the kinematics, gas content, stellar mass, metallicity, and other properties of the galaxy and its environment (e.g., \citealt{law09,wisn11}, see also discussion in \citealt{valo20}). Indeed, cosmological simulations show that high-redshift galaxies are expected to interact with their environments through merging and accretion (e.g., \citealt{pall17,koha19,graz20,zane21}). So while these classes are worthwhile indicators of the general morpho-kinematic state of a galaxy, they may always be improved with higher-resolution, higher-sensitivity observations.

%\textbf{As detailed in Section \ref{gm20}, our data lacks the spatial resolution for morphological classification using a G-M$\rm _{20}$ diagram. Similarly, the majority of the W15 criteria are not useful for our classification (see Section \ref{w15sec}). We note that criteria 1 and 2 (i.e., the presence of a velocity gradient and a rotation-dominated best-fit model) are met by all of the galaxies classified as rotators.}

Of the 29 galaxies that were fit with $\ge2$ rings by \BB, 6, 5, 3, and 15 were found to be ROT, MER, DIS, and UNC, respectively. As discussed in Section \ref{subpro}, these 29 galaxies constitute a representative subset (i.e., in redshift and SFR) of the high-M$_{*}$  portion of the ALPINE sample. The significance of this diversity is explored in the next Section.

%With only 14 sources confidently classified, this analysis lacks the sample size to declare any statistical properties of the morpho-kinematics of galaxies at $z\sim4-6$. In addition, our various assumptions have introduced strong biases towards strong, extended line emitters. Future in-depth analyses of each object, and high-resolution, high-sensitivity follow-up observations are needed to meet this need. Nevertheless, the three analyses detailed here (i.e., \BB, G-M$_{\rm 20}$, and W15) show that there are indeed rotators, mergers, and dispersion-dominated galaxies in the ALPINE sample, adding further credence to the \citet{lefe20} conclusion of morpho-kinematic diversity of main sequence galaxies in the $\sim1-1.5$\,Gyr old Universe.

\subsection{Morpho-Kinematic Diversity Comparison}

To examine the evolution of morpho-kinematic diversity with redshift, we may compare the fractions of galaxies in various kinematic classes in the ALPINE sample found in this work to those of lower-redshift samples. In addition, we will place the results of \citet{lefe20} in this framework for the first time.

In the relatively low-redshift universe (i.e., $z=0.4-0.75$), the IMAGES survey observed 68 intermediate-mass galaxies with VLT/FLAMES-GIRAFFE, resulting in [OII]$\lambda\lambda3726,3729$\,\AA{} data cubes for each \citep{yang08}. Using the resulting moment 1 \& 2 maps, they separate 39 galaxies into 9 ($23\%$) rotating disks, 11 ($28\%$) perturbed rotators, 16 ($41\%$) having complex kinematics, and 3 ($8\%$) unclassifiable. Note that the `perturbed rotator' class contains sources that show disk-like rotation but have a velocity dispersion peak that is not coincident with the galactic centre. While this classification technique contains no possibility of dispersion-dominated sources or mergers, they may be hidden within the `complex kinematics' class, due to the low spatial resolution and field of view of the utilized instrument.

Using VLT/SINFONI, \citet{epin12} examined the H$\alpha$ velocity fields and I-band images of 45 galaxies at $z=0.9-1.6$ from MASSIV (Mass Assembly Survey with SINFONI in VVDS; \citealt{cont12}). This sample is chosen to be representative of SFR$\rm >5\,M_{\odot}\,yr^{-1}$ galaxies at $z\sim1.3$ with $\rm log(M_*/M_{\odot})\sim9-11$. Using a blind group classification method, they find that 19 ($42\%$) are rotating disks, 16 ($36\%$) are non-rotating disks, and 10 ($22\%$) are not robustly classified. Separately, they carried out a search for companions, finding that of the 41 galaxies where this search was able to be performed robustly, 28 ($68\%$) were isolated and 13 ($32\%$) were interacting or merging. 

%This agrees with the pilot sample of MASSIV, where only 9 galaxies at $z=1.2-1.6$ were observed \citep{epin09}. Of this small sample, 3 ($33\%$) are mergers, 2 ($22\%$) are rotating disks, and 4 ($45\%$) are dispersion-dominated systems. Indeed, the fraction of merging systems is in agreement ($33\%$ versus $32\%$ for \citealt{epin12}). However, the ratios of rotators to dispersion-dominated systems are quite discrepant (0.5 versus 1.2 for \citealt{epin12}), highlighting the need for large-number statistics.

%%%%

%The Spectroscopic Imaging survey in the Near-infrared with SINFONI (SINS; \citealt{fors09}) represented a major step towards characterizing the morpho-kinematic diversity of star forming galaxies at $z\sim2.2$. By considering the results of both automatic (i.e., kinemetry; \citealt{shap08}) and qualitative analysis of 62 galaxies at $z=1.3-2.6$ observed in H$\alpha$ emission, they find that the sample contains an even amount of mergers, rotation-dominated galaxies, and dispersion-dominated galaxies (i.e., $\sim33\%$ for all). 

A tremendous amount of work has been done to characterize the morpho-kinematics of galaxies around cosmic noon (i.e., $z\sim2$). By taking deep NIR adaptive optics-assisted (AO) observations from the Spectroscopic Imaging survey in the Near-infrared with SINFONI (SINS; \citealt{fors09}) and zCOSMOS (\citealt{lill07,lill09}) samples,  \citet{fors18} classified a representative subset of 35 sources from $z=1.35-2.38$ (i.e., SINS/zC-SINF). This sample contains $70\%$ rotation-dominated disks and $30\%$ other classes. This value agrees with the results of KMOS$^{\rm 3D}$ \citep{wisn19}, who found that $77\%$ of their $0.6<z<2.7$ galaxies were rotation-dominated. 
%highlights the amazing potentional of high-resolution observations to reveal the intrinsic kinematics of a galaxy.

%%%%

At higher redshift ($z\sim2.6-3.8$), \citet{gner11} observed a sample of galaxies from the AMAZE \citep{maio08} and LSD \citep{mann09} programs with VLT/SINFONI, resulting in a set of [OIII]$\lambda\lambda7007,4959$\,\AA{} and H$\alpha$ data cubes. By fitting each of their 32 velocity fields with a simple plane model, they find that 11 ($34\%$) are rotators, 14 ($44\%$) are non-rotators, and 7 ($22\%$) are unclassifiable\footnote{Based on the results reported in their Table 2.}. This percentage of rotators agrees well with that of the KMOS Deep Survey (KDS; \citealt{turn17}), which examined 77 $z\sim3.5$ SFGs and found that $34\pm8\%$ were rotation dominated.
%This results in a slightly higher ratio of rotators to dispersion-dominated systems ($\sim0.79$) than at lower redshift.

\citet{lefe20} used a similar blind group classification approach as \citet{epin12}: a group of collaborators examined moment maps, multi-wavelength photometry, spectra, and channel maps of the 75 ALPINE galaxies detected in [CII] at $>3.5\sigma$ by \citet{beth20}. They find 9 ($12\%$) rotators, 31 ($41\%$) mergers, 15 ($20\%$) extended dispersion-dominated galaxies, 8 ($11\%$) compact dispersion-dominated galaxies, and 12 ($16\%$) galaxies that were too weak to classify. There are two notes of interest here: the fraction of sources classified as mergers is higher than at lower redshifts ($49\%$, excluding the unclassifiable galaxies), and the ratio of rotators to dispersion-dominated systems is low ($0.39$). When compared to the $z\sim1$ kinematic diversity of \citet{epin12}, this suggests that mergers were more common for SFGs at high redshift. A more in-depth study of these merging systems in ALPINE is presented in \citet{roma21}.

By combining our analyses (see Section \ref{synth}), we are able to robustly classify 14 high-M$_*$ ALPINE galaxies, finding 6 ($43\%$) rotators, 5 ($36\%$) mergers, and 3 ($21\%$) dispersion-dominated sources. Surprisingly, this massive subset of the ALPINE sample contains a higher number of rotators than mergers.  The differences between these fractions and those of \citet{lefe20} are likely primarily due to two reasons: our exclusion of unresolved sources, and our stricter criteria for classification (see Section \ref{methsec}). The latter results in different classifications for some galaxies compared to \citet{lefe20}. As an example, \citet{lefe20} classified DC432340 as a merger (class 2) due to the presence of a northern extension and an asymmetric PVD. In this work, we classify it as UNC, as it is too asymmetric to be a ROT or DIS, and only shows a single source, excluding it from MER.

In general, we find that for the ALPINE sample, both the full set of [CII]-detected galaxies and the massive subset studied in this work feature significant morpho-kinematic diversity, containing rotators, mergers, and dispersion-dominated systems. This first statistical test of the diversity of star-forming main sequence galaxies only 1-1.5\,Gyr after the Big Bang suggests a higher number of mergers compared to lower redshift sources, and a number of massive rotators. This agrees with cosmological simulations, which show that while high-redshift galaxies are expected to interact with their environments through merging and accretion, the central galaxy may still feature strong rotation (e.g., \citealt{koha19,kret21}). 

\subsection{Main Sequence Rotators at z>4}\label{MSR}
A vast sample of rotating galaxies has been well studied at $z<4$ (e.g., \citealt{epin08,fors09,gner11}), revealing dynamical masses and empirical scaling relations. While a number of clumpy or merging galaxies have been detected at $z>4$ (e.g., \citealt{carn18,pave18,diaz18}), and a handful of rotating starburst or quasar host galaxies at $z\sim4-6$ have been well-modelled (e.g., \citealt{jone17,pens20,tada20,frat20,lell21}), only two unlensed rotating main sequence galaxies have been observed at $z\sim4-6$: HZ9 (\citealt{capa15}) and J0817 (\citealt{neel17,neel20})\footnote{Note that we exclude the galaxy HZ10 (likely a close-separation merger, \citealt{pave16}) and the strongly gravitationally lensed rotators of \citet{rizz20} and \citet{rizz21}.}. In this section, we add six new $z\sim4-6$ rotators from the ALPINE sample to this class: CG32, DC396844, DC494057, DC552206, DC881725, and VC.7875. 

\subsubsection{Previous Data}

To compare each source on equal footing, we re-analyse the [CII] kinematics of the two previous rotators by fitting tilted ring models using the \BB analysis detailed in Section \ref{methsec}. Both of these sources were previously fit with the 2-D tilted ring fitting code GIPSY \textlcsc{rotcur} \citep{jone17}, resulting in decent reproductions of each velocity field. However, the two-dimensional nature of this previous work made it impossible to account for beam smearing or velocity dispersion, resulting in fits that were not quite physical.

\begin{figure*}
\centering
\includegraphics[width=0.49\textwidth]{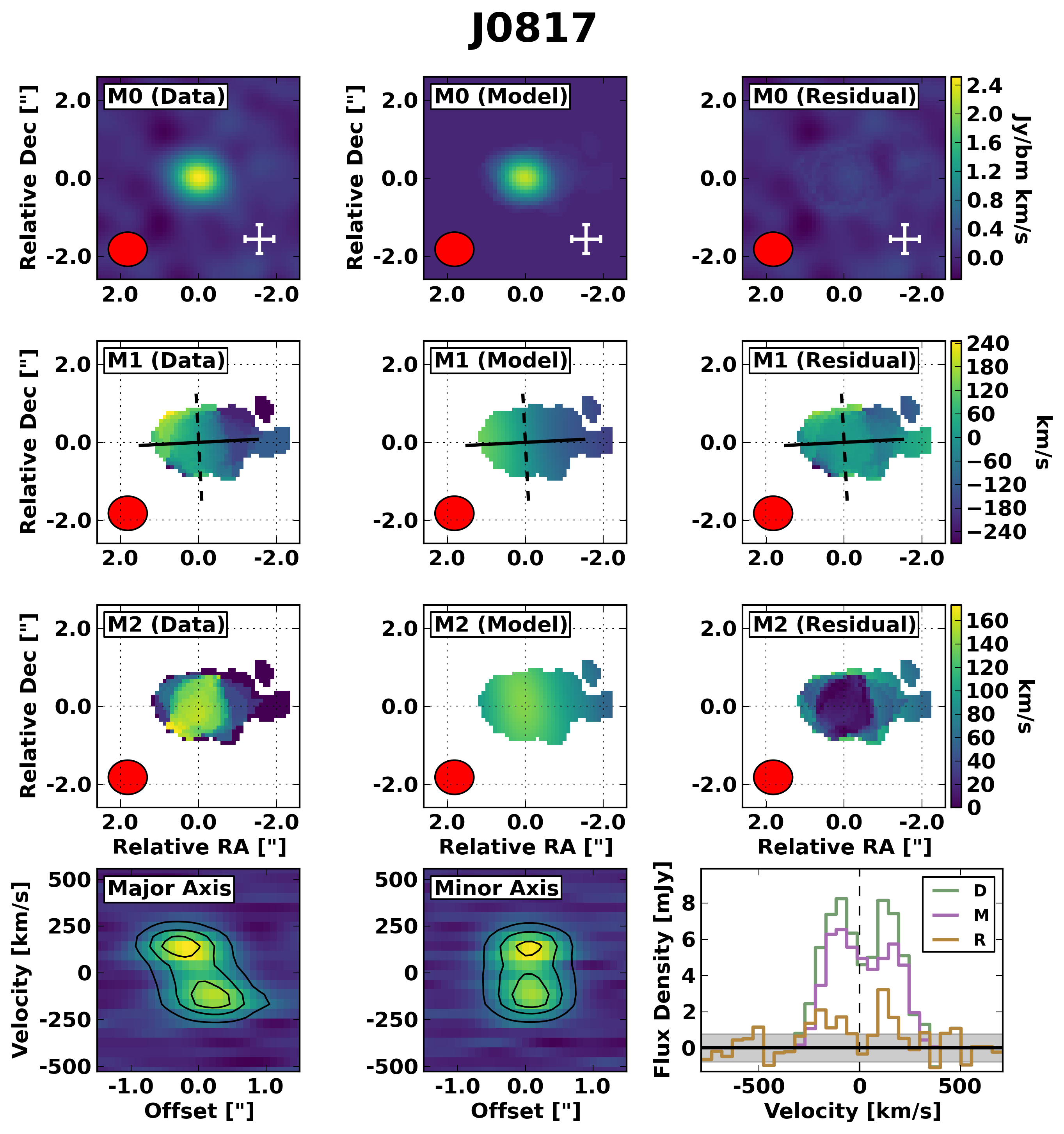}
\includegraphics[width=0.49\textwidth]{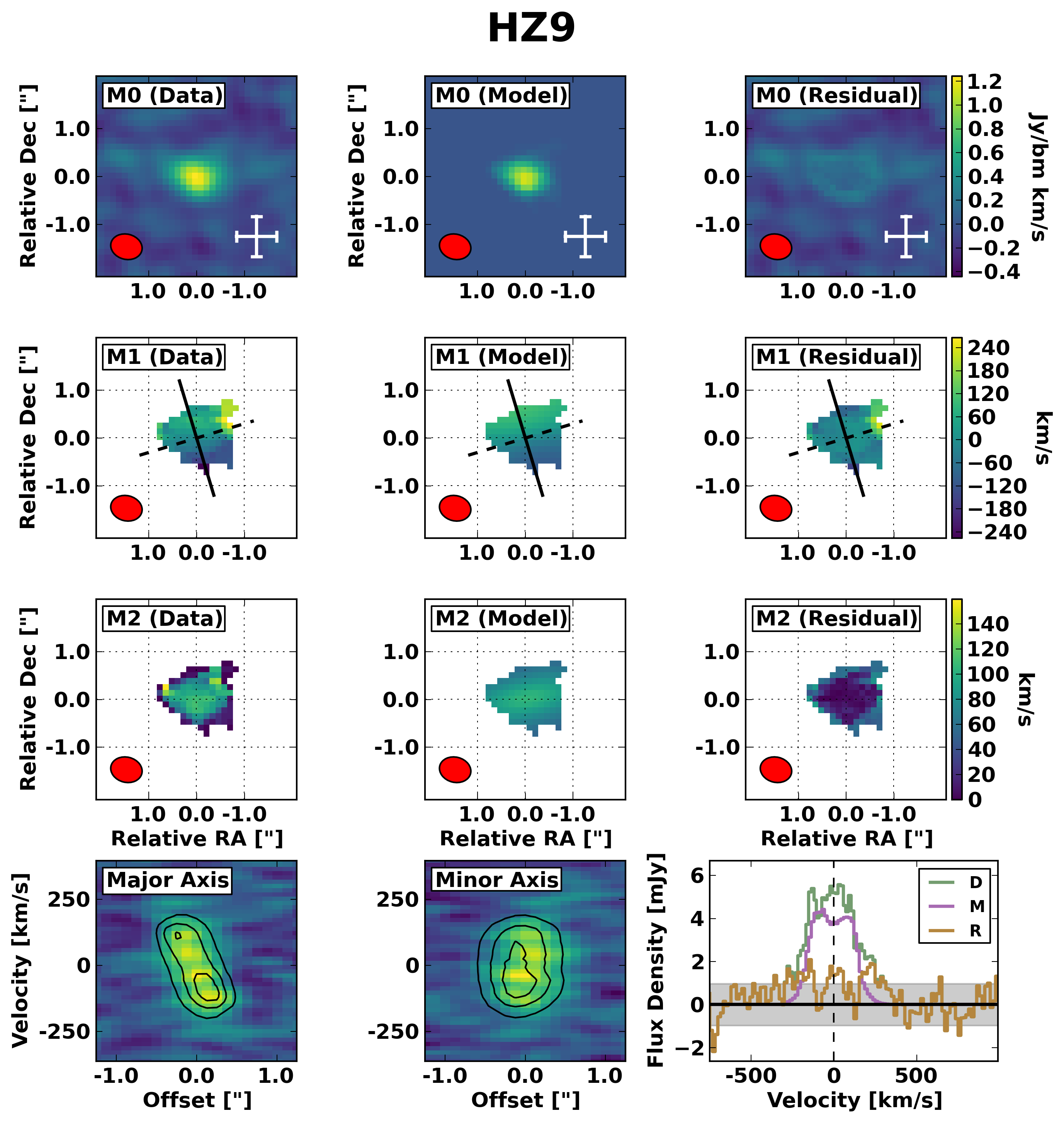}
\caption{Moment maps, PVDs, and spectra for observed data, model, and residual for the two previously detected $z>4$ main sequence, unlensed rotators not in the ALPINE sample. In each of these figures, the first three rows show (from top to bottom) the moment 0 (integrated intensity), moment 1 (velocity field), and moment 2 (velocity dispersion field; see \ref{PE} for details of moment map creation). For these rows, the three columns denote (from left to right) the observed data cubes, model cubes, and the corresponding residuals. The white crosses in the lower right corner of each panel in the first row show a 5\,kpc$\times$5\,kpc physical scale.The solid lines in the second row represent the kinematic major axis, while the dashed lines represent the minor axis. The bottom row shows (from left to right) the major axis PVD, minor axis PVD, and extracted spectra. For each PVD, the observed data are shown by the background colour, while the contours represent the model. The data (D), model (M), and residual (R) spectra are depicted by the green, purple, and orange lines, respectively. The $1\sigma$ uncertainty, calculated as (average RMS noise level per channel)$\rm \times\sqrt{number\,\,of\,\,beams\,\,in\,\,source}$, is shown by the shaded grey area. }
\label{Extra1}
\end{figure*}

The data for HZ9 comes from \citet{capa15}, and is the same data cube analysed by \citet{jone17}. The galaxy J0817 has been observed in [CII] emission with ALMA at two different resolutions: $\sim1''$ \citep{neel17} and $\sim0.2''$ \citep{neel20}.  To have similar spatial resolution to the ALPINE data, we choose to analyse the $1''$ J0817 [CII] data cube of \citet{neel17}, and will examine the $\sim0.2''$ observation of \citet{neel20} in Section \ref{08172}. Each of these cubes was passed through the same \BB analysis as the ALPINE sources, resulting in the best-fit parameters listed in Table \ref{bbtab} (see also Figure \ref{Extra1}).

Both of these data cubes are well-fit by \BB, with small residuals in each moment map and agreement between the model and data PVDs.

\subsubsection{Addition of ALPINE Sources}

\begin{figure}
\centering
\includegraphics[width=\columnwidth]{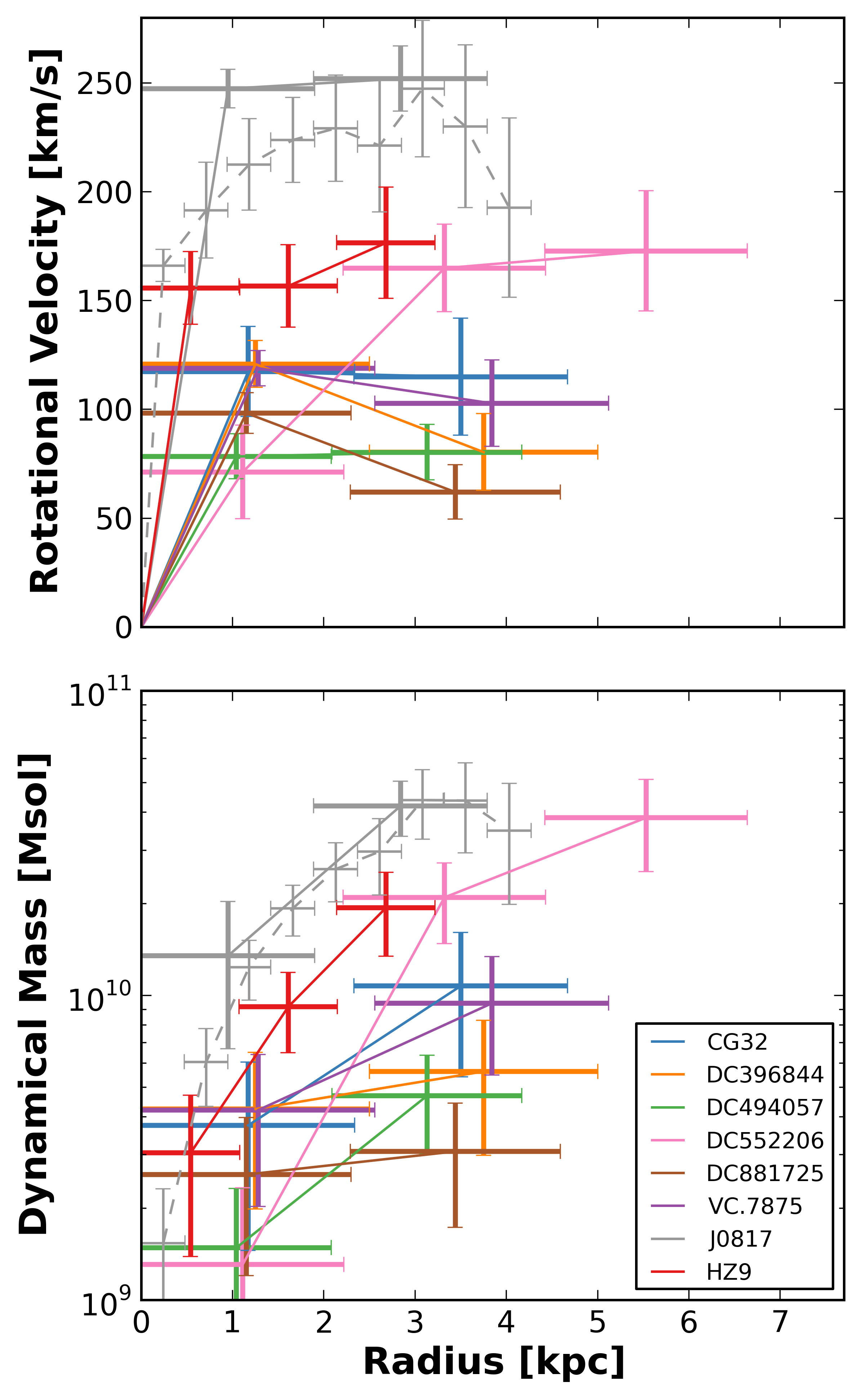}
\caption{Comparison of the best-fit rotation curves (top) and dynamical mass profiles (bottom) for the six new unlensed, main sequence $z\sim4-6$ rotators from ALPINE and the two previous such objects. For J0817, we show the results for both the $1''$ (solid grey lines) and $0.2''$ data (dashed grey lines; see Section \ref{08172}). Vertical extent of each point in the top panel shows the fit uncertainty output by \BB (i.e., not including systematic uncertainty on the inclination), while the vertical extent of each bar in the lower panel is the dynamical mass uncertainty (see equation \ref{mdeq}) accounting for the fit uncertainty in rotational velocity output by \BB and a radius uncertainty of half a ring width. The horizontal extent of each point shows the width of a ring. }
\label{ROTCOMP}
\end{figure}

We compare each of the best-fit rotation curves and dynamical mass profiles of the eight unlensed, main sequence $z\sim4-6$ rotators in Figure \ref{ROTCOMP}. The dynamical mass is estimated using the radius and circular velocity of each ring, assuming perfectly circular rotation:
\begin{equation}\label{mdeq}
\rm M_{dyn}=\frac{v^2r}{G}
\end{equation}
where G is the gravitational constant (e.g., \citealt{wang10,debr14,lang20}). The values used to create the rotation curves, velocity dispersion profiles, and dynamical mass profiles are listed in Appendix \ref{rottabsec}.

Due to the highly nonlinear nature of the kinematic fitting method used in \BB, the inclusion of this inclination uncertainty in the rotation curves or dynamical mass profiles is nontrivial. Since \BB fits each ring individually, a different inclination will result in both a different spatial distribution and kinematic properties of each ring, vastly changing the model. Due to this complexity, we simply note that the uncertainty in the best-fit inclination induces an additional source of uncertainty in each rotation curve and dynamical mass profile, which is not depicted in Figure \ref{ROTCOMP}.

The dynamical profiles of these sources may be compared to those derived through 2D-modelling by \citet{jone17}, which includes both main sequence and starburst galaxies. We find smaller spatial extents (i.e., maximum model ring radii), due to our accounting for beam smearing, resulting in smaller dynamical masses. This may be seen in HZ9, where the 2-D approach yielded a maximum dynamical mass of $\sim4\times10^{10}$\,M$_{\odot}$ and a spatial extent of $3.8$\,kpc, while we find $\rm M_{dyn}\sim2\times10^{10}$\,M$_{\odot}$ and a spatial extent of $2.7$\,kpc. This confirms the well-known fact that beam smearing has severe impacts on the recovered kinematic parameters of marginally resolved objects, and must be taken into account (e.g., \citealt{bege89,debl08,moli19}).

When examining the rotation curves in Figure \ref{ROTCOMP}, it is evident that the rotators in this sample show high inclination-corrected rotation speeds ($\sim50-250$\,km\,s$^{-1}$) and modest spatial extents ($r<7$\,kpc). The dynamical mass profiles for these sources show some variety, with some rising rotation velocity profiles (i.e.,  DC552206, J0817, HZ9),  approximately flat profiles (i.e., CG32, DC494057, VC.7875), and declining profiles (i.e., DC396844, DC881725).

At first glance, these declining rotation curves could be interpreted as evidence for high baryon fractions (or a weak dark matter halo), as suggested in some analyses of $z\sim1-3$ galaxies (e.g., \citealt{lang17,genz17}). However, there are several main points that do not allow such a strong conclusion. Primarily, the [CII] observations of these sources have synthesized beam sizes of $\sim7$\,kpc, and thus smooth over large swathes of each galaxy (i.e., beam smearing). This effect of low resolution is also seen in the small number of rings in each model. Similarly, while the cores of these galaxies are very well detected, the weaker outskirts may be confused by noise contributions. Finally, each of these tilted ring models includes a nonzero velocity dispersion for each ring, which is not accounted for in these dynamical masses. 

Interestingly, half of the ALPINE rotators also show significant [CII] halos (DC396844, DC881725, VC.7875; \citealt{fuji20}). The coincidence of enriched (i.e., not pristine) halos and ordered rotation may be caused by a number of phenomena, including star formation-driven outflows, AGN activity (unlikely in the ALPINE sample, which was chosen to exclude type 1 AGN), and past merger activity. Indeed, cosmological simulations (e.g., \citealt{koha19}) have shown that galaxies may undergoing major mergers and revert to a rotation-dominated state a short time (i.e., tens of Myr) later. 

Here, we are able to show that our current data strongly argues for order, disk-like rotation in each of these high-redshift galaxies, and that their rotation curves show a variety of shapes over $r\sim5$\,kpc. The detailed comparison of how these rotators compare to low-redshift rotating galaxies (e.g., $v/\sigma$, M$_{\rm dyn}$ vs M$_*$, resolved gas fractions) will be presented in a future work (Rizzo et al. in prep). In addition, high-resolution observations will allow us to delineate the small-scale kinematics of these rotating galaxies, and low-resolution observations will reveal the kinematics of their surrounding gaseous halos.

\subsection{Effects of Higher Spatial Resolution}\label{08172}

Throughout this work, we have focused on [CII] datasets from the ALPINE survey, which feature spatial resolutions of $\sim1''$. For the ALPINE sources (i.e., $4.4<z<5.9$), this spatial scale corresponds to $\sim6-7$\,kpc. This resolution is sufficient to identify a number of rotators, mergers, and dispersion-dominated objects, and to estimate the dynamical mass of each. This morpho-kinematic characterization may be made even more precise and informative by increasing the spatial resolution of observations. 

To illustrate the effects of increased spatial resolution on the recovered kinematics, we apply our \BB analysis to the recent high-resolution ALMA [CII] observation of J0817 ($z=4.2603$; $0.2''\sim1.4$\,kpc; \citealt{neel20}). The $1''\sim6.8$\,kpc [CII] observations \citep{neel17} of this source were analysed in Section \ref{MSR}.

\begin{figure}
\centering
\includegraphics[width=\columnwidth]{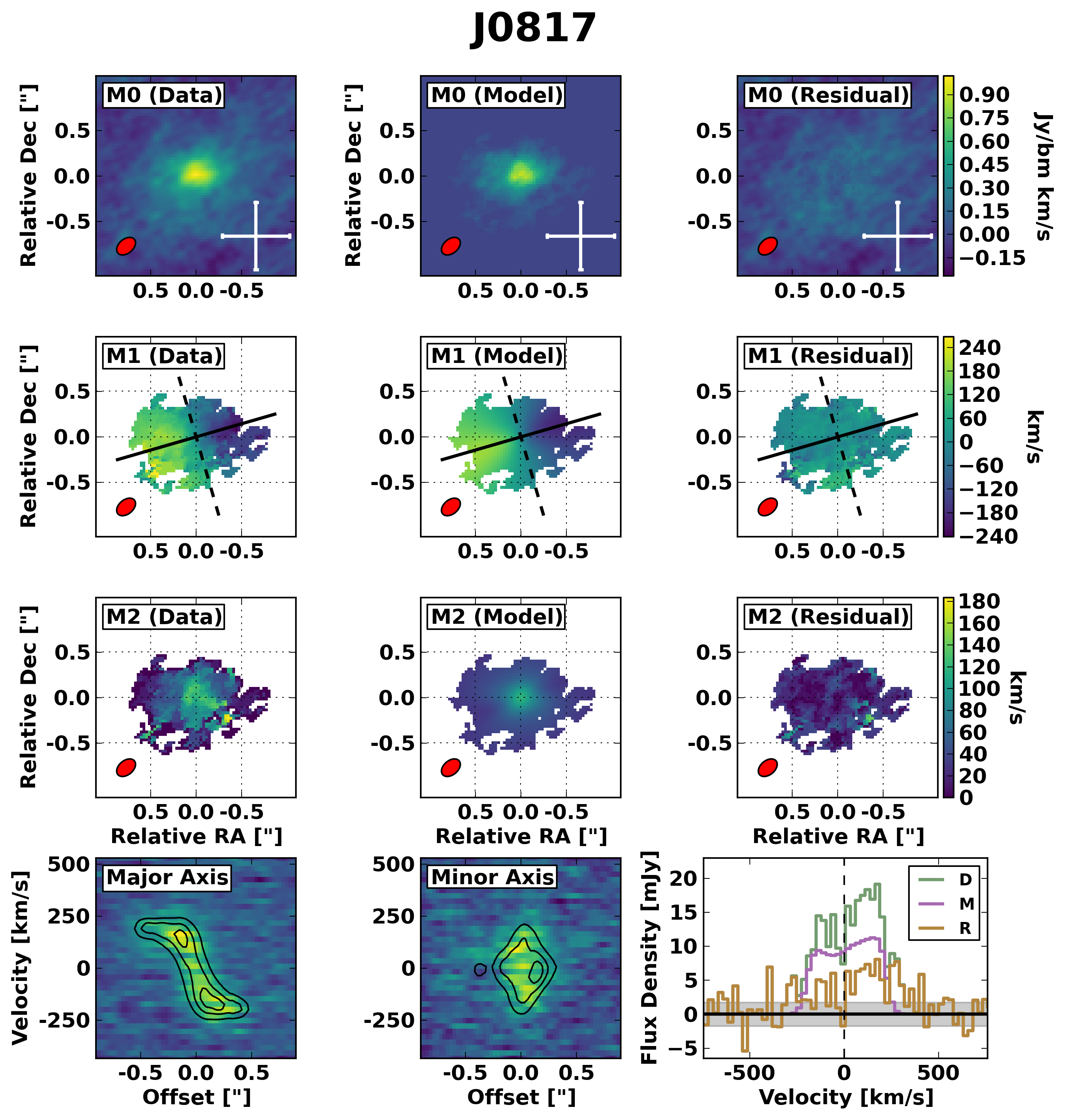}
\caption{Moment maps, PVDs, and spectra for observed data, model, and residual for the high-resolution ($0.2''$) data of J0817 from \citet{neel20}. In each of these figures, the first three rows show (from top to bottom) the moment 0 (integrated intensity), moment 1 (velocity field), and moment 2 (velocity dispersion field; see \ref{PE} for details of moment map creation). For these rows, the three columns denote (from left to right) the observed data cubes, model cubes, and the corresponding residuals. The white crosses in the lower right corner of each panel in the first row show a 5\,kpc$\times$5\,kpc physical scale. The solid lines in the second row represent the kinematic major axis, while the dashed lines represent the minor axis. The bottom row shows (from left to right) the major axis PVD, minor axis PVD, and extracted spectra. For each PVD, the observed data are shown by the background colour, while the contours represent the model. The data (D), model (M), and residual (R) spectra are depicted by the green, purple, and orange lines, respectively. The $1\sigma$ uncertainty, calculated as (average RMS noise level per channel)$\rm \times\sqrt{number\,\,of\,\,beams\,\,in\,\,source}$, is shown by the shaded grey area.}
\label{BBresultsJ0817B}
\end{figure}

We present the results of applying the \BB analysis (Section \ref{methsec}) to the high-resolution J0817 [CII] data in Figure \ref{BBresultsJ0817B}. The $\sim1.4$\,kpc resolution of these data enables us to resolve the galaxy over many beams (i.e., $>5$), revealing the presence of a central bulge (see moment zero map), a complex velocity field, and an non-axisymmetric velocity dispersion map. \BB is able to reproduce the moment 1 and 2 maps, as well as the PVDs. The spectrum features a significant red residual, which suggests that these high-resolution data expose non-rotating components of the J0817 system.

We may compare the observed morpho-kinematics of J0817 at $0.2''$ (Figure \ref{BBresultsJ0817B}) and $1''$ (Figure \ref{Extra1}) spatial resolution. The effects of beam smearing are drastically reduced in the high-resolution data, as seen by the more compact moment zero map, complex velocity field and sharp pair of PVDs. The rotator nature of this source is discernible from the low-resolution data, but high-resolution observations allow for more detailed characterization.

Finally, we may compare the recovered rotation curves and dynamical mass profiles for this galaxy at each resolution (Figure \ref{ROTCOMP}). Using the $0.2''$ data, we are able to explore the rotation curve at higher spatial resolution. The observed curve is in agreement with a steep initial rise in circular velocity, followed by a flattening at large radii (as seen by \citealt{neel20}). The recovered velocities are slightly lower than those recovered from the $1''$ resolution data, suggesting a different kinematic inclination angle. However, the two dynamical mass profiles are in complete agreement, suggesting that the $1''$ resolution data are sufficient for this analysis.

\section{Summary \& Conclusions}\label{CONC}

In this work, we have investigated the morpho-kinematics of the star forming main sequence galaxies in the ALMA ALPINE survey using three quantitative analyses: a tilted ring model fitting code (\BB), a morphological classifier (\textit{G}-M$_{20}$), and the disk-like criteria of \citet{wisn15}.

\begin{itemize}

\item For each galaxy that we fit with \BB 3DFIT (29 sources), we derive the best-fit morphological and kinematic parameters (e.g., position angle, inclination, rotation curve, velocity dispersion profile, and dynamical mass) and present moment maps, PVDs, and spectra, yielding a more detailed view into their kinematic and morphological nature. By examining the placement of this subset of 29 galaxies on the M$_*$-SFR plane, we find that it contains high-M$_*$ ($>10^{9.5}\,$M$_{\odot}$) objects that are representative of the star-forming main sequence at this redshift. The morpho-kinematic diversity of these $z\sim4-6$ galaxies is further supported, with six rotators, five mergers, and three dispersion-dominated galaxies in our high-M$_*$ sample. The morpho-kinematic diversity of the ALPINE sample is placed in context with low-redshift studies, suggesting an increase in number of mergers at $z>4$.  This diversity implies for a number of pathways to building up the mass of these sources (e.g., mergers, filamentary accretion). 

\item To examine the morphology of the ALPINE sample in an additional quantitative manner, we perform one of the first investigations of the distribution of high-redshift ($z>4$) galaxies on the Gini-M$_{\rm 20}$ plane (e.g., \citealt{lotz04}) using maps of the integrated [CII] and rest-frame FIR emission. We find that our sources do not separate by morpho-kinematic class and feature low $G$ (signifying a large presence of diffuse emission. To test whether this scatter is due to the low resolution of our observations (i.e., 1-3 beams per source), we simulate the scatter in this plane due to point sources of different SNR, finding that the scatter in \textit{G} is primarily due to the low resolution of our observations (i.e., $\rm\sim1\,kpc/px$). This groundbreaking test suggests that future, high-resolution ALMA observations of high-$z$ galaxies will enable the use of this diagnostic plot.

\item We also apply the disk criteria of \citet{wisn15} to the 29 galaxies where \BB 3DFIT modelling was possible due to sufficient resolution and S/N. We find that the first two \citet{wisn15} criteria are useful for identifying rotating galaxies, but these criteria are not suitable for the current data when used alone.

\item We compare the six confirmed main sequence $z\sim4-6$ rotators in the ALPINE sample to the two previously detected unlensed sources, finding high inclination-corrected rotational velocities, little evidence for declining rotation curves, and comparable dynamical mass profiles. While future high-sensitivity observations will be required to place stronger constraints on the behaviour of these rotation curves at large radii, these current results show strong evidence for ordered rotation in a sample of $z\sim4-6$ SFGs.

\item To test the reliability of our recovered morpho-kinematics, we apply the same tilted ring model fitting procedure as above to both a low-resolution ($\sim1''$) and high-resolution ($\sim0.2''$) [CII] observation of the $z\sim4.26$ SFG J0817. The low-resolution observation is well-fit by such a model, revealing the intrinsic rotation of the galaxy. Thus, our $\sim1''$ resolution ALPINE data are appropriate for morpho-kinematic characterization and dynamical mass derivation. By using higher-resolution data, we may confirm these, and derive more detailed kinematics of the system (e.g., deviations from circular rotation).

\end{itemize}
By applying these analyses to the statistical sample of ALPINE, we may gain insight into the morpho-kinematics of main sequence galaxies at $z>4$. Future high-resolution [CII] observations will allow us to precisely determine the merger rate, rotation properties, turbulence support, and dynamical mass of each source. In parallel, additional detailed investigations of single targeted (e.g., \citealt{jone20,gino20b}) and serendipitous (e.g., \citealt{roma20}) galaxies will reveal the morpho-kinematics of early galaxies. Together, these statistical and single observations will inform current cosmological simulations (e.g., \citealt{koha19}) and allow us to precisely determine the kinematics (e.g., merger rate, relative populations of kinematic classes, baryon cycle) of galaxies in the early Universe.

\section*{Acknowledgements}
We thank the anonymous referee for useful feedback that resulted in an improved manuscript.
This paper is based on data obtained with the ALMA Observatory, under Large Program 2017.1.00428.L. We also analyse archival data from ALMA programs 2012.1.00523.S and 2017.1.01052.S. ALMA is a partnership of ESO (representing its member states), NSF (USA) and NINS (Japan), together with NRC (Canada), MOST and ASIAA (Taiwan), and KASI (Republic of Korea), in cooperation with the Republic of Chile. The Joint ALMA Observatory is operated by ESO, AUI/NRAO and NAOJ. Based on data products from observations made with ESO Telescopes at the La Silla Paranal Observatory under ESO programme ID 179.A-2005 and on data products produced by TERAPIX and the Cambridge Astronomy Survey Unit on behalf of the UltraVISTA consortium.
This program is supported by the national program Cosmology and Galaxies from the CNRS in France. 
G.C.J. and R.M. acknowledge ERC Advanced Grant 695671 ``QUENCH'' and support by the Science and Technology Facilities Council (STFC). 
J.D.S. was supported by the JSPS KAKENHI Grant Number *JP18H04346*, and the World Premier International Research Center Initiative (WPI Initiative), MEXT, Japan.
A.C. and M.T. acknowledge the support from grant PRINMIUR 2017 $-$ 20173ML3WW$\_$001.
E.I. acknowledges partial support from FONDECYT through grant N$^\circ$\,1171710.
This paper is dedicated to the memory of Olivier Le F\`evre, PI of the ALPINE survey.

\section*{Data Availability}
The data analysed in this work are available from the ALMA data archive (\url{https://almascience.nrao.edu/asax/}) under project code 2017.1.00428.L. Catalogs and data products are available from the ALPINE Data Release page (\url{https://cesam.lam.fr/a2c2s/data_release.php}).

\bibliographystyle{mnras}
\bibliography{ALPBB}
\label{lastpage}

\appendix

\section{Additional Notes on Tilted Ring Modelling}\label{morebb}

In Section \ref{methsec}, we present an overview of our signal isolation and tilted ring model fitting procedure, as well as the results and properties of the analzyed subsample. Here, we discuss several assumptions that we have made in developing this process.

We have excluded all serendipitous (i.e., non-target) sources, which are analysed in \citet{loia20}. By definition, each of these sources is not at the phase centre of our observations, and thus require a separate primary beam correction for accurate morpho-kinematic analysis. Since they were not selected as part of the ALPINE sample, they do not necessarily represent SFGs at $z\sim4.4-5.9$. For example, they include sources at $z\sim1$ (S5110377875, S460378) as well as a well-studied SMG at $z=4.54$ (S842313, also known as AzTEC/C17;  \citealt{aret11}, COSMOS J100054+02343; \citealt{cari08,capa08,schi08}). Because of this variety, they will not be used in our current goal of probing the kinematic diversity of this specific class of galaxies. However, a number of these sources were successfully fit with our tilted ring model analysis (e.g.,  S842313), so their morpho-kinematics may be analysed in a future work.

As part of the fitting procedure, we only allow inclination to vary from $10-80^{\circ}$, excluding face-on ($i<10^{\circ}$) and edge-on ($i>80^{\circ}$) disks. Face-on disks feature large spatial extents, but low line-of-sight velocities, resulting in dispersion-dominated, Gaussian line profiles (e.g., \citealt{koha19}). These sources are indistinguishable from true dispersion-dominated galaxies. On the other hand, edge-on disks are dominated by velocities along the line of sight, but have small axis ratios, and are likely to be poorly resolved along their minor axis. Because of these features, \BB has been found to return poor fits for extreme inclinations (e.g., \citealt{dite15}).

Our analysis is based on signal isolation using \BB SEARCH, rather than a threshold-based 2D or 3D mask. This has the benefit of excluding low-level noise within the mask, but the relatively high lower SNR threshold (SNR$_{\rm lower}=2.5$) causes the \BB fits to miss diffuse, low-level emission, as seen in the nonzero residuals in the spectra and moment zero maps. Indeed, recent ALMA studies (e.g., \citealt{fuji19,fuji20,gino20a}) have revealed extended [CII] emission around $z>4$ galaxies, implying the presence of an diffuse, enriched circumgalactic medium (CGM). Because this CGM emission is not captured with our signal isolation technique, the scope of the kinematic analysis in this paper is focused on the relatively compact interstellar medium (ISM) of the galaxy. 

The \BB 3DFIT routine allows the user to choose from two methods of normalizing the model. The first (`local') rescales the flux density of each pixel in the model data cube so that the moment zero map of the model is equal to the moment map of the masked input data cube. The second ('azimuthal') performs a similar rescaling procedure, but requires the model to have azimuthal symmetry, resulting in an elliptical model. To account for the complex morphologies of the ALPINE sources, we have assumed `local' normalization. This choice has no effect on the recovered moment 1 or moment 2 maps, but allows our model to fit models more complex than a simple symmetric disk. In addition, objects that are obviously not simple rotators (such as the merger VC9780 in Figure \ref{BBresults1A}) return best-fit moment 0 maps that have multiple components or disturbed morphologies, adding additional evidence to their merger classifications.

The majority of our sources have only $2-3$\,rings per model (Figure \ref{rcs}). This is much fewer than low-redshift studies (e.g., $>20$ for \citealt{debl08,lang20}), but is comparable to other $z>4$ studies (e.g., \citealt{jone17,frat20}). This low number of rings is a direct result of the relatively low spatial resolution of these observations (i.e., $\sim1''\sim7\,$kpc), as this dictates the spatial scale that we may examine. Each of these tilted ring models is able to reproduce the morphology and kinematics of each galaxy (see Figures \ref{BBresults1A} and \ref{BBresults1} through \ref{BBresults6}). So while our current analysis is not adversely affected by a limited number of free parameters (or ring number), future high-resolution observations will allow for the examination of small-scale morpho-kinematics (e.g., \citealt{neel20}).

\section{Galaxy Details}\label{appdesc}
The diversity of this sample makes it worthwhile to comment on each source successfully fitted with \BB individually. Here, we present moment maps, PVDs, and spectra created from the data, \BB model, and residual for each source. In addition, we discuss the reasoning behind each morpho-kinematic classification (see Section \ref{synth}) and compare them to the classifications of \citet{lefe20}. The rotation curves and velocity dispersion profiles of each source are depicted in Figure \ref{rcs}, while the best-fit parameters and W15 criteria are listed in Table \ref{bbtab}.

\begin{figure*}
\centering
\includegraphics[width=0.49\textwidth]{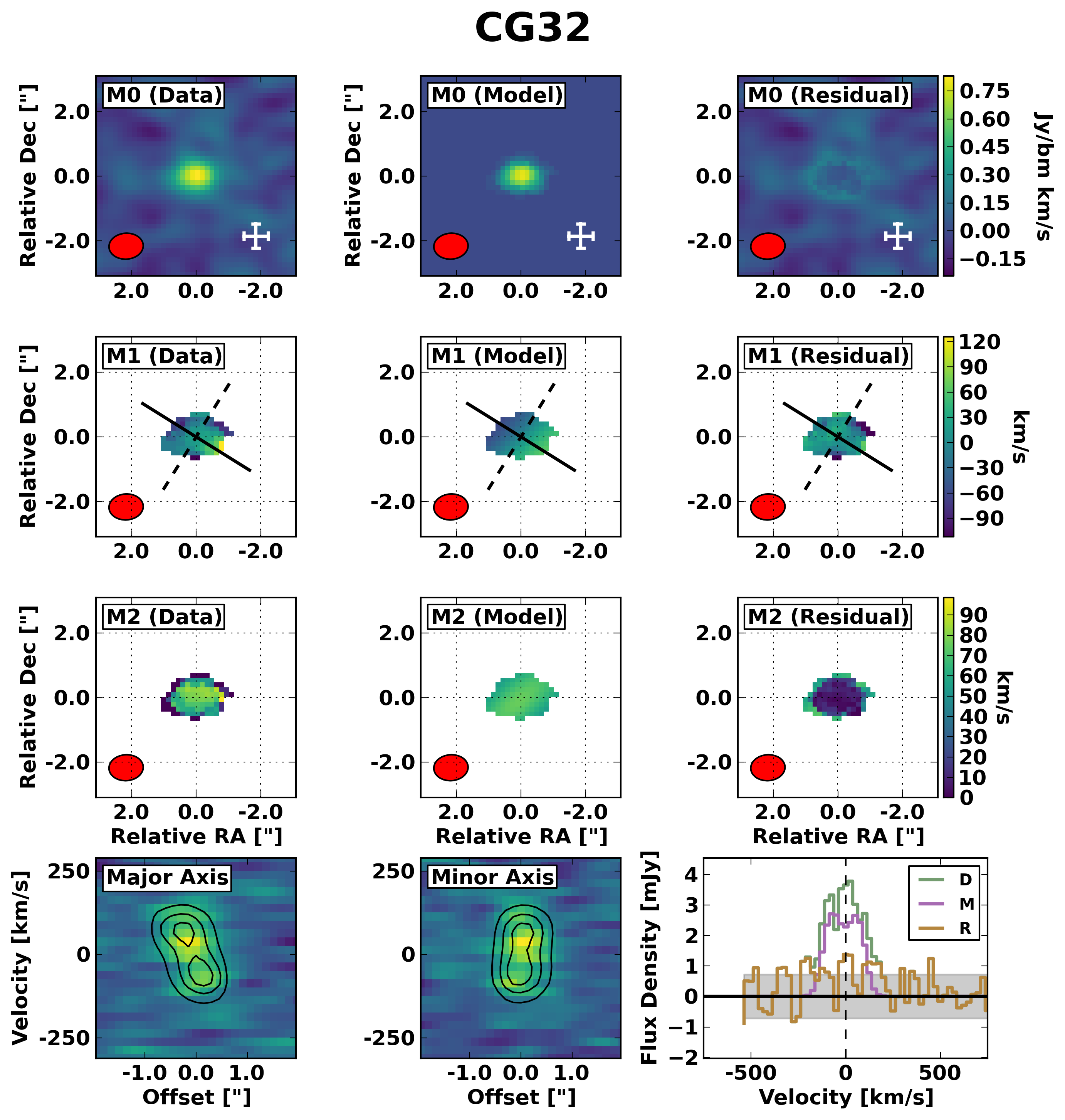}
\includegraphics[width=0.49\textwidth]{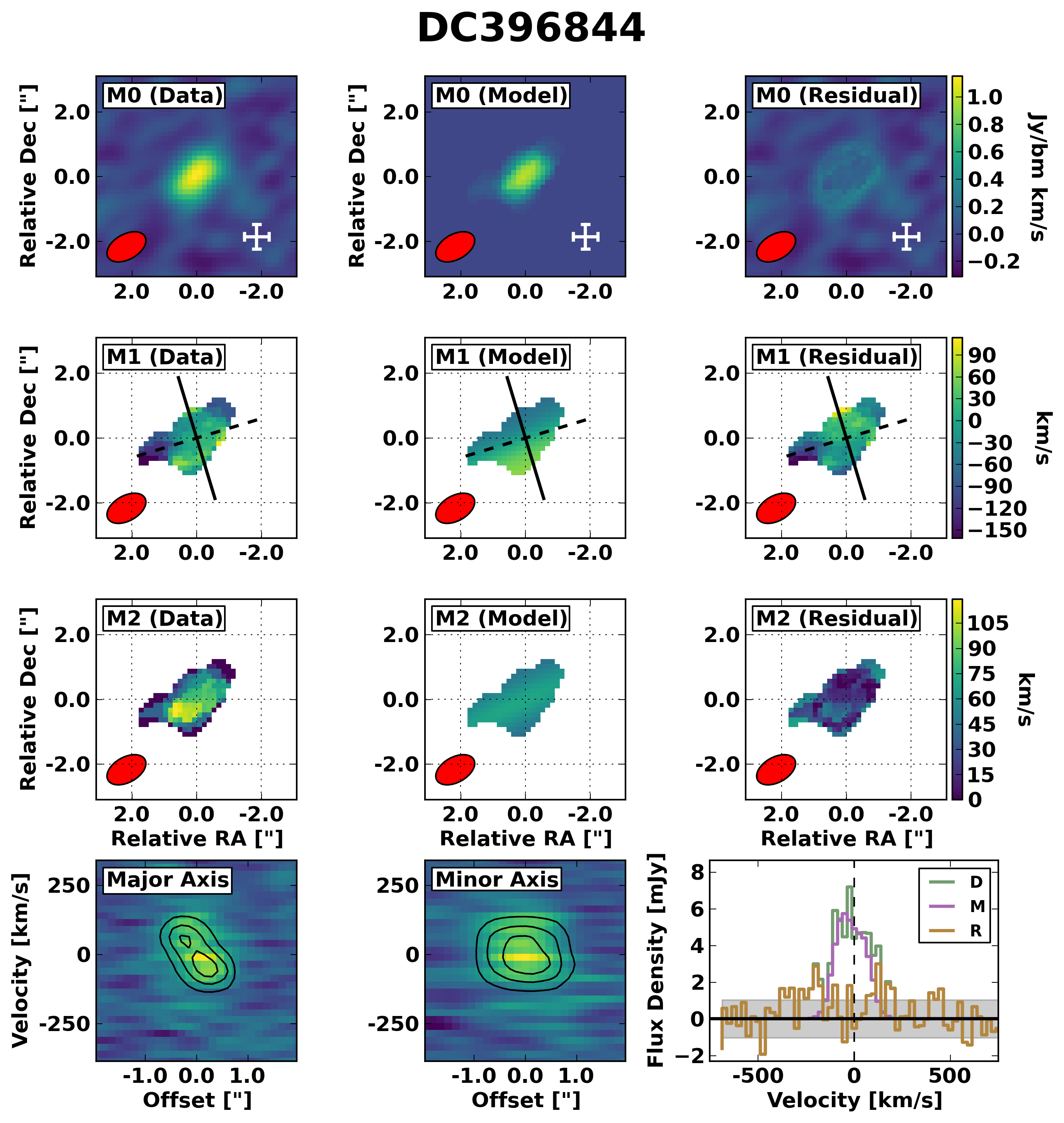}
\includegraphics[width=0.49\textwidth]{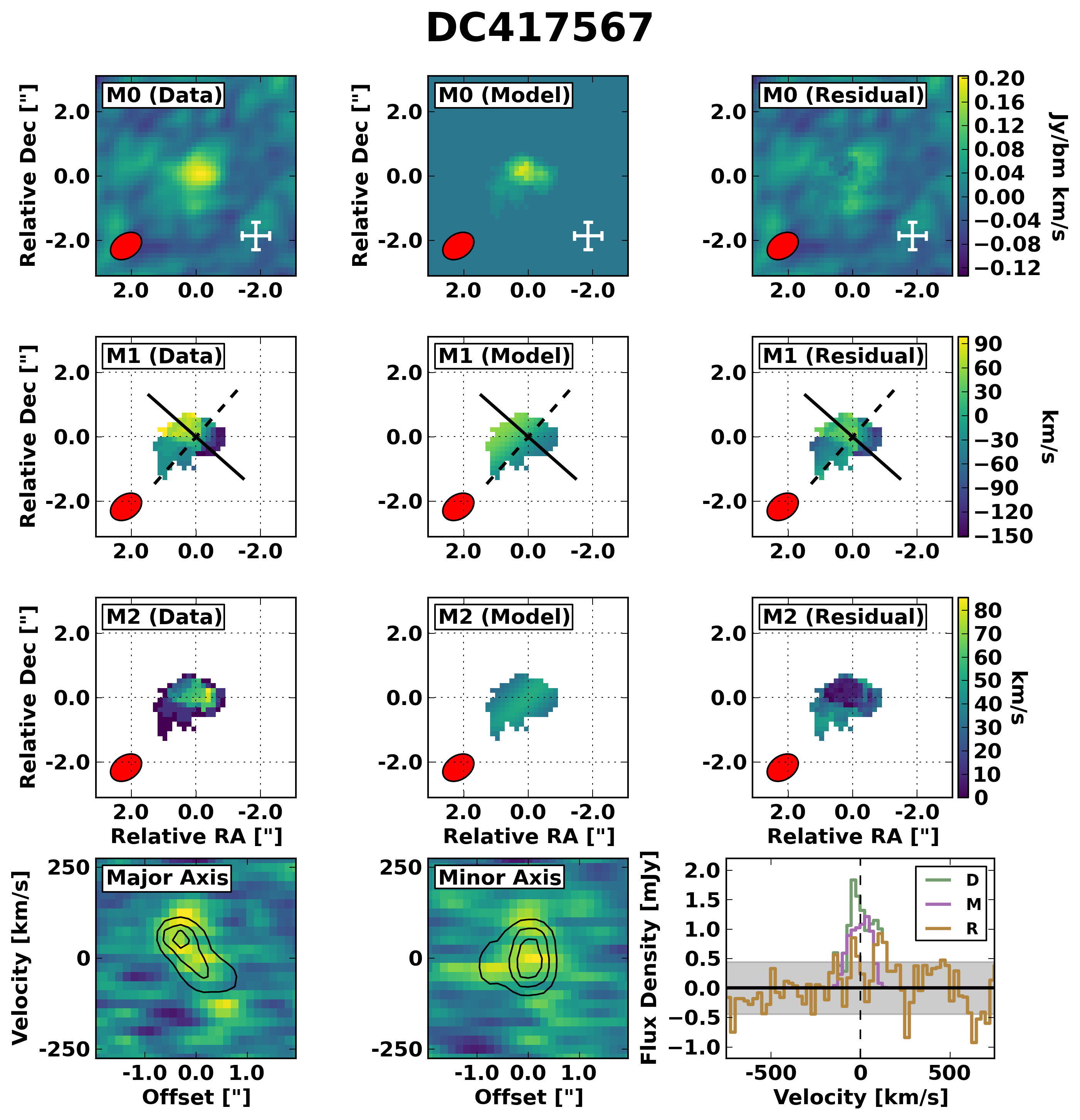}
\includegraphics[width=0.49\textwidth]{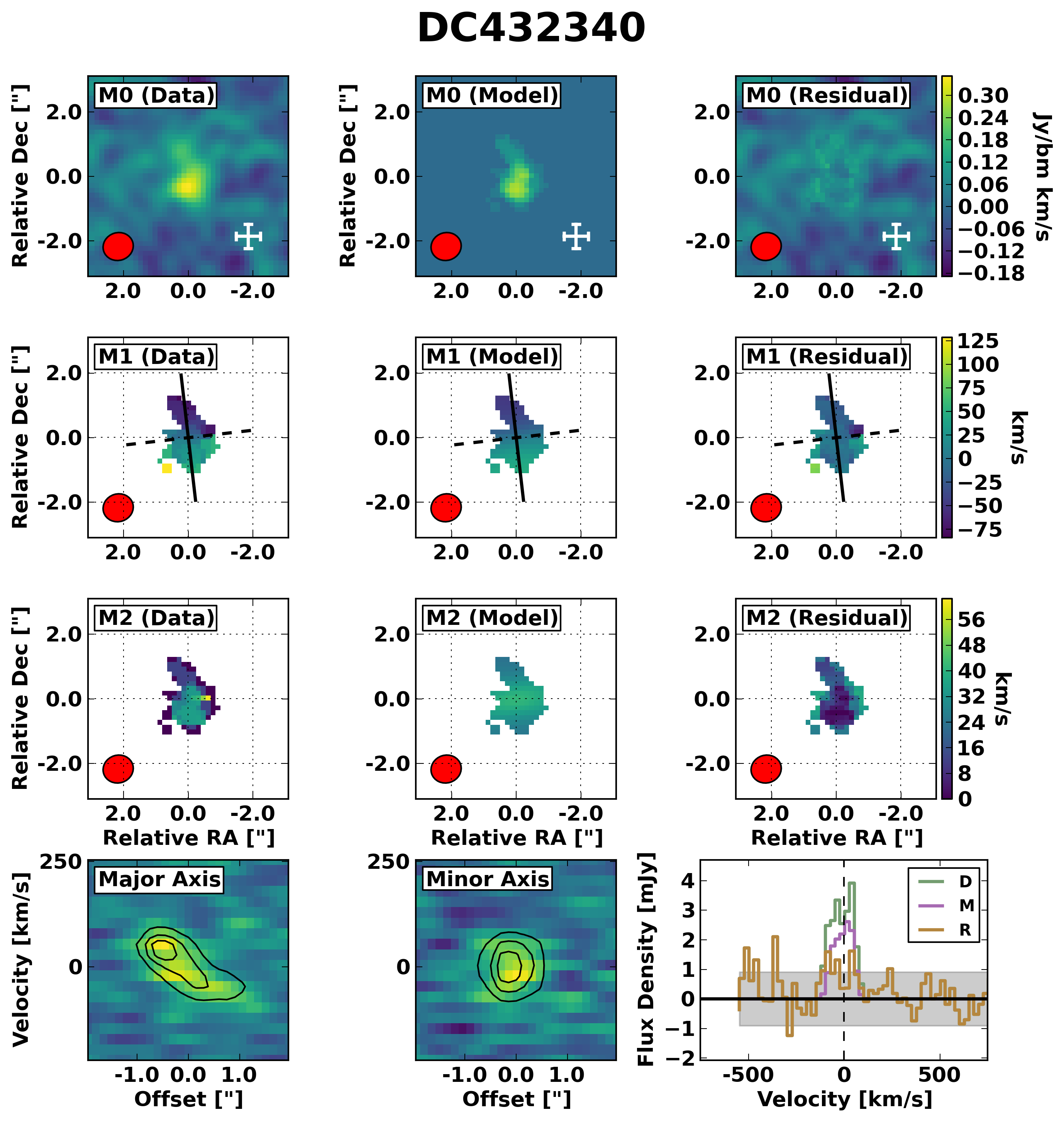}
\caption{Moment maps, PVDs, and spectra for observed data, model, and residual. In each of these figures, the first three rows show (from top to bottom) the moment 0 (integrated intensity), moment 1 (velocity field), and moment 2 (velocity dispersion field; see \ref{PE} for details of moment map creation). For these rows, the three columns denote (from left to right) the observed data cubes, model cubes, and the corresponding residuals. The white crosses in the lower right corner of each panel in the first row show a 5\,kpc$\times$5\,kpc physical scale. The solid lines in the second row represent the kinematic major axis, while the dashed lines represent the minor axis. The bottom row shows (from left to right) the major axis PVD, minor axis PVD, and extracted spectra. For each PVD, the observed data are shown by the background colour, while the contours represent the model. The data (D), model (M), and residual (R) spectra are depicted by the green, purple, and orange lines, respectively. The $1\sigma$ uncertainty, calculated as (average RMS noise level per channel)$\rm \times\sqrt{number\,\,of\,\,beams\,\,in\,\,source}$, is shown by the shaded grey area.}
\label{BBresults1}
\end{figure*}

\begin{figure*}
\centering
\includegraphics[width=0.49\textwidth]{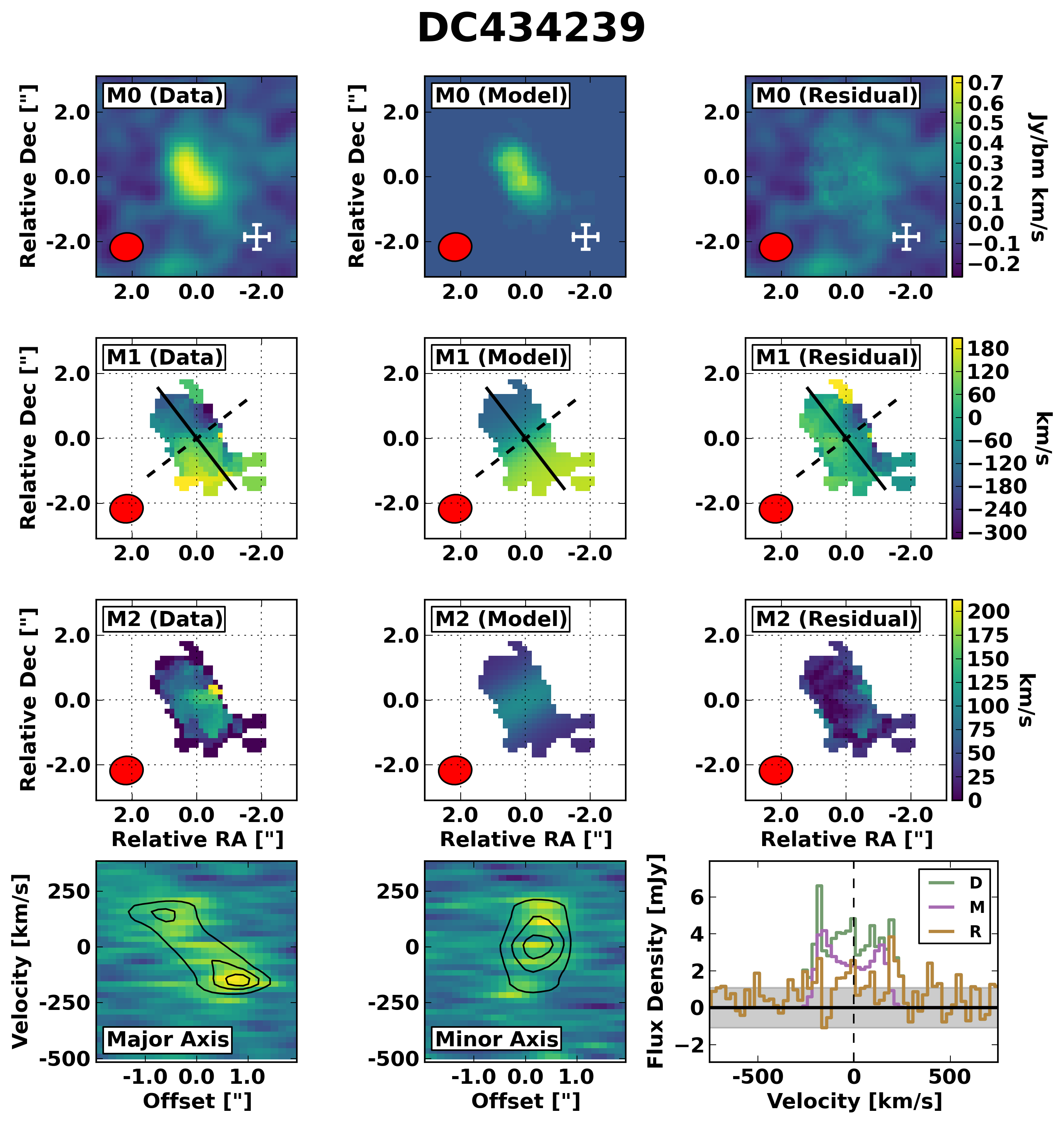}
\includegraphics[width=0.49\textwidth]{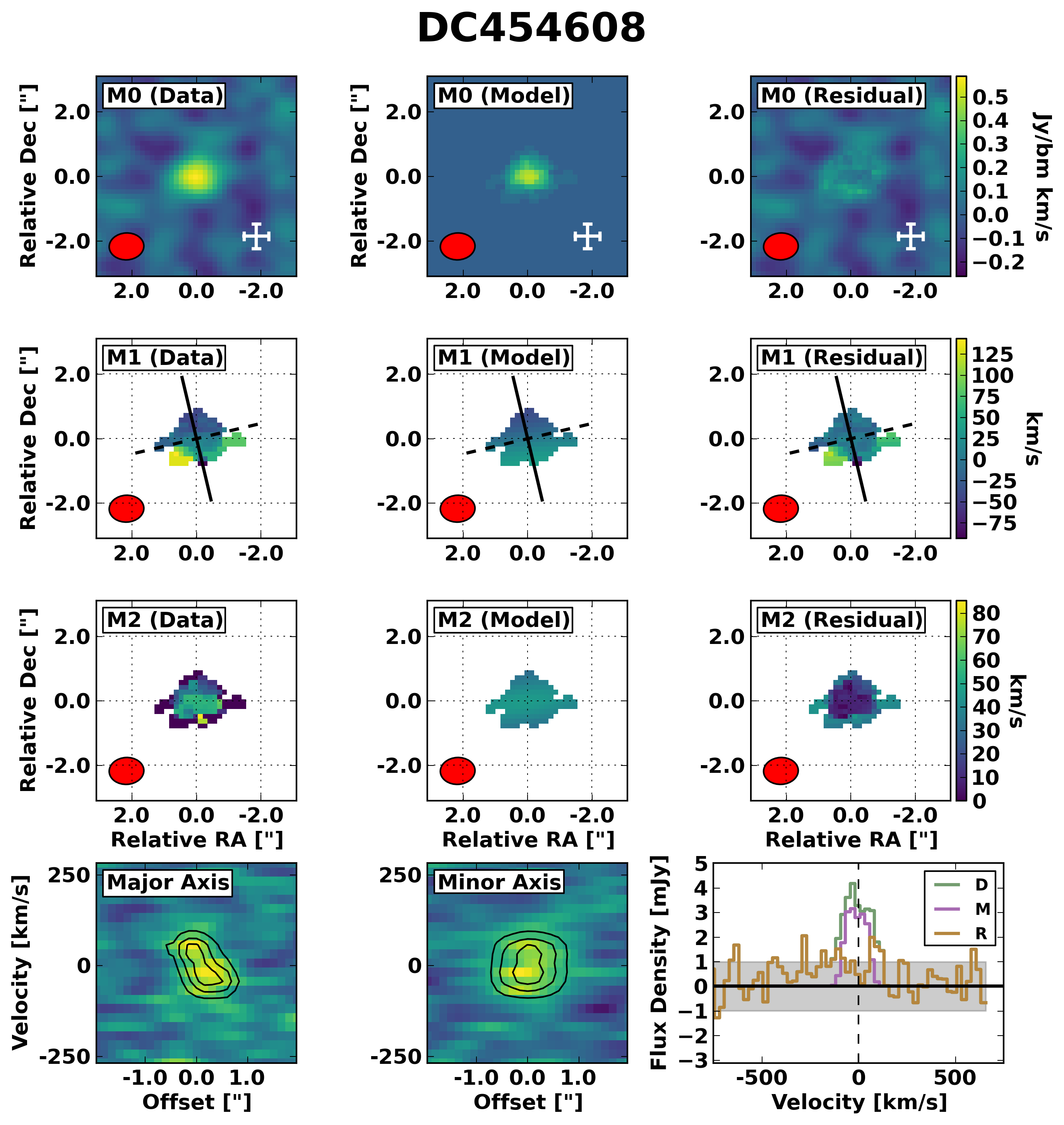}
\includegraphics[width=0.49\textwidth]{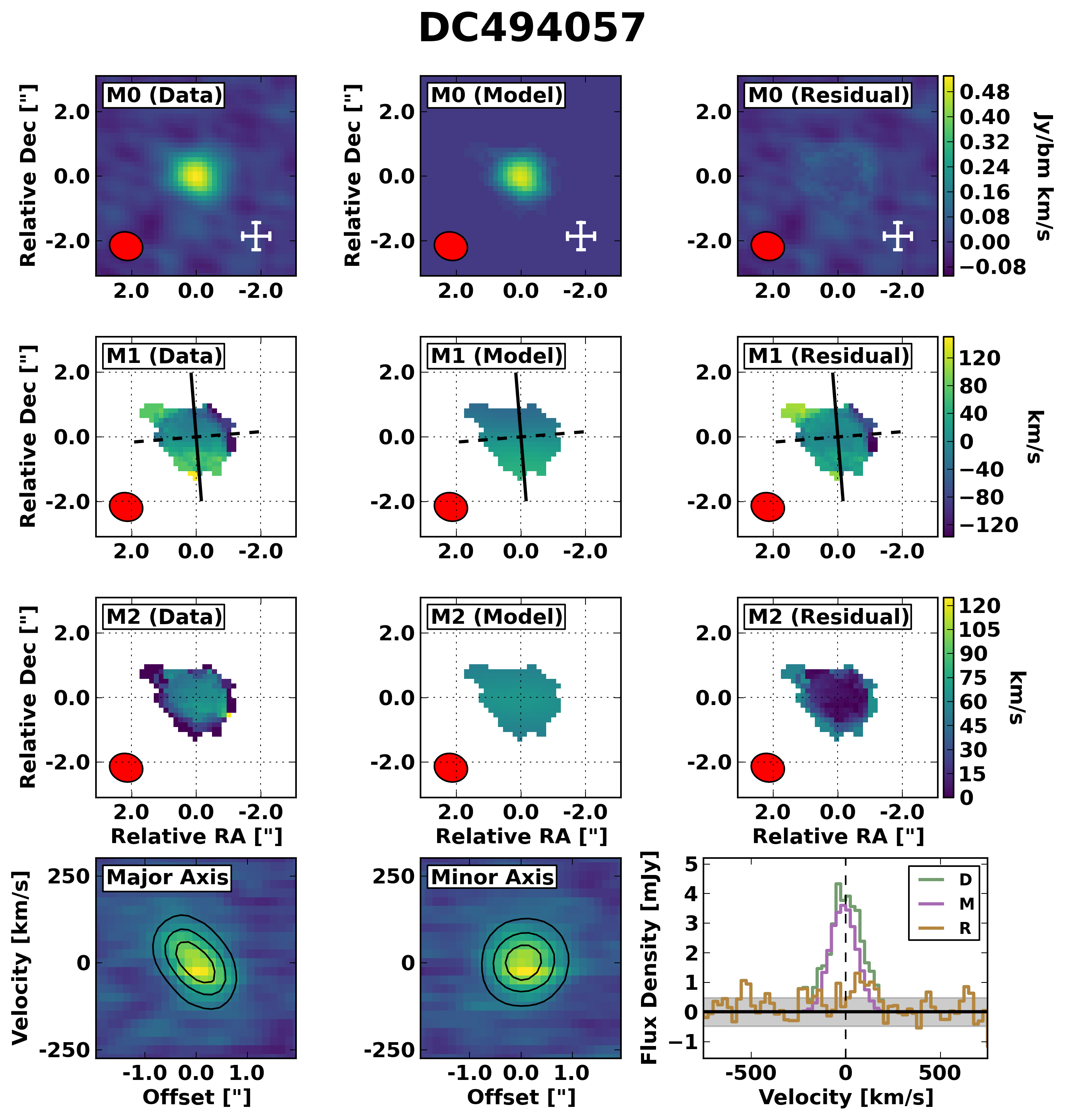}
\includegraphics[width=0.49\textwidth]{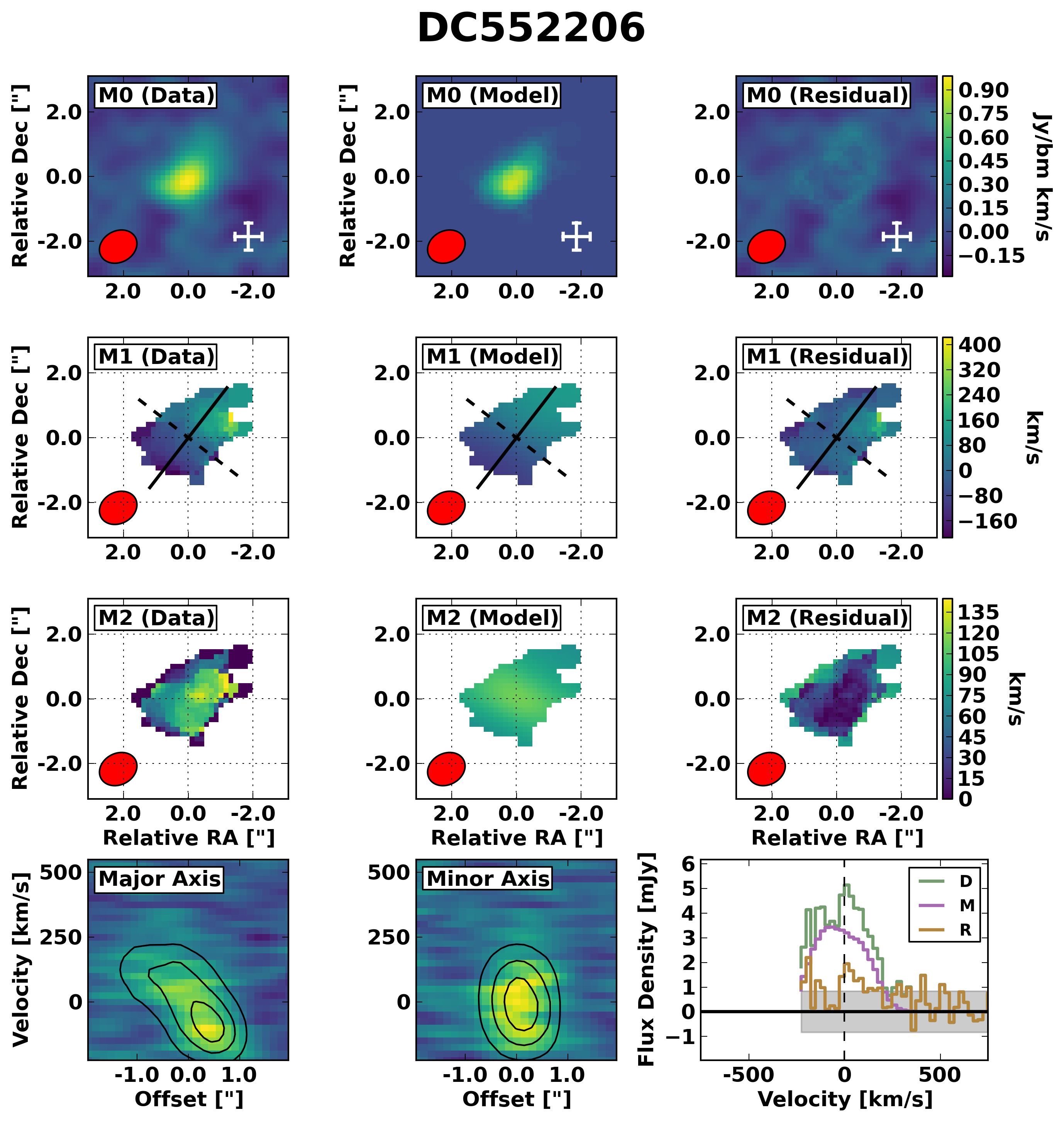}
\caption{Moment maps, PVDs, and spectra for observed data, model, and residual. See Figure \ref{BBresults1} for details.}
\label{BBresults2}
\end{figure*}

\begin{figure*}
\centering
\includegraphics[width=0.49\textwidth]{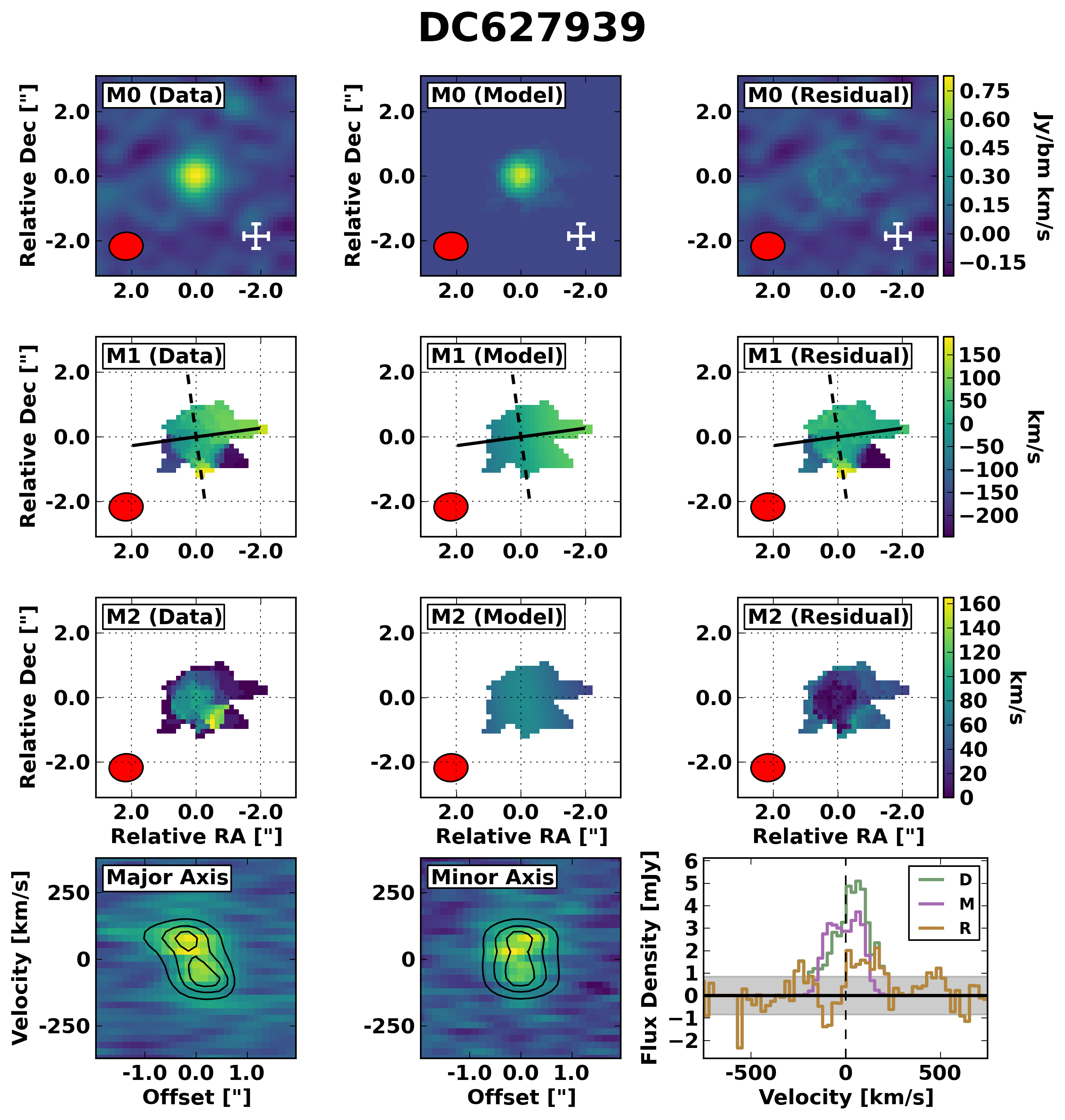}\includegraphics[width=0.49\textwidth]{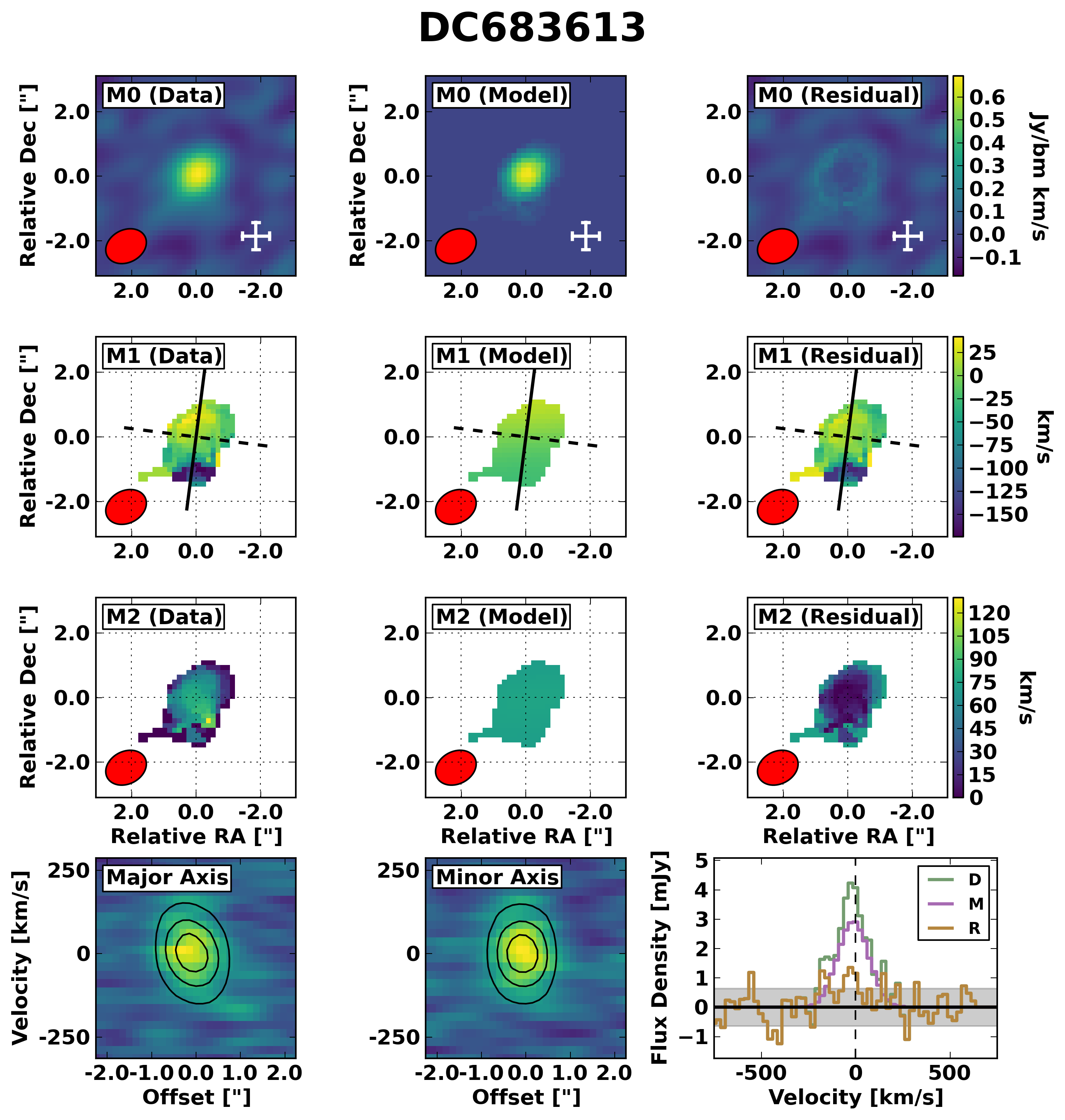}
\includegraphics[width=0.49\textwidth]{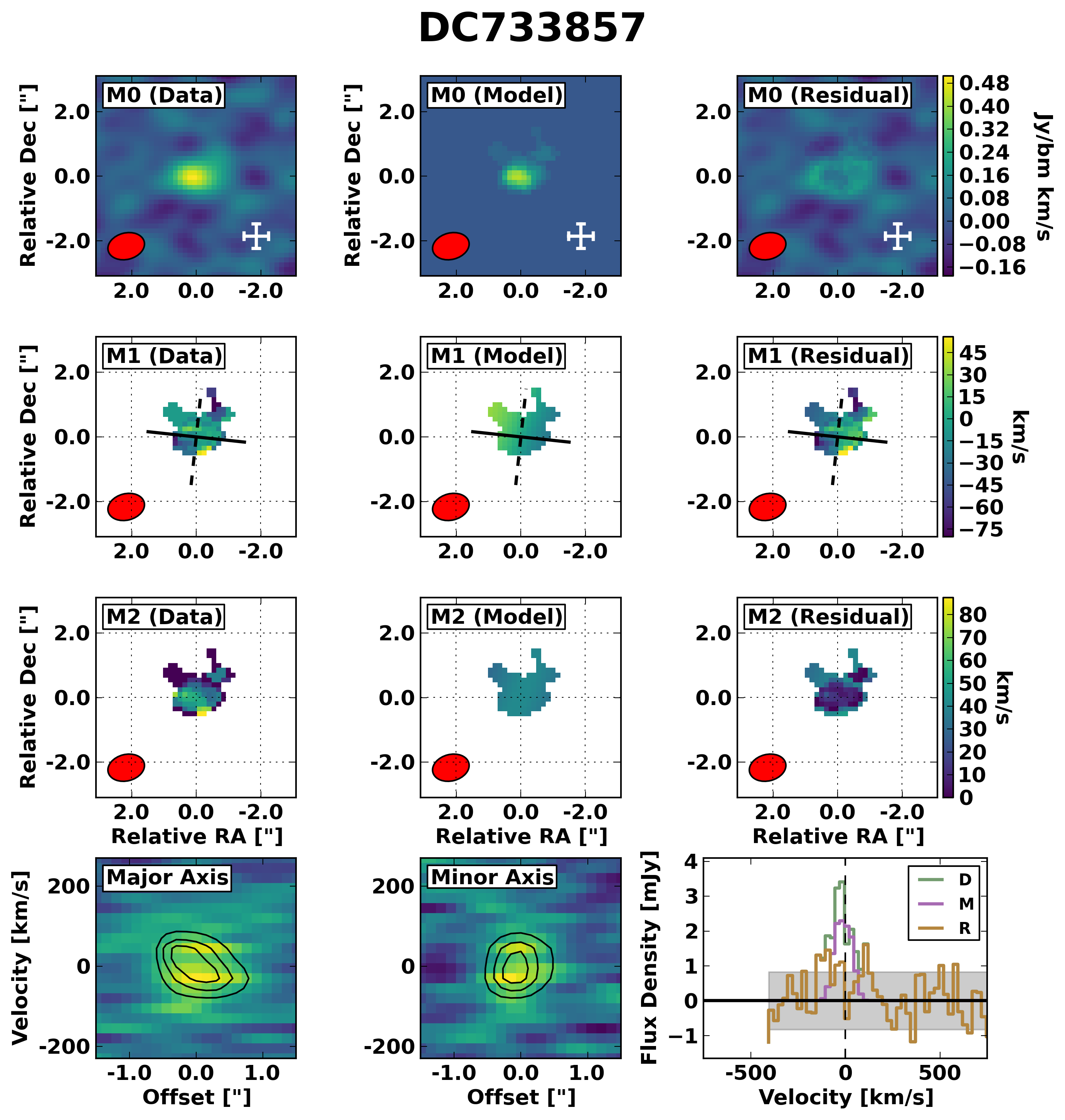}
\includegraphics[width=0.49\textwidth]{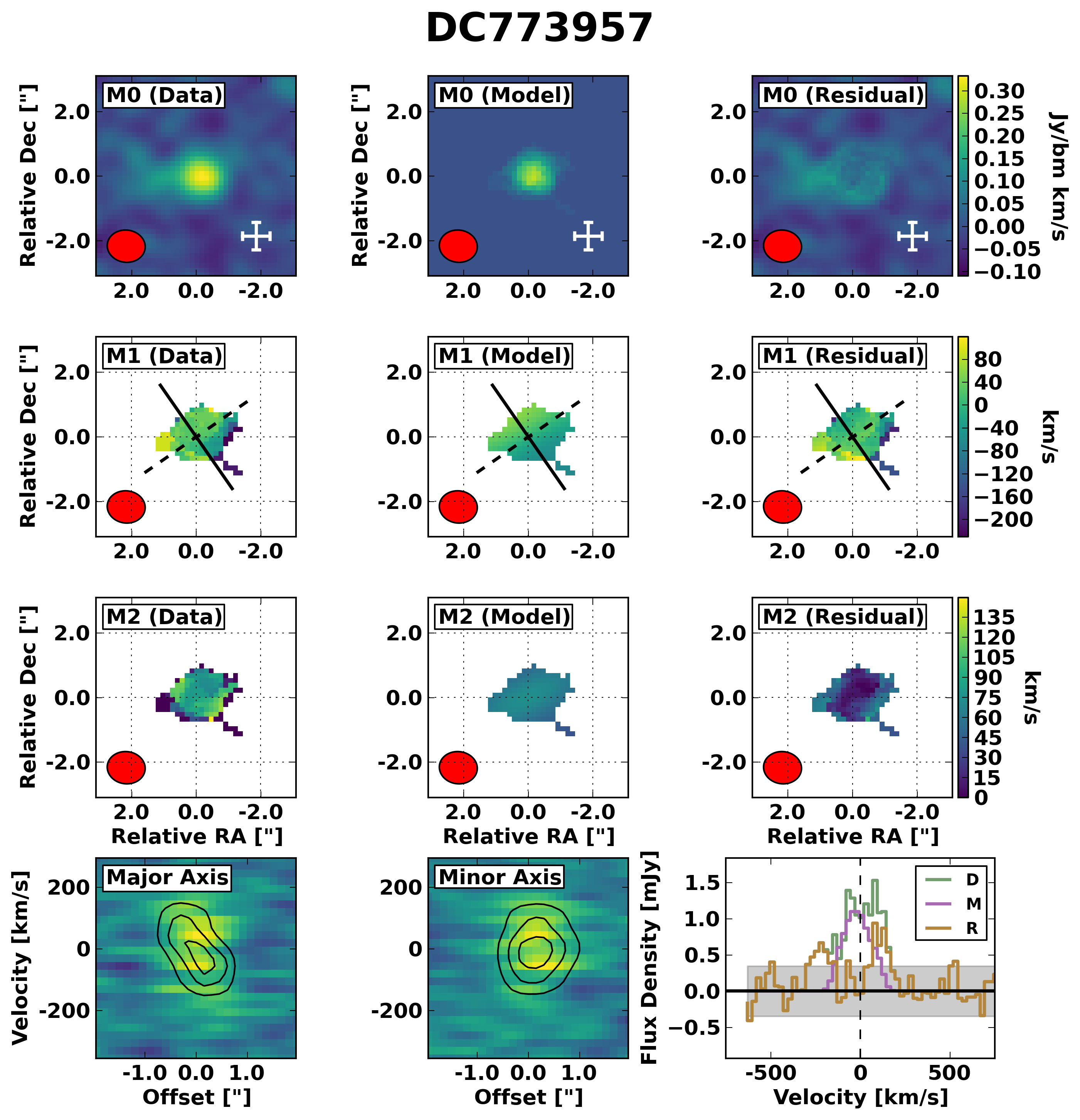}
\caption{Moment maps, PVDs, and spectra for observed data, model, and residual. See Figure \ref{BBresults1} for details.}
\label{BBresults3}
\end{figure*}

\begin{figure*}
\centering
\includegraphics[width=0.49\textwidth]{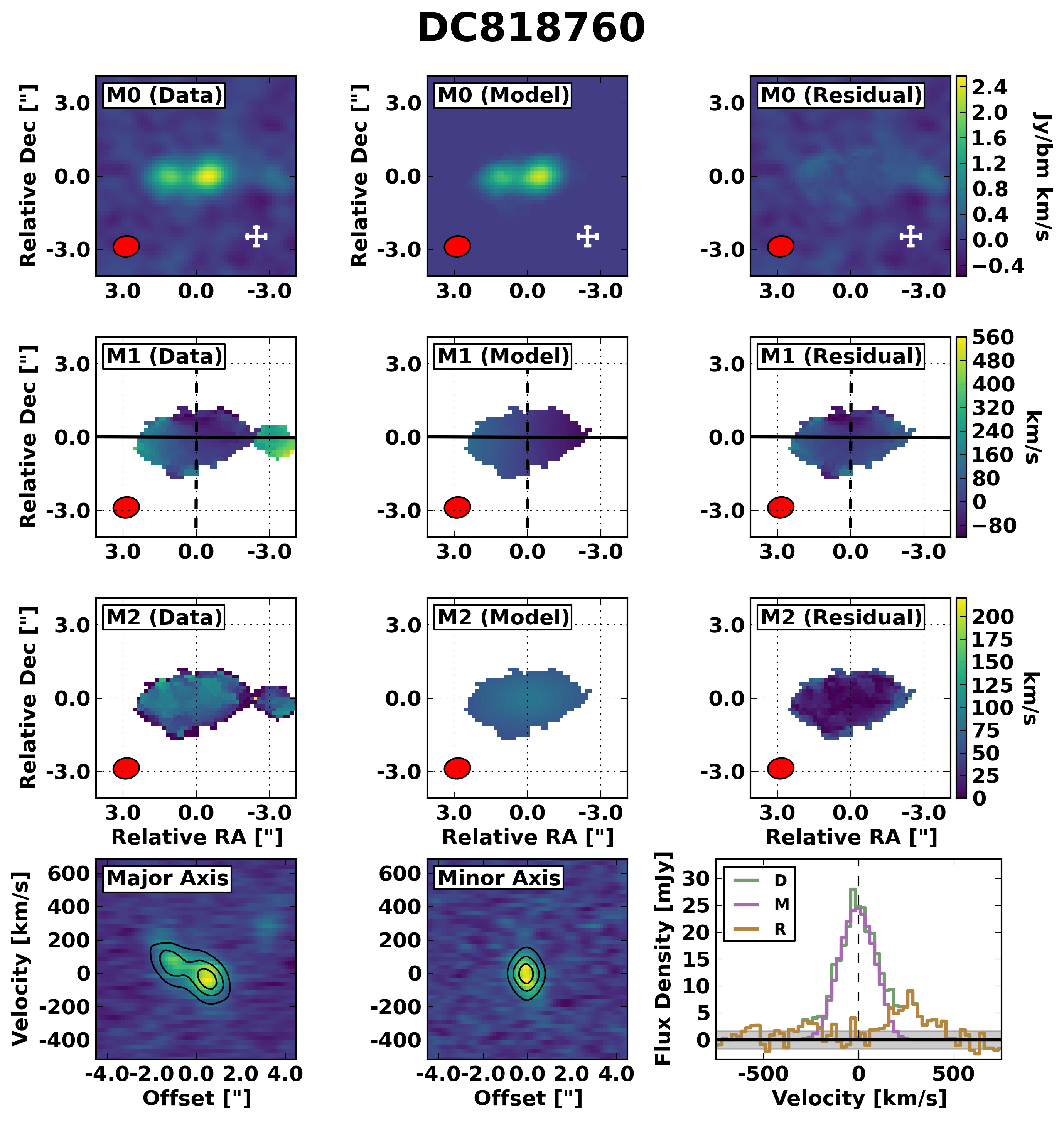}
\includegraphics[width=0.49\textwidth]{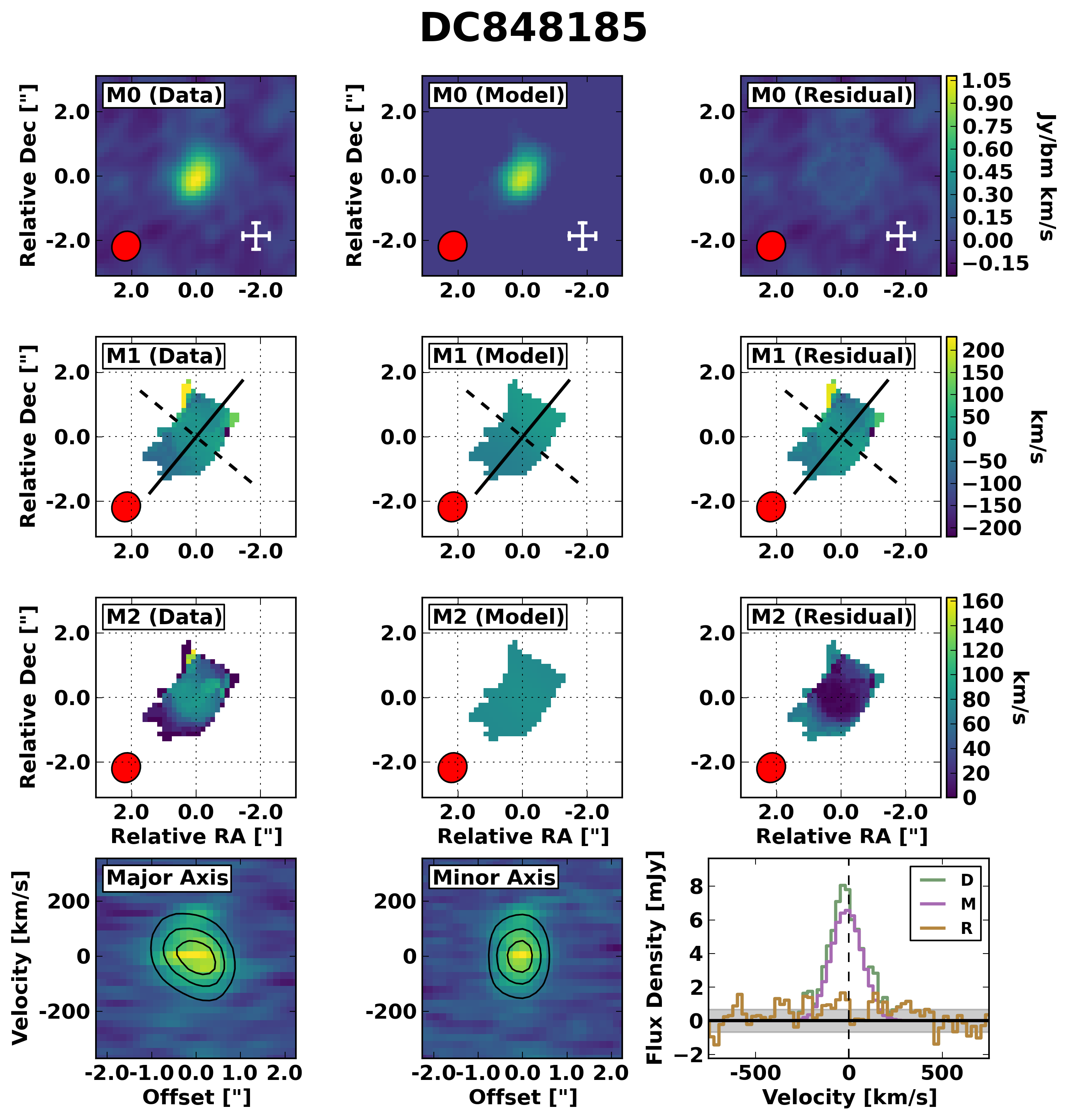}
\includegraphics[width=0.49\textwidth]{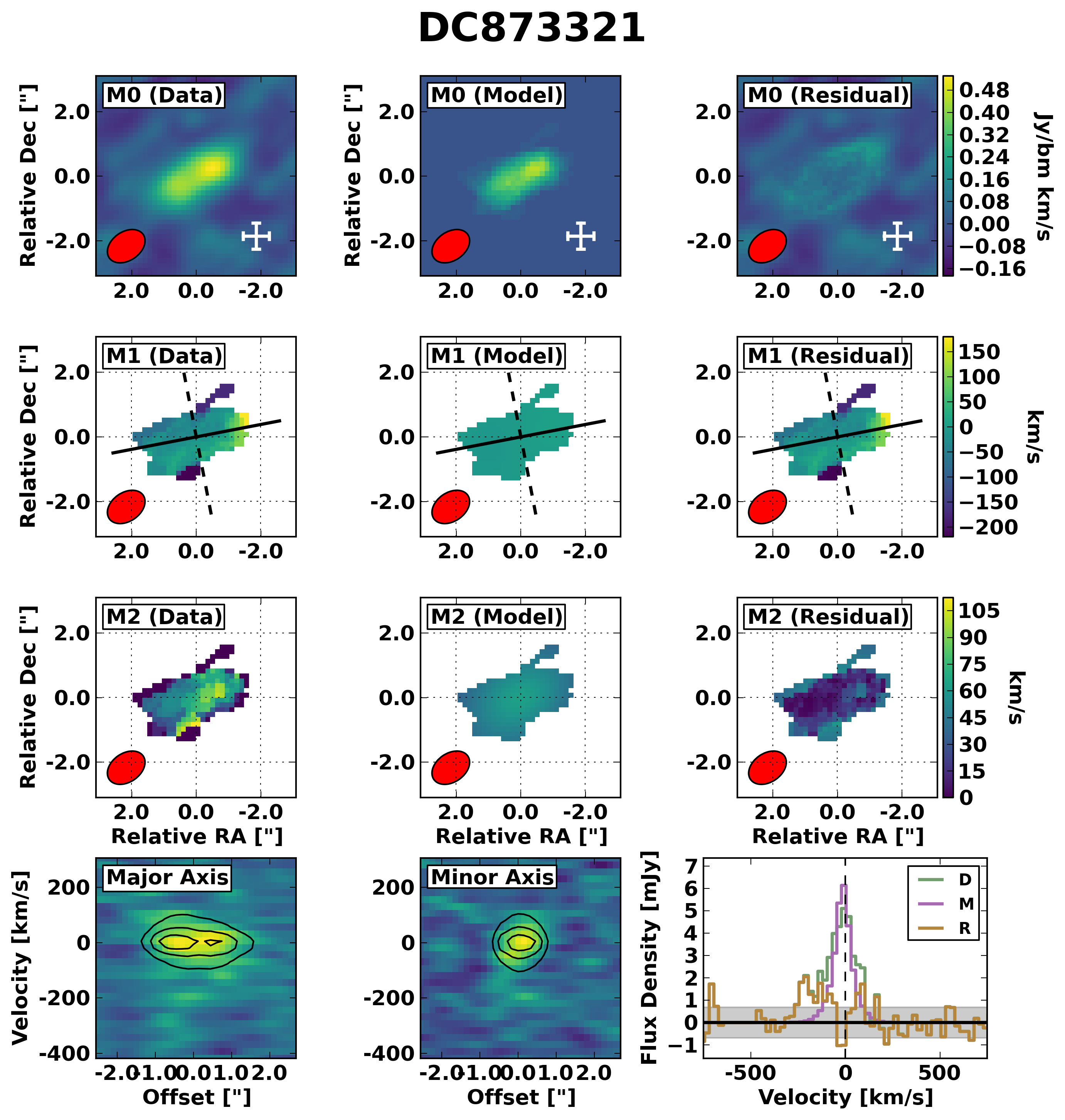}
\includegraphics[width=0.49\textwidth]{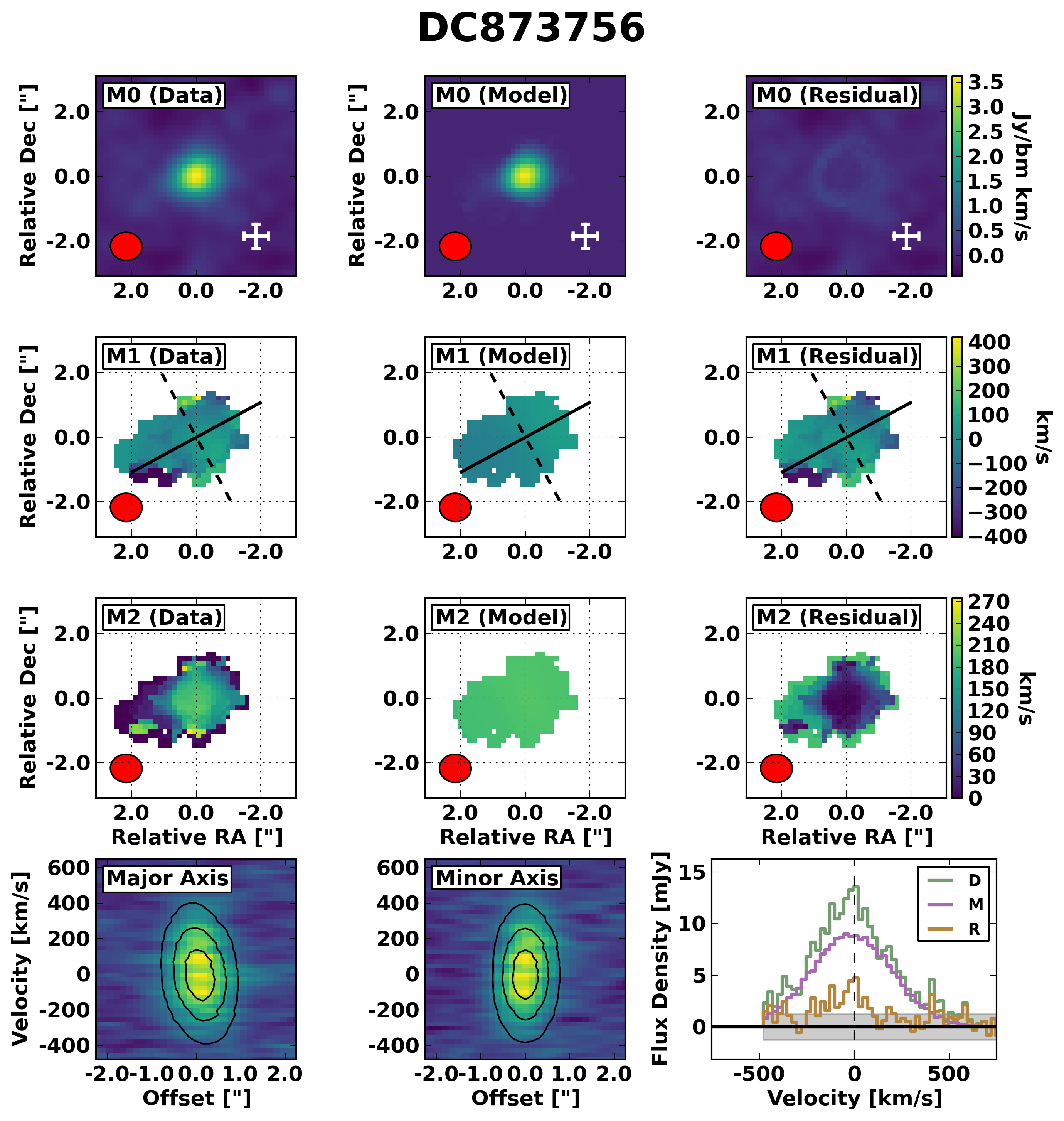}
\caption{Moment maps, PVDs, and spectra for observed data, model, and residual. See Figure \ref{BBresults1} for details.}
\label{BBresults4}
\end{figure*}

\begin{figure*}
\centering
\includegraphics[width=0.49\textwidth]{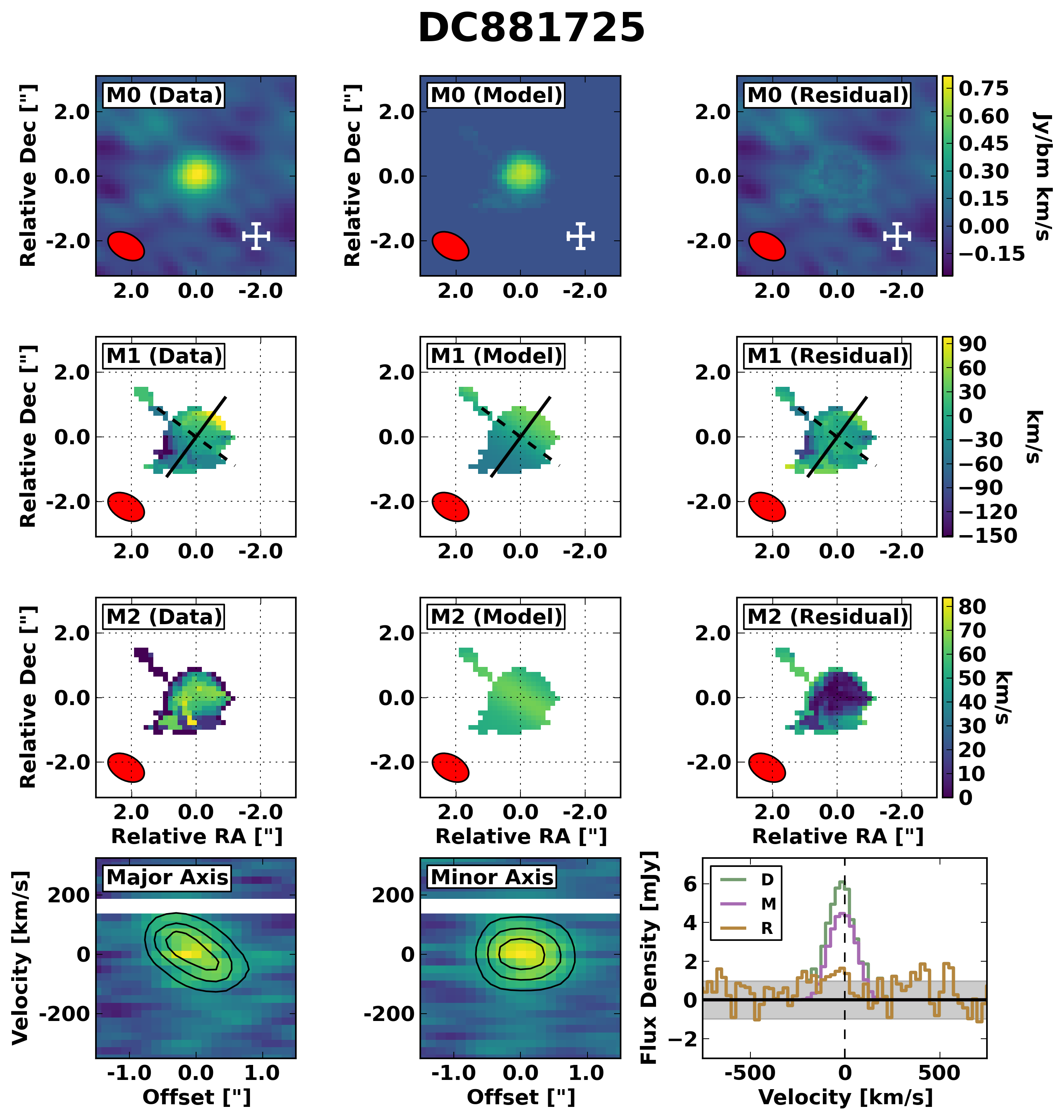}
\includegraphics[width=0.49\textwidth]{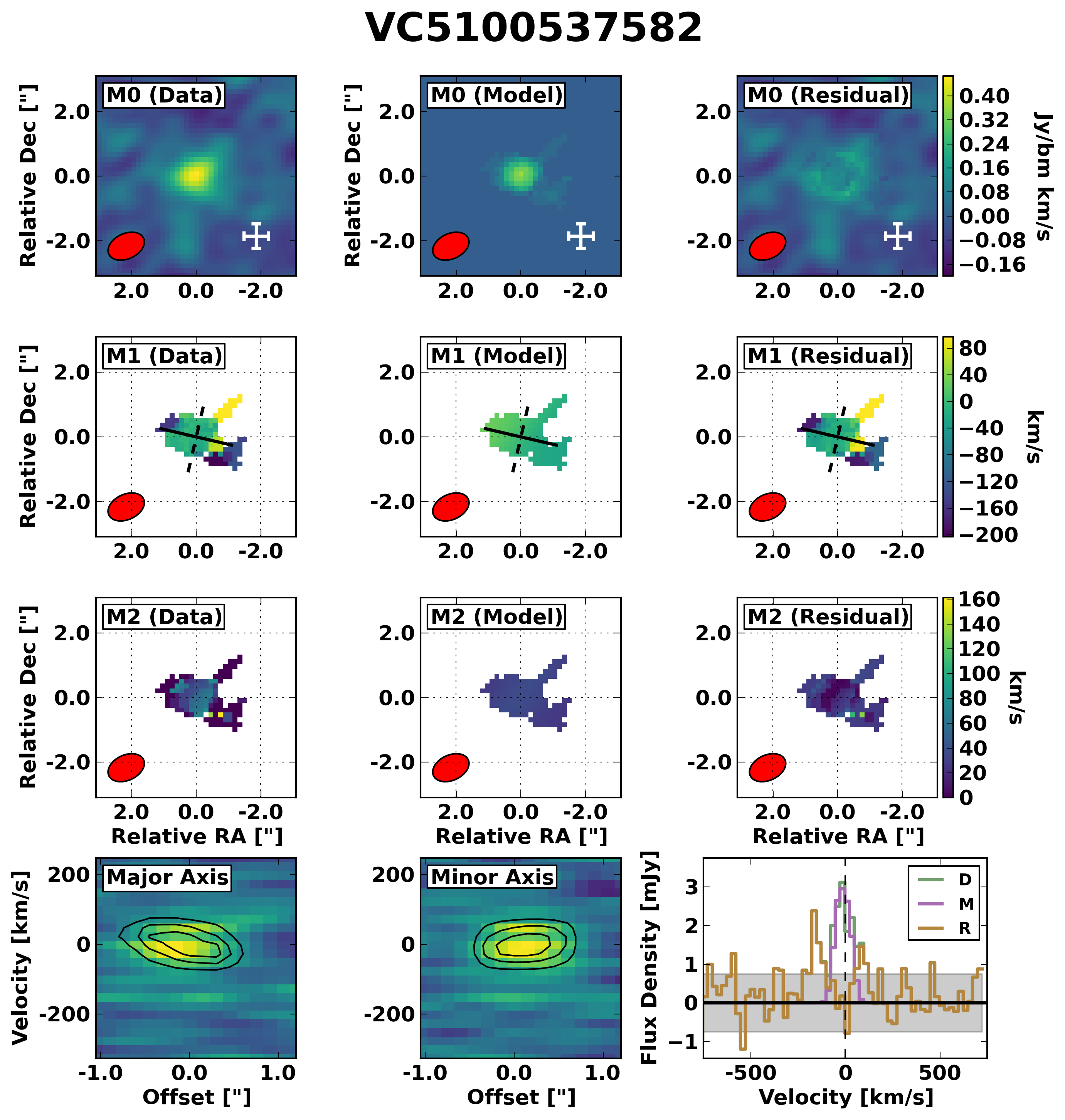}
\includegraphics[width=0.49\textwidth]{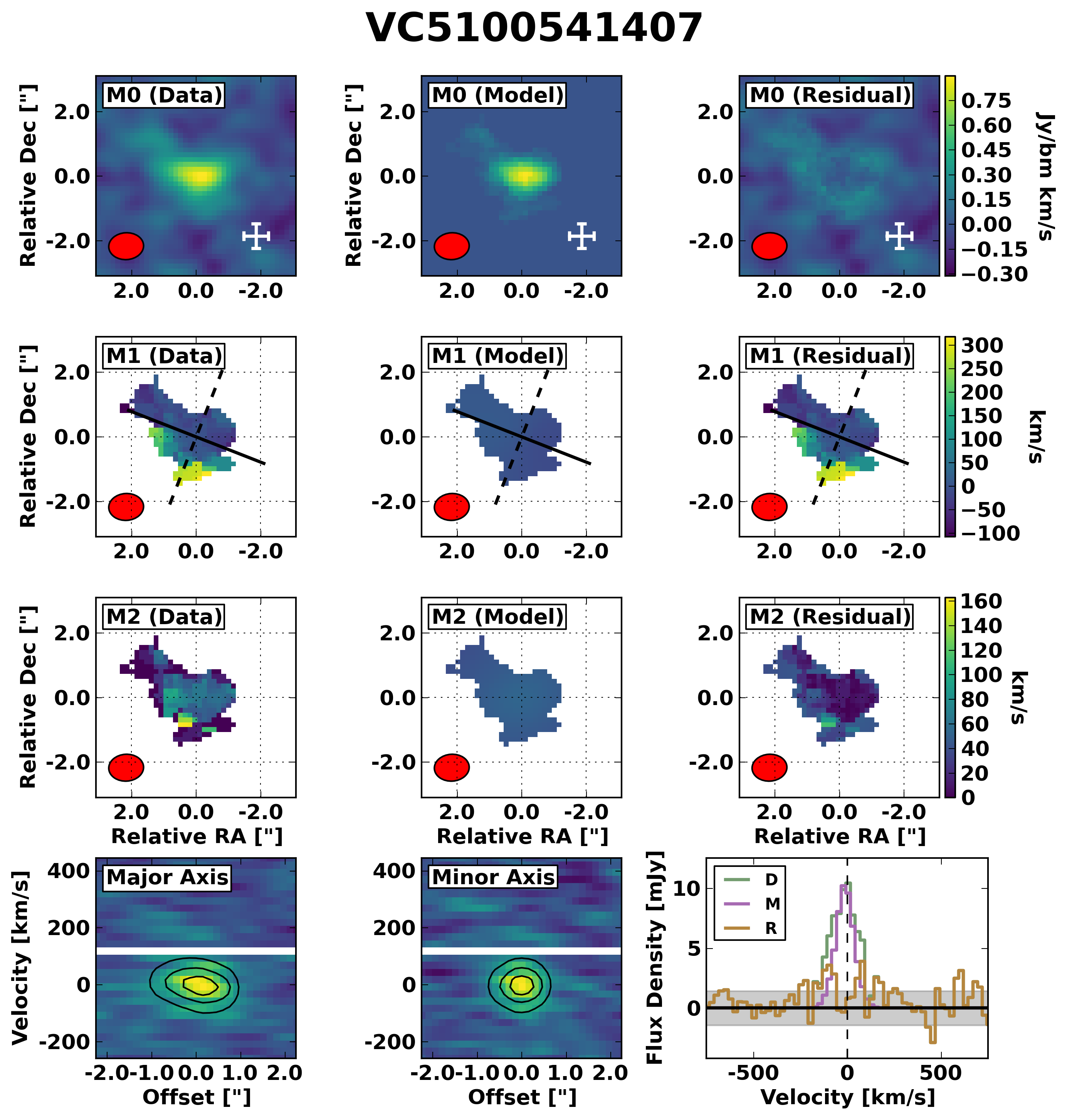}
\includegraphics[width=0.49\textwidth]{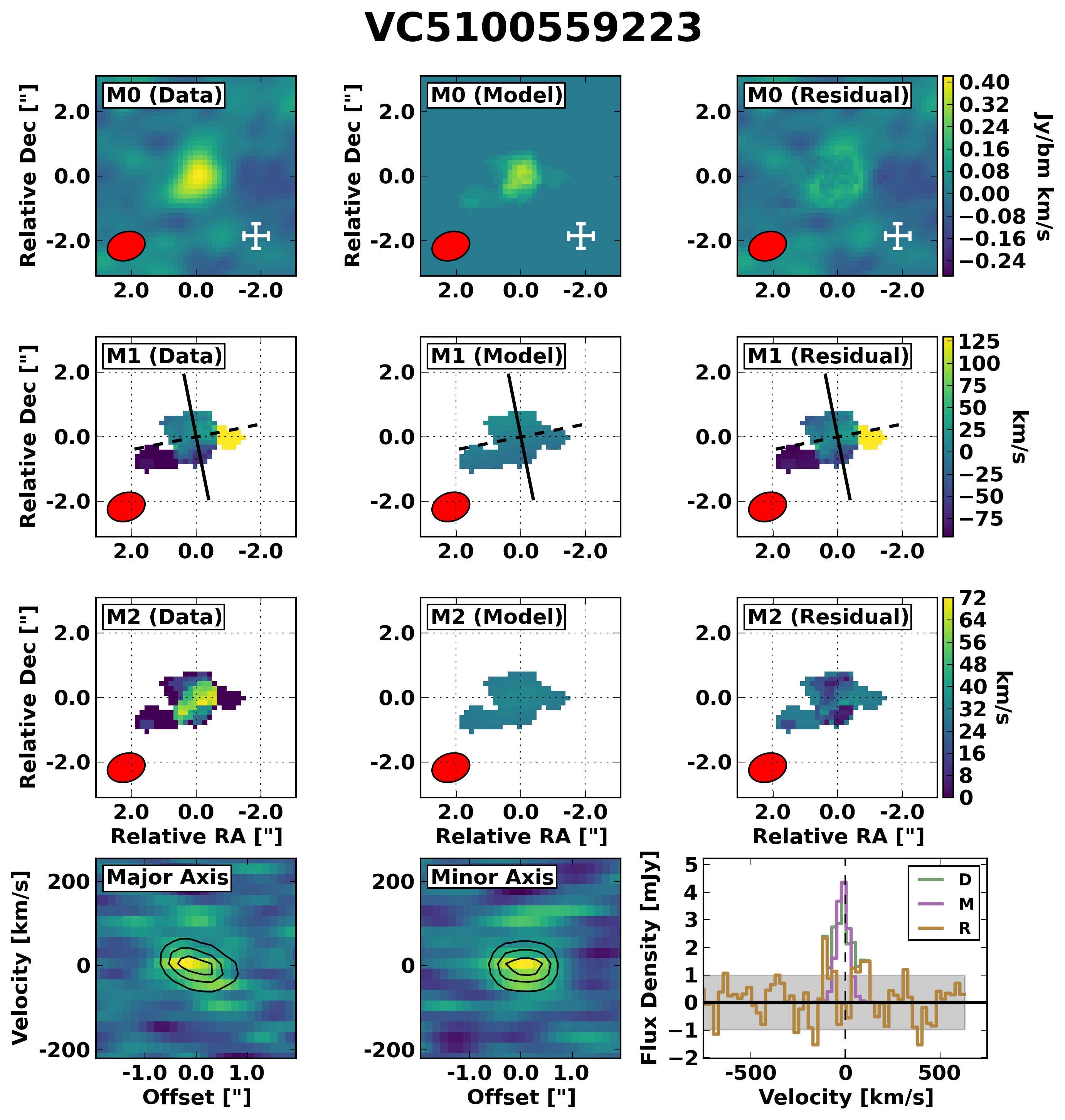}
\caption{Moment maps, PVDs, and spectra for observed data, model, and residual. See Figure \ref{BBresults1} for details.}
\label{BBresults5}
\end{figure*}

\begin{figure*}
\centering
\includegraphics[width=0.49\textwidth]{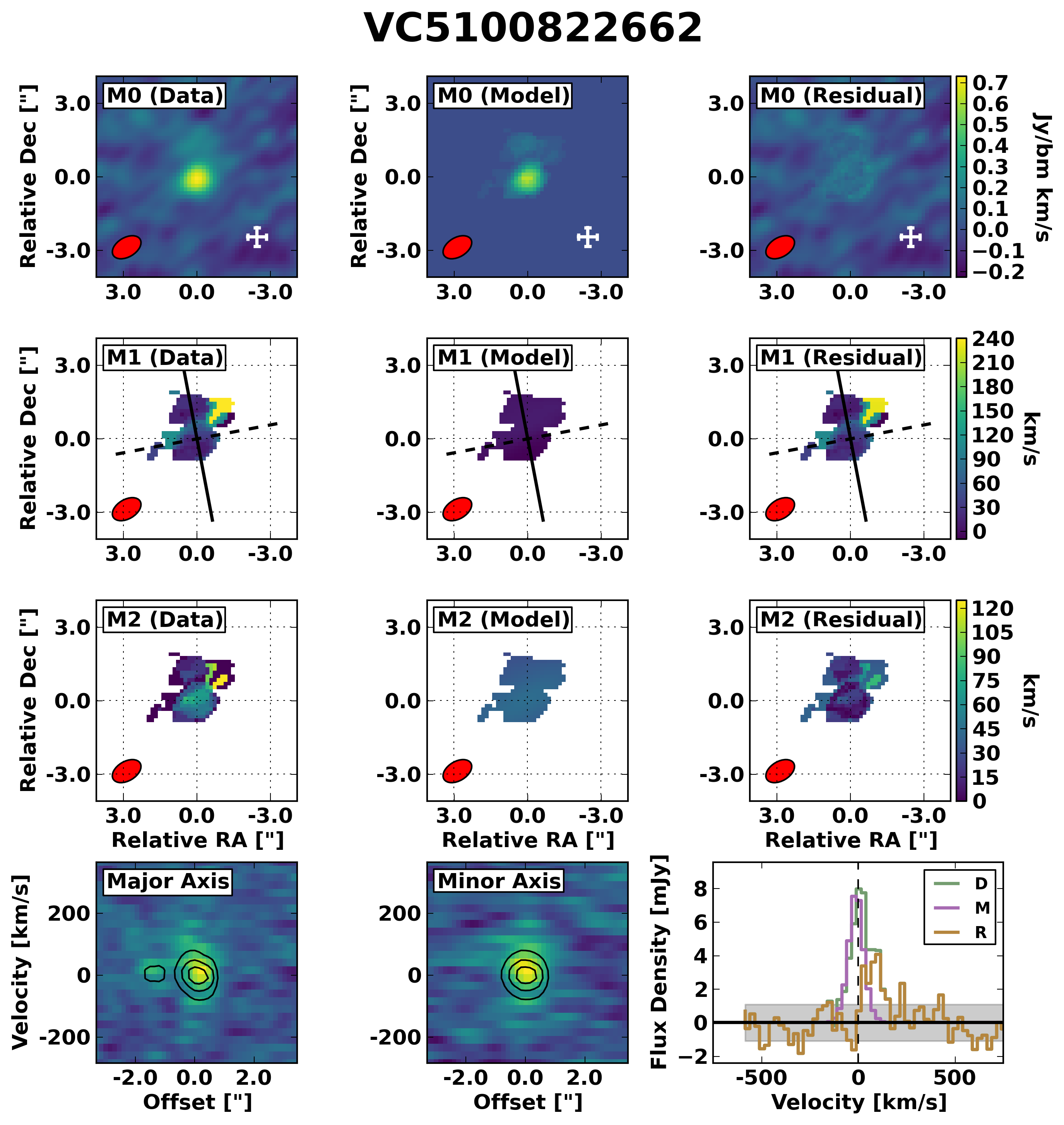}
\includegraphics[width=0.49\textwidth]{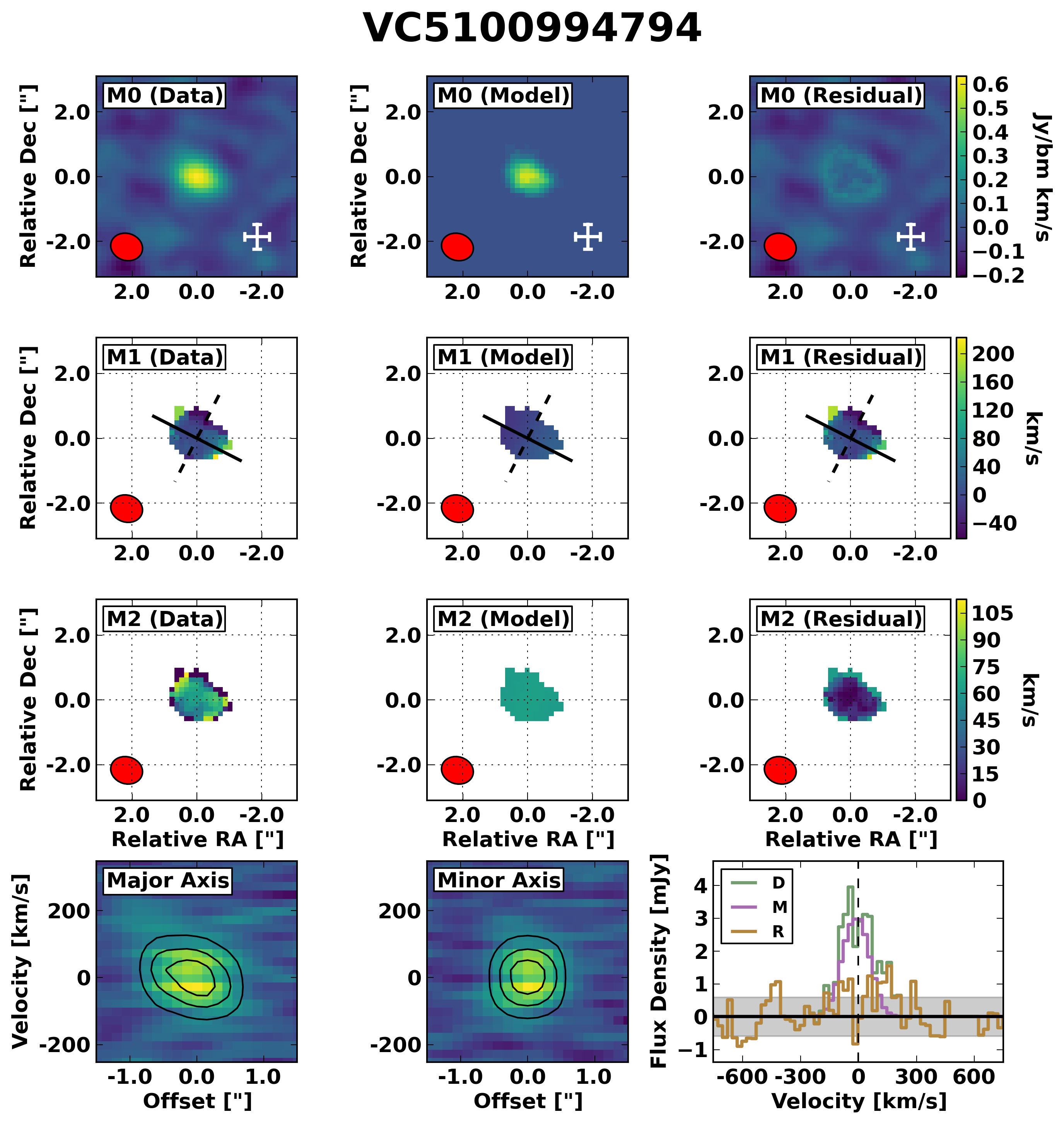}
\includegraphics[width=0.49\textwidth]{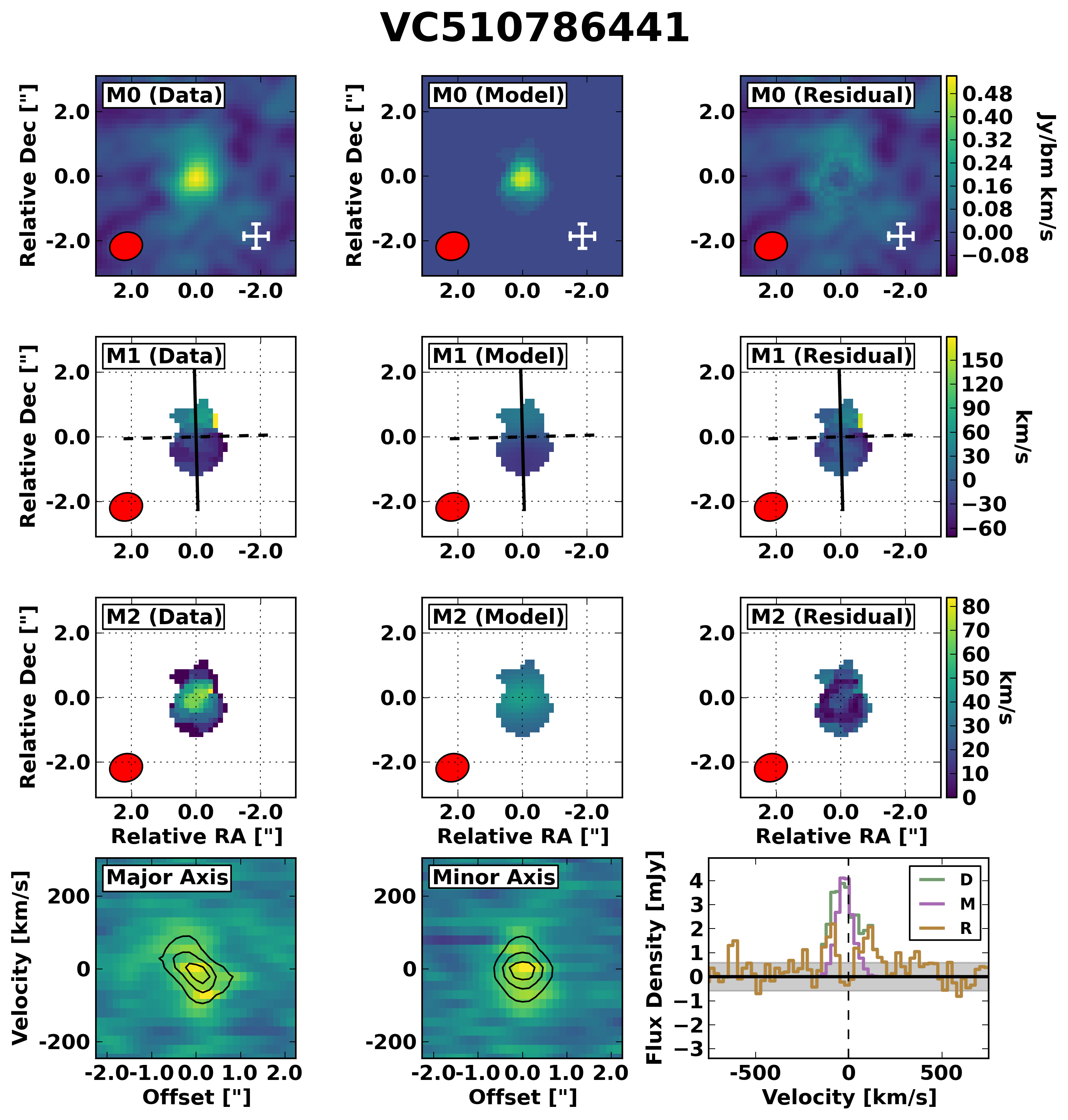}
\includegraphics[width=0.49\textwidth]{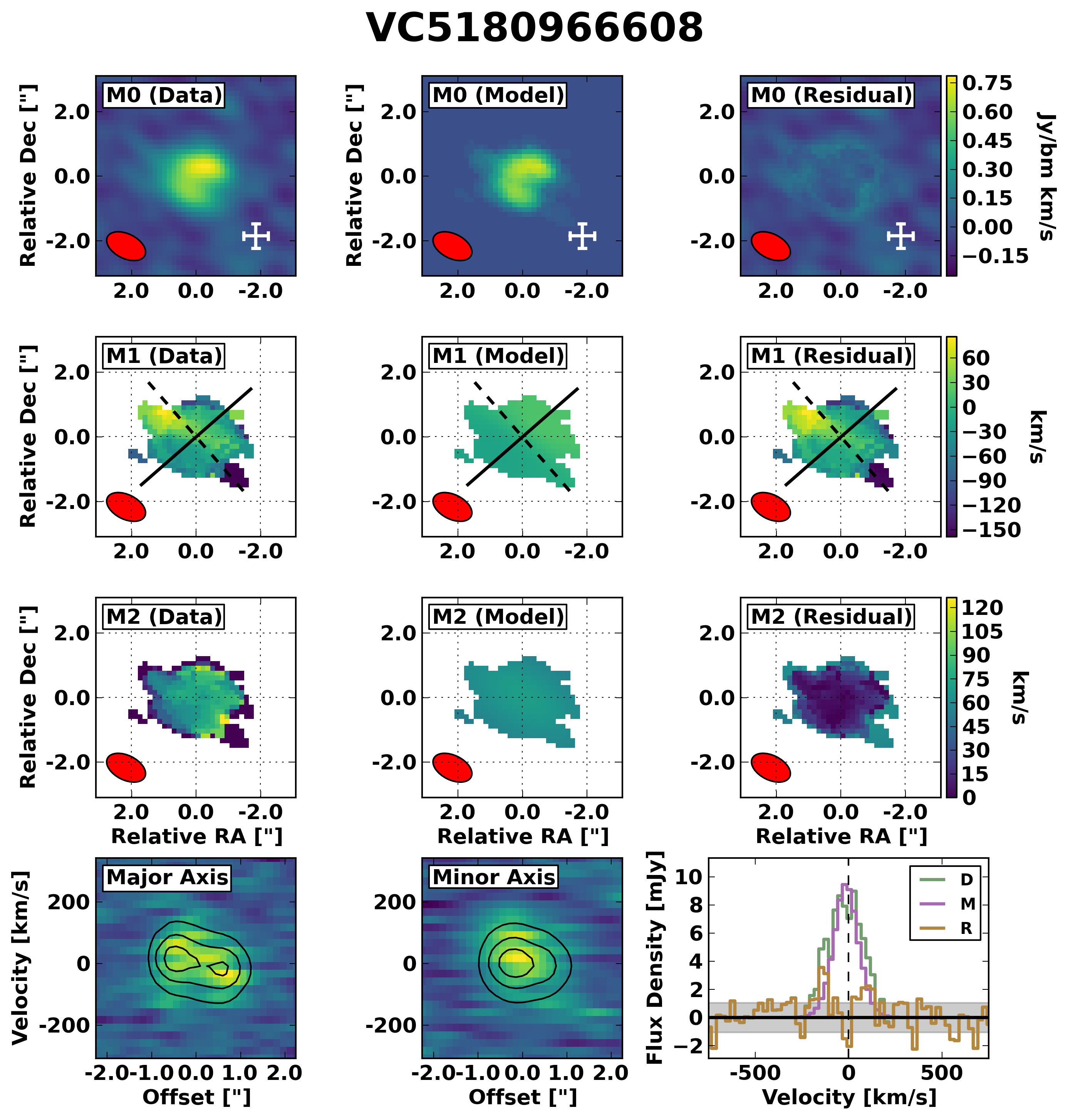}
\caption{Moment maps, PVDs, and spectra for observed data, model, and residual. See Figure \ref{BBresults1} for details.}
\label{BBresults6}
\end{figure*}

\begin{figure*}
\centering
\includegraphics[width=0.49\textwidth]{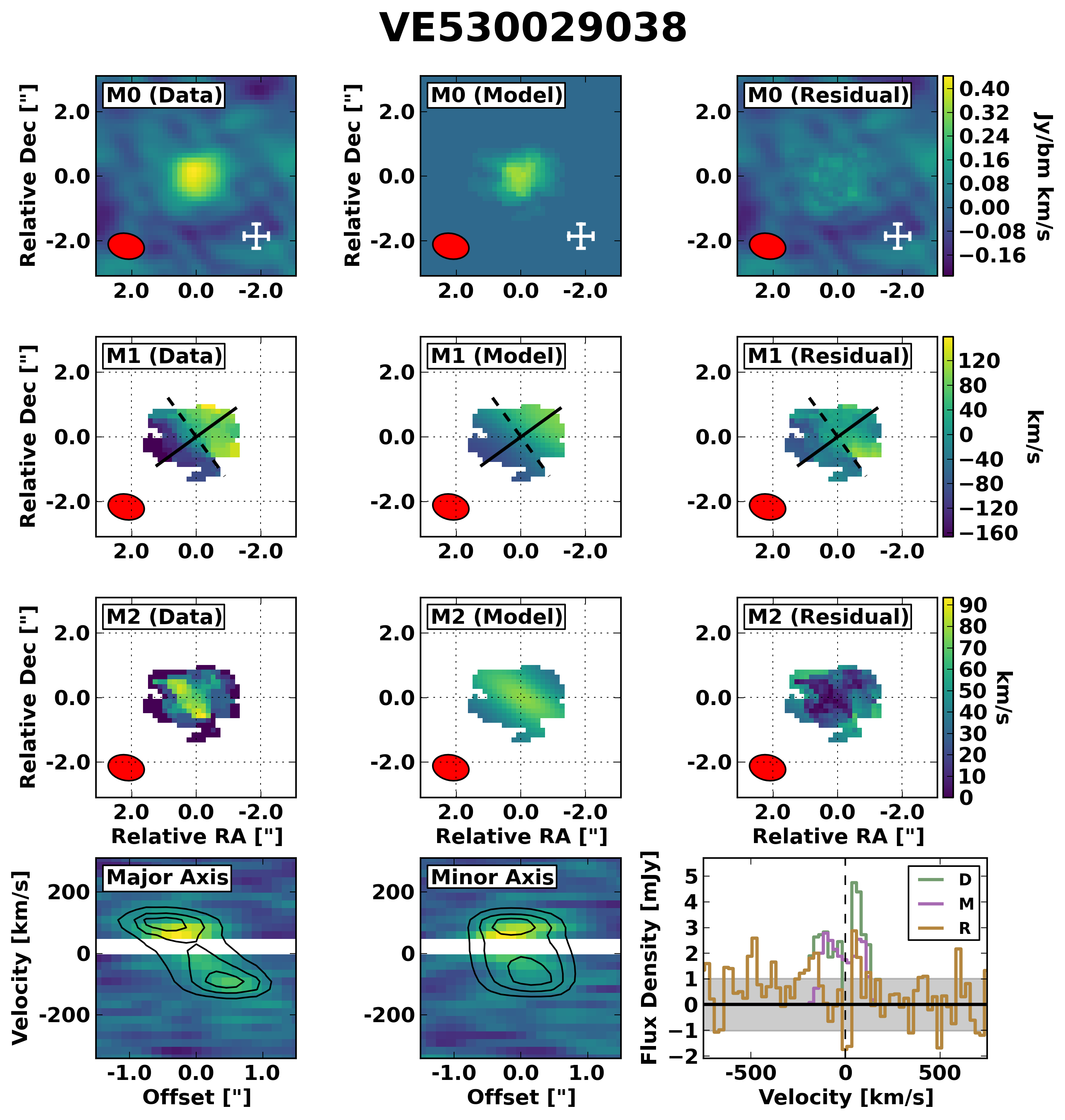}
\caption{Moment maps, PVDs, and spectra for observed data, model, and residual. See Figure \ref{BBresults1} for details.}
\label{BBresults7}
\end{figure*}
 
\noindent
\textbf{CANDELS GOODS 32 (ROT)}: Originally classified as a merger, we can see that the observed velocity field shows a strong, but not perfectly symmetric, gradient (Figure \ref{BBresults1}). The spectral residuals are quite low, the best-fit model is rotation-dominated, and four of the W15 criteria are met, so there is strong evidence that this source is a rotator.

\noindent
\textbf{DEIMOS COSMOS 396844 (ROT)}: Since the major axis of the restoring beam is nearly orthogonal to the observed velocity gradient of this source (Figure \ref{BBresults1}), its morphological and kinematic position angles are not in agreement (see failure of W15 criterion 4). The model is mostly well fit, adding additional credence to its original classification as a rotator. However, a broad region of residual emission is apparent at $\sim-250$\,km\,s$^{-1}$. At first glance, this residual emission may be interpreted as evidence for a broad blueshifted outflow or low-level satellite. But since it is so weak, we suggest that it is only noise. Due to the low residuals of a rotation-dominated model, we agree with the \citet{lefe20} classification of a rotator.

\noindent
\textbf{DEIMOS COSMOS 417567 (UNC)}: Due to its complex PVD behaviour (i.e., multiple peaks, no overall trends), this source was originally classified as a merger. This is supported by the poor \BB fit, where significant residuals exist in every plot (Figure \ref{BBresults1}). However, we note that this source is one of the weakest in our present sample, and we may simply be experiencing the negative effects of poor SNR. While all W15 criteria besides PA agreement are met, the large residuals and odd appearance of each PVD earn this source a `uncertain' classification.

\noindent
\textbf{DEIMOS COSMOS 432340 (UNC)}: The curved blue half of this galaxy was originally interpreted as evidence for an ongoing merger. \BB returns a reasonable rotation-dominated fit, and all five W15 criteria are met. However, the morphology of this source is quite disturbed, and the minor axis PVD shows a `C'-shape (Figure \ref{BBresults1}). The SNR of this source is too low to definitely classify it as an ongoing merger, a galaxy with an outflow, or a low-SNR well-ordered rotating disk, so we classify this source as `uncertain'.

\noindent
\textbf{DEIMOS COSMOS 434239 (MER)}: With a double intensity peak and unique PVD, this source was originally classified as a merger. We note that this source returns the same W15 criteria as DC552206 (a rotator), while the two galaxies show drastically different morpho-kinematics, highlighting the need to examine the morpho-kinematics of each source in multiple ways. The large residuals in the velocity field and spectrum, as well as the poor match in the major axis PVD (Figure \ref{BBresults2}), cause us to classify this source as a merger.

\noindent
\textbf{DEIMOS COSMOS 454608 (UNC)}: The double-peaked major axis PVD originally earned this galaxy the classification of merger, but our \BB fits are inconclusive (Figure \ref{BBresults2}). The tilted ring model returns large moment one residuals, and a positive-velocity clump is included by SEARCH, making interpretation difficult. This clump may be excluded by adopting a higher SNR$_{\rm lower}$ threshold, but then this source is too compact to be fit by $\ge2$ rings, and so the fit fails. Due to the low-SNR and low spatial resolution, we classify this source as `uncertain'.

\noindent
\textbf{DEIMOS COSMOS 494057 (ROT)}: This source is nearly a perfect example of a rotator, with only small residuals in the moment zero and two maps and spectrum, and well-fit PVDs (Figure \ref{BBresults2}). We do note that the moment one map shows residuals around the perimeter of the source, but this is likely due to the low-SNR of these regions. In addition, the morphological and kinematic position angles disagree (resulting in a failure of W15 criterion four), but this is likely due to the near-unity axis ratio (or low inclination) of the source.

\noindent
\textbf{DEIMOS COSMOS 519281 (UNC)}: While most sources in the present sample show monotonic velocity gradients, DC519281 features a central blueshifted region flanked by two redshifted components (Figure \ref{BBresults1A}). This may be interpreted as evidence for a counterrotating merger (see Q1623-BX528 in \citealt{fors06}). However, the small spatial extent of this source makes physical interpretation impossible, and \BB returns a poor fit. Of the five \citet{wisn15} criteria, this source only meets the second (v$_{\rm rot}>\sigma_{\rm v}$), agreeing that this is not a rotator. We also note that when applying SEARCH, the standard lower SNR of 2.5 returned a noise peak to the southwest, so we applied a SNR$_{\rm lower}$ of 2.7 for this source. While the complex velocity field of this source could be interpreted as evidence for ongoing merging activity, the low resolution and SNR of this source cause us to classify it as uncertain.
 
\noindent
\textbf{DEIMOS COSMOS 552206 (ROT)}: The redshift of each source in the ALPINE sample was taken from previous UV spectroscopic observations. However, the velocity difference between these redshifts and those of [CII] may be substantial (i.e., up to $\sim500$\,km\,s$^{-1}$, see \citealt{cass20} for details and physical interpretation). In the case of DC552206, this velocity shift caused the [CII] line to fall on the edge of an ALMA sideband, resulting in an incomplete line profile (Figure \ref{BBresults2}). However, since the observed profile features a prominent velocity gradient (seen in the moment 1 map and major axis PVD), \BB is still able to fit an agreeable model, suggesting that this source is truly a rotator, rather than a merger, as suggested previously. Of course, a follow-up observation containing the blue edge of this line should be conducted to confirm this.

\noindent
\textbf{DEIMOS COSMOS 627939 (UNC)}: This source was originally classified as a merger, mainly due to its complex PVDs. \BB returns a reasonable fit, which reproduces the primary velocity gradient, as well as the overall shape of each PVD (Figure \ref{BBresults3}). The returned W15 criteria are identical to those of DC396844 and DC494057, which are both classified as rotators. However, substantial residuals are present in both the velocity map and spectrum, hinting at ongoing merging below our spatial resolution. Since we can neither fit a rotating disk model nor distinguish multiple components, we classify this source as `uncertain'.

\noindent
\textbf{DEIMOS COSMOS 683613 (UNC)}: While this source was originally classified as an extended dispersion-dominated galaxy, further examination of the [CII] moment maps and PVDs presented in Figure \ref{BBresults3} show that the [CII] appears to mainly emanate from a central source, with a weaker, blue, compact source to the south. The dominant source is moderately well fit by a tilted ring model, and appears to be a weaker version of the dispersion-dominated source DC848185. However, since it is not clear if the southern component is evidence of merging or simply noise, we classify this as uncertain.

\noindent
\textbf{DEIMOS COSMOS 733857 (UNC)}: With no obvious velocity gradient (resulting in a failure of W15 criterion one) and a poorly fit rotation-dominated model, (Figure \ref{BBresults3}) this source is unlikely to be a rotator. It was originally classified as an extended, dispersion-dominated galaxy, but since it is only  marginally resolved and features a complex major-axis PVD, we classify it as uncertain.

\noindent
\textbf{DEIMOS COSMOS 773957 (UNC)}: As a relatively weak, spatially compact source, this source is difficult to characterize. The integrated spectrum is moderately well fit, but each of the PVDs show that the \BB fit is obviously not sufficient (Figure \ref{BBresults3}). The present observations are insufficient to determine whether this morpho-kinematic complexity is noise or truly signal. Therefore, we classify this as `uncertain'.

\noindent
\textbf{DEIMOS COSMOS 818760 (MER)}: As previously explored in detail \citep{jone20}, DC818760 is most likely a system of three galaxies undergoing a merger. The two dominant galaxies are closely related in both space and velocity, and an east-west monotonic velocity gradient is evident (Figure \ref{BBresults4}). Due to our relatively low resolution, a tilted ring model is able to replicate the flux density distribution, overall velocity gradient, and velocity dispersion of these two merging sources. However, the spectrum shows a significant red residual due to the third galaxy in this system not being captured by the fit, the model velocity dispersion and rotational velocity are nearly identical at all radii (Figure \ref{rcs}), and the spectral width of the eastern galaxy is poorly captured. So while this good fit highlights the ambiguity of close mergers and rotating disks, we conclude that this source is a triple merger.

\noindent
\textbf{DEIMOS COSMOS 848185 (DIS)}: Also known as HZ6 \citep{capa15} or LBG-1 \citep{pave19}, this source exhibits strong [CII] emission, but little evidence for rotation (Figure \ref{BBresults4}). However, \BB still returns an excellent fit, with a large velocity dispersion (see also Figure \ref{rcs}). Thus, we agree that this source is extended and dispersion-dominated.

\noindent
\textbf{DEIMOS COSMOS 873321 (MER)}: Originally described as a merger in \citet[called HZ8]{capa15}, DC873321 features two bright peaks of emission at the same velocity (Figure \ref{BBresults4}). \BB returns a poor fit with comparable rotation velocity and velocity dispersion, strengthening this classification of `merger'.

\noindent
\textbf{DEIMOS COSMOS 873756 (DIS)}: Even with the brightest \cii luminosity in the ALPINE sample, this source shows very little evidence for ordered rotation (Figure \ref{BBresults4}). Indeed, \BB is able to fit a dispersion-dominated model, resulting in only small residuals. The two PVDs are identical, with no evident gradient. There is some evidence for extended emission to the west at $\rm v\sim0$\,km\,s$^{-1}$, which is also present as blue residuals in the spectrum. It is not clear if this represents a minor merger, an outflow perpendicular to the line of sight, or another feature. Since the main galaxy dominates the field, we conclude that this source is extended and dispersion-dominated. 

\noindent
\textbf{DEIMOS COSMOS 881725 (ROT)}: With a strong velocity gradient and central velocity dispersion peak, this source is well fit by \BB, resulting in low spectral residuals (Figure \ref{BBresults5}). The moment one residuals are complex, but are mostly nonzero around the low-SNR perimeter. Thus, we agree with the \citet{lefe20} classification of a rotator.

\noindent
\textbf{VUDS COSMOS 5100537582 (UNC)}: This source shows little to no velocity gradient, and PVDs along both axes are nearly identical (Figure \ref{BBresults5}), earning it a \citet{lefe20} classification of 3 (extended dispersion-dominated). However, \BB fits it with a rotation-dominated model, and its W15 values are the same as those of DC733857, an UNC source. Visual inspection of the moment maps and PVDs shows that we lack the sensitivity and velocity resolution to determine the true morpho-kinematics of this source. Due to this ambiguity, we classify this source as `uncertain'.

\noindent
\textbf{VUDS COSMOS 5100541407 (UNC)}: With a significant extension to the northeast (Figure \ref{BBresults5}) and a number of W15 criteria failures, VC.1407 is difficult to characterize. \BB fits a moderately dispersion-dominated model, which results in significant residuals. Its extended morphology earned it a merger classification by \citet{lefe20}, but we classify it as `uncertain'.

\noindent
\textbf{VUDS COSMOS 5100559223 (UNC)}: This source is quite weak, as our SEARCH technique includes two single-channel clumps (Figure \ref{BBresults5}, similarly to DC454608). It was originally classified as an extended dispersion-dominated source, but it is poorly fit with a dispersion-dominated \BB model and shows a possible velocity gradient in the major axis PVD. Until higher-sensitivity observations are taken, we classify this source as `uncertain'.

\noindent
\textbf{VUDS COSMOS 5100822662 (MER)}: Similarly to DC873321, VC.2662 features two components at the same velocity (Figure \ref{BBresults6}). But unlike this other source, one component (which is the targeted galaxy) dominates the moment zero map. \BB returns a poor fit, adding further evidence that this is a merging system.

\noindent
\textbf{VUDS COSMOS 5100994794 (UNC)}: \BB returns a rotation-dominated model (see passed W15 criterion two), and the residuals are low, so it is possible that this source is a rotator, rather than the previous classification of an extended dispersion-dominated galaxy. In addition, the set of W15 criteria are identical to those of DC552206, a rotator. However, its major-axis PVD shows a broad region of red emission that dominates the velocity gradient (Figure \ref{BBresults6}), and while the model is rotation-dominated at small radii, it is dispersion-dominated at large radii. Thus, we classify it as `uncertain'

\noindent
\textbf{VUDS COSMOS 5101209780 (MER)}: The moment zero map of this source is qualitatively similar to that of VC.2662: a dominant source to the southwest, with a secondary source to the northeast (Figure \ref{BBresults1A}). However, here the targeted ALPINE source is the subdominant [CII] emitter. The [CII] emission is relatively well-fit with a rotation-dominated tilted ring model, but the discrepancies in each PVD reveal the underlying merger nature of this source. Its kinematics and other properties are investigated in depth in \citet{gino20b}.

\noindent
\textbf{VUDS COSMOS 5101218326 (DIS)}: Similarly to DC873756, this source is quite luminous in [CII], and the \BB fit strongly suggests an extended dispersion-dominated galaxy (Figure \ref{BBresults1A}). In addition, nearly all W15 criteria fail, strongly arguing against rotation. 

\noindent
\textbf{VUDS COSMOS 510786441 (UNC)}: With an elongated morphology and disturbed major axis PVD, VC.6441 was originally classified as a merger. However, \BB fits a marginally rotation-dominated model to this source (Figure \ref{BBresults6}), and the W15 criteria are similar to rotators in the sample (e.g., DC552206). Examination of the PVDs shows that the underlying morpho-kinematics are too complex to be classified as a rotator, but our observations are too low-resolution to unambiguously classify this as a merger. We thus classify it as `uncertain'.

\noindent
\textbf{VUDS COSMOS 5110377875 (ROT)}: This source is perhaps the best case of an ordered rotator in the current sample, but it is not quite ideal (Figure \ref{BBresults1A}). While the spectral and moment zero residuals are low, the velocity dispersion peak is not coincident with the intensity peak, the major axis PVD is not symmetric across $v=0$\,km\,s$^{-1}$, and the velocity field residuals are significant. As suggested by \citet{fuji20}, these discrepancies may be caused by interaction with an extended halo or an extended outflow. Despite these imperfections, we classify this as a rotator.

\noindent
\textbf{VUDS COSMOS 5180966608 (UNC)}: VC.6608 is one of the most mysterious sources in the ALPINE sample. On one hand, its spectrum is well fit by a high-velocity dispersion tilted ring model, suggesting a single, extended source (Figure \ref{BBresults6}). However, the two intensity peaks evident in the major axis PVD are nearly coincident with point sources in HST /ACS F814W, Subaru i+, and Subaru r+ images \citep{fais20}. These point sources may indicate the bright cores of closely interacting galaxies or may be caused by a dense screen of dust blocking the central rest-frame optical continuum emission. The FIR continuum emission peaks between these two point sources, but extends over both of them. High-resolution spectral line imaging of this source is required to determine its nature. Until then, we classify this as `uncertain'.

\noindent
\textbf{VUDS ECDFS 530029038 (UNC)}: Similarly to other sources, VE.9038 features a significant velocity offset between its [CII] and UV spectral line redshifts \citep{cass20}. Instead of shifting the [CII] profile to the edge of a sideband, this emission is shifted to the intersection of the two SPWs that compose the sideband (Figure \ref{BBresults7}). Due to the bandpass shape of ALMA SPWs, a few edge channels at the edge of each SPW must be flagged. This extra flagging occasionally creates gaps in frequency coverage between bands, as seen here, where the central few channels of the [CII] profile are flagged. These missing channels at v$\sim0$\,km\,s$^{-1}$ have an interesting effect, as the resulting moment 1 map is less smoothed in frequency, and the velocity gradient therefore appears more pronounced than it would with full velocity coverage. Despite this mostly cosmetic feature, \BB is able to fit a rotating disk model to the [CII] emission. However, the large velocity residual on the red half, and the incomplete velocity coverage mean that additional observations must be made before this source is characterized.

\section{Best-Fit Profiles of Rotators}\label{rottabsec}

\begin{table*}
\centering
\caption{Best-fit rotational velocities, velocity dispersions, and dynamical masses for each model ring, as output by \BB. In addition to the six confirmed rotators in the ALPINE sample, we also present the results of fitting the two previously detected $4<z<6$ main sequence unlensed rotators. Radius values are assumed ring radii (see Section \ref{PE}. Dynamical masses are estimated using equation \ref{mdeq}.}
\label{bbtab2}
\begin{tabular}{c|cccc}
 & R & $\rm v_{rot}$ & $\rm \sigma_v$ & $\rm M_{dyn}$ \\
Source & [kpc] & [km\,s$^{-1}$] & [km\,s$^{-1}$] & [$\rm M_{\odot}$] \\ \hline
CG32 & 1.17 & $117.43\pm20.71$ & $39.06\pm14.95$ & $(3.8\pm2.3)\times10^{9}$\\
 & 3.5 & $115.04\pm26.96$ & $19.14\pm12.47$ & $(1.1\pm0.5)\times10^{10}$\\\hline
DC396844 & 1.25 & $120.88\pm10.74$ & $44.42\pm7.75$ & $(4.2\pm2.3)\times10^{9}$\\
 & 3.75 & $80.42\pm17.66$ & $19.84\pm11.75$ & $(5.6\pm2.6)\times10^{9}$\\\hline
DC494057 & 1.04 & $78.44\pm10.25$ & $53.24\pm6.21$ & $(1.5\pm0.8)\times10^{9}$\\
 & 3.13 & $80.31\pm12.75$ & $44.98\pm7.1$ & $(4.7\pm1.7)\times10^{9}$\\\hline
DC552206 & 1.11 & $71.28\pm21.74$ & $116.65\pm11.85$ & $(1.3\pm1.0)\times10^{9}$\\
 & 3.32 & $164.98\pm20.23$ & $67.4\pm13.24$ & $(2.1\pm0.6)\times10^{10}$\\
 & 5.53 & $172.84\pm27.63$ & $65.34\pm14.61$ & $(3.8\pm1.3)\times10^{10}$\\\hline
DC881725 & 1.15 & $98.37\pm9.54$ & $40.45\pm6.94$ & $(2.6\pm1.4)\times10^{9}$\\
 & 3.44 & $62.07\pm12.54$ & $48.4\pm7.77$ & $(3.1\pm1.3)\times10^{9}$\\\hline
VC.7875 & 1.28 & $118.97\pm8.1$ & $47.04\pm5.69$ & $(4.2\pm2.2)\times10^{9}$\\
 & 3.84 & $102.85\pm19.84$ & $60.84\pm12.74$ & $(9.4\pm4.0)\times10^{9}$\\\hline
J0817 & 0.95 & $247.46\pm8.87$ & $32.17\pm11.36$ & $(1.4\pm0.7)\times10^{10}$\\
 & 2.84 & $252.09\pm14.94$ & $35.98\pm10.47$ & $(4.2\pm0.9)\times10^{10}$\\\hline
HZ9 & 0.54 & $155.85\pm16.84$ & $71.1\pm8.38$ & $(3.0\pm1.7)\times10^{9}$\\
 & 1.61 & $156.77\pm19.02$ & $75.12\pm9.22$ & $(9.2\pm2.7)\times10^{9}$\\
 & 2.68 & $176.63\pm25.45$ & $4.82\pm8.0$ & $(1.9\pm0.6)\times10^{10}$\\\hline
\bottomrule
\end{tabular}
\end{table*}

\end{document}